\documentclass{ecsarticle}     
\usepackage[brazilian]{babel}
\usepackage[latin1]{inputenc}
\usepackage[T1]{fontenc}
\usepackage{lscape}
\usepackage{graphicx,url}
\usepackage{graphics}
\usepackage[usenames,dvipsnames]{color,colortbl}
\usepackage{multirow}
\usepackage{graphicx}
\usepackage[inline]{enumitem}
\usepackage{tabularx}
\usepackage{multirow}
\usepackage[table]{xcolor}

\usepackage[none]{hyphenat} 
\emergencystretch=\maxdimen
\usepackage{natbib}            

\newcommand{\BibTeX}{{\rm B\kern-.05em{\sc i\kern-.025em b}\kern-.08em T\kern-.1667em\lower.7ex\hbox{E}\kern-.125emX}}

\newcounter{address}
\newcounter{alg}

\renewcommand\appendix{\par
  \setcounter{section}{0}
  \setcounter{subsection}{0}
  \setcounter{figure}{0}
  \setcounter{table}{0}
  \renewcommand\thesection{Apêndice \Alph{section}}
  \renewcommand\thefigure{\Alph{section}\arabic{figure}}
  \renewcommand\thetable{\Alph{section}\arabic{table}}
}

\begin{document}
\frontmatter
\title      {Caracterização do Estado da Arte sobre as Metodologias que utilizam como base a técnica DVFS Intra-Tarefa}
\authors    {
				\texorpdfstring{\href{mailto:rawlinson.goncalves@gmail.com}{Rawlinson S. Gonçalves}}{Rawlinson S. Gonçalves} e 
				\texorpdfstring{\href{mailto:rbarreto@icomp.ufam.edu.br}{Raimundo da Silva Barreto}}{Raimundo da Silva Barreto}
            }
\department {Instituto de Computação - IComp}
\group      {Grupo de Interesse em Sistemas Embarcados - GISE}
\university	{Universidade Federal do Amazonas - UFAM}
\addresses  {\groupname\\\deptname\\\univname}
\date       {\today}
\subject    {}
\maketitle

\begin{abstract}
Nos últimos anos tem havido uma crescente utilização de sistemas embarcados devido os avanços da tecnologia, a redução dos custos dos equipamentos eletrônicos e, principalmente, a popularização dos dispositivos móveis. Muitos desses sistemas implementam políticas de baixo consumo de energia para prolongar ao máximo a sua autonomia, pois possuem uma quantidade reduzida de recursos e a grande maioria deles são alimentados por baterias.
Um modo de minimizar o consumo de energia desses dispositivos são através das aplicações de técnicas de baixo consumo de energia. Dentre as inúmeras técnicas presentes na literatura, a técnica de escalonamento dinâmico de tensões e frequências (em inglês, \textit{Dynamic Voltage and Frequency Scaling} - DVFS) intra-tarefa tem desempenhado um papel importante, pois permite que cada tarefa gerencie os recursos mínimos necessários para que haja redução do consumo de energia do processador e seus \textit{deadlines} sejam respeitados, quando considerado um contexto de sistema de tempo real.
%
%
Portanto, este trabalho tem como objetivo principal a aplicação de uma revisão sistemática da literatura com o intuito de identificar e conhecer os principais métodos que utilizam a técnica DVFS intra-tarefa, aplicado no contexto de sistemas de tempo real, para reduzir o consumo de energia do processador. Por fim, serão exibidos relatórios contendo as principais características extraídas, assim como as vantagens e desvantagens de cada abordagem.
\end{abstract}




\mainmatter

\section{Introdução}
\label{Section:Introducao}

Nos últimos anos, o consumo de energia passou a ser uma métrica importante de qualidade para o projeto de sistemas embarcados. Assim, a otimização do consumo de energia tornou-se uma grande linha de pesquisa, principalmente devido à crescente demanda do mercado por melhorias na autonomia dos dispositivos embarcados móveis sem fio \citep{Cohen_2012_a}. Além disso, tem se tornado um dos principais fatores que podem decidir o valor de mercado do produto. Por outro lado, as pesquisas em otimização de energia não tem recebido investimentos suficientes, devido à sua crescente escala e complexidade  \citep{Takase_2011_a}. Isso vem ocorrendo devido a necessidade cada vez maior de incorporar novos recursos e tecnologias a estes tipos de dispositivo, enquanto que o desenvolvimento de novas técnicas de otimização do consumo não tem acompanhado esse crescimento. Exemplos desses recursos são: GPS(em inglês,\textit{Global Position System}), sensores de batimento cardíaco, câmeras mais robustas, sensores de temperatura, processadores com vários núcleos, entre outros.

Dentre todos esses recursos, o processador é um dos componentes que mais consomem recursos energéticos provenientes da bateria, o que implica dizer que quanto mais rápido é o processador, maior será o seu consumo de energia \citep{Yang_2009_a}. Isso ocorre, devido a maioria dos processadores utilizam a tecnologia CMOS (em inglês, \textit{Complementary Metal Oxide Semiconductor}), onde o consumo de energia ocorre principalmente durante os pulsos de \textit{clock} da CPU. Assim, a tensão aplicada sobre ele (e, correspondentemente, a frequência) está diretamente relacionada com o consumo de energia final \citep{Cohen_2012_a}.

Sendo assim, essa relação do consumo de energia fica mais clara quando analisamos a Equação~\ref{eq:Energia_Consumida_Processador}, que é um modelo simplificado do consumo de energia de um processador, mostrado no trabalho de \cite{Shin_Kim_2001_a}.

\begin{equation}
    E \propto C_{l}\;\times\;N_{cycle}\;\times\;V_{dd}^{2}
    \label{eq:Energia_Consumida_Processador}
\end{equation}

Onde \( C_{l} \) é a capacitância de carga, \( N_{cycle} \) é o número de ciclos executados e \( V_{dd}^{2} \) é a tensão fornecida. Analisando mais detalhadamente a  Equação~\ref{eq:Energia_Consumida_Processador}, temos que a tensão aplicada sobre o processador, por ser um termo quadrático, irá demandar bastante energia. Dessa forma, desenvolver um controle mais refinado sobre essa variável implicará diretamente na diminuição quadrática do consumo de energia do dispositivo \citep{Aboughazaleh_2003_b}. Esse argumento tem sido base para vários trabalhos presentes na literatura, principalmente para justificar o uso da técnica de escalonamento dinâmico de tensões e frequências (em inglês, \textit{Dynamic Voltage and Frequency Scaling} - (DVFS)).

As técnicas DVFS existentes são divididos em dois grupos: DVFS intra-tarefa e DVFS inter-tarefa. No primeiro, a tensão é ajustada dentro de limites individuais da própria tarefa, enquanto que o segundo, a tensão é ajustada tarefa por tarefa a cada instante de atuação do escalonado do sistema \citep{Tatematsu_2011_a}. No entanto, o foco dessa pesquisa está nas técnicas DVFS intra-tarefa.

O principio básico de funcionamento das técnicas DVFS intra-tarefa está na análise estática do fluxo de execução da aplicação, feita através do grafo de fluxo de controle (em inglês, \textit{Control-Flow Graph} - CFG) \citep{Insup_2008_a}. São através dessas análises que serão definidos os pontos do código que irão realizar os chaveamentos das tensões e frequências a serem aplicadas sobre processador  \citep{Shin_Kim_2001_a}. 

\begin{figure}[ht]
	\centering
	\includegraphics[scale=0.50]{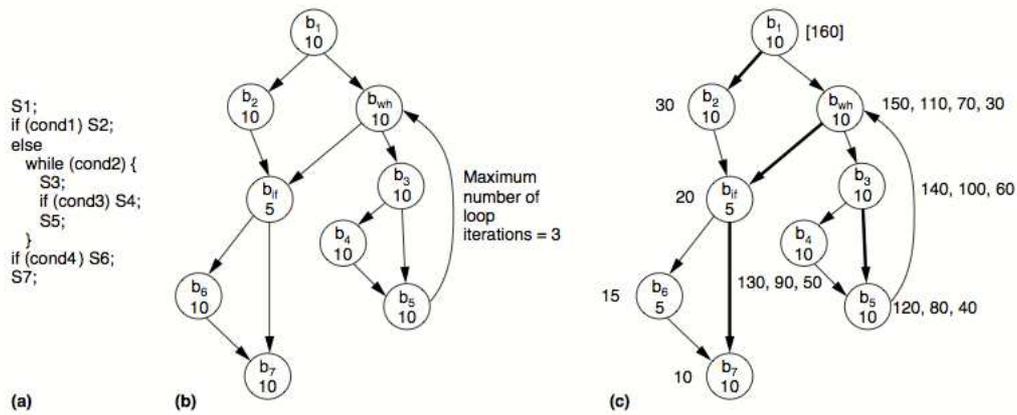}
	\caption{Exemplo de uma tarefa, onde: (a) Mostra o seu código fonte, (b) O CFG extraído a partir do código fonte e (c) Mostra o processo de análise da quantidade de ciclos de execução da tarefa no pior caso \citep{Shin_Kim_2001_a}.}
	\label{fig:Exemplo_Grafo_Fluxo_Controle}
\end{figure}

\begin{figure}[ht]
	\centering
	\includegraphics[scale=0.50]{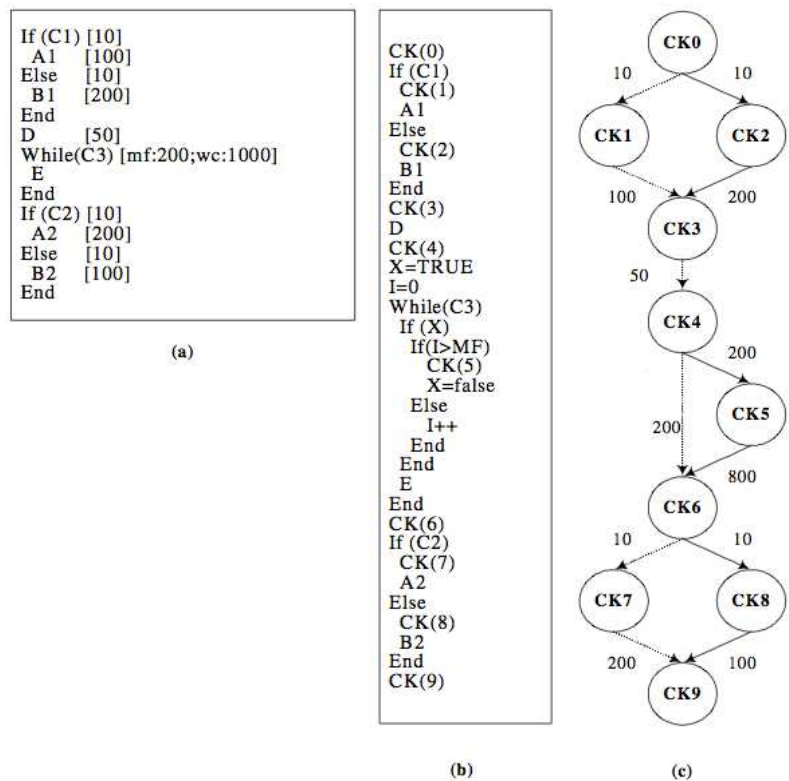}
	\caption{Exemplo de uma tarefa, onde: (a) Mostra o seu código fonte, (b) O CFG extraído a partir do código fonte e (c) Mostra o processo de inserção de pontos de controle \citep{Yi_Yang_Chen_2005_a}.}
	\label{fig:Exemplo_Grafo_Fluxo_Controle_2}
\end{figure}

As Figuras~\ref{fig:Exemplo_Grafo_Fluxo_Controle} e ~\ref{fig:Exemplo_Grafo_Fluxo_Controle_2} mostram exemplos de como são feitas as extrações do grafo de fluxo de controle de uma aplicação a partir do seu código fonte, a análise da quantidade de ciclos necessários para a execução da tarefa no seu pior caso e um exemplo da inserção de pontos de controle.

A partir das análises do código e do grafo de fluxo de controle, vários trabalhos deram significativas contribuições, principalmente quanto a metodologias de inserção de pontos de controle no condigo fonte da tarefa, com o intuito diminuir o consumo de energia do processador \citep{Takase_2011_a, Tatematsu_2011_a, Ishihara_2009_a, Yi_Yang_Chen_2005_a}. Essas metodologias trabalham em tempo de compilação, aumentando significativamente a complexidade e a inserção de \textit{overheads} dentro da aplicação \citep{Chen_Hsieh_Lai_2008_b}. A grande problemática nessas metodologias, que utilizam apenas a técnica DVFS intra-tarefa, está em dar suporte a preempções, visto que alguns trabalhos presentes na literatura consideram somente modelos de tarefas não preemptivos\footnote{Segundo \cite{Tanenbaum_2001}, preempção é o recurso que permite ao sistema operacional melhor gerenciar as fatias de tempo do processador entre os processos que estão em execução no sistema.} \citep{Oh_Kim_Kim_Kyung_2008_a}. Geralmente, as metodologias que dão suporte a preempção, necessitam incorporar outras técnicas, como por exemplo a técnica DVFS inter-tarefa \citep{Cohen_2012_a, Takase_2011_a, Chen_Hsieh_Lai_2008_a, Chen_2008_a, Zitterell_2008_a, Xian_Lu_2006_a}.


Diante dos conceitos apresentados, decidimos investigar na literatura trabalhos / pesquisas relevantes que utilizem a técnica DVFS intra-tarefa como base para a construção de suas metodologias. Para alcançar resultados com valor científico, foi decidido realizar uma revisão sistemática baseado no trabalho \cite{Kitchenham_2004}, que introduziu o conceito em Engenharia de Software Baseada em Evidência (ESBE). Essa abordagem surgiu na medicina e foi trazida para a engenharia de software com o objetivo de fornecer meios pelos quais as melhores evidências atuais de pesquisa pudessem ser integradas com experiências práticas e valores humanos no processo decisório relativo ao desenvolvimento e manutenção de software. Uma revisão sistemática ``é um meio de identificar, avaliar e interpretar toda pesquisa disponível e relevante sobre uma questão de pesquisa, um tópico ou um fenômeno de interesse, e tem por objetivo apresentar um avaliação justa de um tópico de pesquisa, usando uma metodologia confiável, rigorosa e auditável'' \citep{Kitchenham_2004}.

A aplicação da revisão sistemática da literatura requer que seja seguido um conjunto bem definido e sequencial de passos, segundo um protocolo de pesquisa desenvolvido apropriadamente. Este protocolo é construído considerando um tema específico que representa o elemento central da investigação. Os passos da pesquisa, as estratégias definidas para coletar as evidências e o foco das questões de pesquisa são definidos explicitamente, de tal forma que outros pesquisadores sejam capazes de reproduzir o mesmo protocolo de pesquisa e, também, de julgar a adequação dos padrões adotados no estudo \citep{biolchini_2005}.

Em razão disso, foi conduzido uma revisão sistemática com o objetivo de identificar e conhecer as metodologias que utilizam a técnica DVFS intra-tarefa, dentro do contexto de sistemas de tempo real, para diminuir o consumo de energia do processador.

\textbf{Organização do trabalho.} Este relatório apresenta esta revisão e discute seus resultados. O texto está dividido em seis seções. A Seção~\ref{section:Revisao_Sistematica} abordará sobre a metodologia utilizada. A Seção~\ref{section:Planejamento_Revisao_Sistematica} relata o planejamento da revisão e o protocolo preparado para a mesma. A Seção~\ref{section:Conducao_Revisao_Sistematica} descreve como esta revisão foi conduzida, enquanto que a Seção~\ref{section:Analise_Resultados_Revisao_Sistematica} apresenta e discute a análise e publicação dos resultados. A Seção~\ref{section:consideracoes_finais} apresentam as considerações finais. Além destas seis seções, são apresentados três apêndices. O \ref{section:Apendice_Conducao_Expressao_Busca} como se deu o processo de construção da expressão de busca, descreve a lista de publicações tomadas como base para validação da expressão de busca (também chamada de lista de controle), além de mostrar informações adicionais sobre o processo de condução da revisão sistemática. O \ref{section:Apendice_Filtros_1_e_2} mostram as publicações selecionadas após a execução do 1\textordmasculine e 2\textordmasculine filtros. Por fim, o \ref{section:Apendice_Base_Dados_Revisao_Sistematica} mostra a base de dados, da revisão sistemática, criada a partir dos dados extraídos das publicações selecionadas após a execução do 2\textordmasculine \: filtro.

\section{Revisão Sistemática}
\label{section:Revisao_Sistematica}

A revisão sistemática requer um esforço considerável quando comparado com uma revisão de literatura informal. Enquanto que a revisão de literatura informal é conduzida de forma \textit{ad-hoc}, sem planejamento e critérios de seleção estabelecidos sem nenhuma metodologia pré-definida, a revisão sistemática segue um protocolo formal para conduzir uma pesquisa sobre um determinado tema, com uma sequência bem definida de passos metodológicos \citep{mafra_2006}.

O processo para a condução de revisões sistemáticas envolve três etapas \citep{mafra_2006}:

\begin{enumerate}
\item \textbf{Planejamento da Revisão}: os objetivos da pesquisa são listados e o protocolo da revisão é definido.

\item \textbf{Condução da Revisão}: nesta atividade, as fontes para a revisão sistemática são selecionadas, os estudos primários são identificados, selecionados e avaliados de acordo com os critérios de inclusão, exclusão e de qualidade estabelecidos durante o protocolo da revisão.

\item \textbf{Análise dos Resultados}: os dados dos estudos são extraídos e sintetizados para análise e apresentação dos resultados.
\end{enumerate}

Conduzimos a revisão sistemática deste trabalho baseado nas três etapas citadas anteriormente e de acordo com as diretrizes definidas por \cite{biolchini_2005}, \cite{mafra_2006} e \cite{Kitchenham_2004}. Entretanto, como o objetivo deste trabalho é realizar um estudo exploratório de caracterização do estado da arte, podemos
dizer que esta revisão sistemática se caracteriza como uma quasi-sistemática \citep{Travassos_2008}, pois segue o mesmo processo da revisão sistemática e preserva o rigor e mesmo formalismo para as fases metodológicas de elaboração de protocolo e execução da revisão, mas sem a aplicação de uma meta-análise a princípio, que pode ser aplicada posteriormente.


\section{Planejamento da Revisão Sistemática}
\label{section:Planejamento_Revisao_Sistematica}

O protocolo utilizado para o estudo foi derivado do trabalho produzido por \cite{Santos_2008} e \cite{Kitchenham_2007}. Para cada uma das subseções a seguir serão apresentados o que se espera a partir do protocolo (texto em itálico e entre chaves. Exemplo: "\textit{\{Itálico\}}") e o conteúdo de fato utilizado no estudo em questão.

\subsection{Contexto}
\textit{\{Descrever um breve relato sobre o problema que motivou a realização do estudo; delimitar o problema; identificar o que é importante e o que está fora do escopo; justificar a necessidade de conduzir o estudo para tratar o problema apresentado.\}}

Atualmente, dentre as várias técnicas voltadas para o baixo consumo de energia, a técnica DVFS intra-tarefa é uma das que mais são citadas no meio científico. Portanto, o intuito deste trabalho é de mapear todas as metodologias presentes na literatura, que utilizam a técnica DVFS intra-tarefa para minimizar o consumo de energia do processador dentro do contexto de sistemas de tempo real. A partir desse mapeamento será possível fazer um relatório para cada uma metodologia catalogada, a fim de se obter uma base de dados consistente e atualizada sobre o estado da arte.

\subsection{Objetivo}
\textit{\{Descrever o objetivo do estudo a partir do paradigma GQM (do inglês Goal, Question, and Metric) \citep{Basili_1994}.\}}

A Tabela~\ref{tabela_GQM} mostra o objetivo de estudo a partir do paradigma GQM.

\begin{table}[htbp]
	\centering
	\caption{Objetivo do estudo utilizando o paradigma GQM.}
	\begin{tabular}{|l|p{8cm}|}
	\hline \textbf{Analisar} & Publicações cientificas através de um estudo baseado em revisão sistemática. \\ 
	\hline \textbf{Com o propósito de} & Identificar as técnicas / metodologias que diminuem o consumo de energia do processador. \\ 
	\hline \textbf{Com relação as} & Aplicação da técnica DVFS intra-tarefas. \\ 
	\hline \textbf{Do ponto de vista do} & Pesquisador \\ 
	\hline \textbf{No contexto} & Acadêmico ou industrial voltado para o baixo consumo de energia em sistemas de tempo real. \\ 
	\hline 
	\end{tabular} 
	\label{tabela_GQM}
\end{table}

\subsection{Questões de Pesquisa}
\label{subsection:Questoes_Pesquisa}

\textit{\{Identificar que questões serão respondidas a partir da identificação e caracterização do objeto de estudo. Ou seja, uma vez identificados / caracterizados os objetos de estudo, que questões relevantes ao problema descrito poderão ser respondidas / discutidas?\}}

Buscamos respostas para a seguinte pergunta:

\begin{itemize}
  \item Q1: Quais são as metodologias que utilizam como base a técnica DVFS intra-tarefa para reduzir o consumo de energia do processador dentro do contexto de sistemas de tempo real?
\end{itemize}

\subsection{Escopo da Pesquisa}

\textit{\{Delimitar os tipos de mecanismos que serão utilizados para realizar as buscas, por exemplo, bibliotecas digitais através dos seus respectivos engenhos de busca, bibliotecas setoriais, livros, catálogo especializado de produtos etc.\}}

Para delinear o escopo da pesquisa foram estabelecidos critérios para garantir, de forma equilibrada, a viabilidade da execução (custo, esforço e tempo), acessibilidade aos dados e abrangência do estudo. A pesquisa dar-se-á a partir de bibliotecas digitais através das suas respectivas máquinas de busca e, quando os dados não estiverem disponíveis eletronicamente, através de consultas manuais.

\subsubsection{Critérios Adotados para Seleção das Fontes}

Para as bibliotecas digitais é desejado:

\begin{itemize}
\item Possuir máquina de busca que permita o uso de expressões lógicas ou mecanismo equivalente;
\item Incluir em sua base publicações da área de exatas ou correlatas que possuam relação direta com o tema
a ser pesquisado;
\item As máquinas de busca deverão permitir a busca no texto completo das publicações.
\end{itemize}

Além disso, os mecanismos de busca utilizados devem garantir resultados únicos através da busca de um mesmo conjunto de palavras-chave (ou expressão de busca). Quando isto não for possível, deve-se estudar e documentar uma forma de minimizar os potenciais efeitos colaterais desta limitação.

\subsubsection{Restrições}
\textit{\{Identificar todas as restrições associadas ao estudo. Identificar o intervalo de tempo válido para as buscas. O acesso aos dados, em geral, não deve incorrer em ônus para a pesquisa.\}}

A pesquisa está restrita à análise de publicações obtidas, exclusivamente, a partir das fontes selecionadas com base nos critérios supracitados.

\subsection{Idiomas}
\textit{\{Deve-se identificar os idiomas das publicações que serão aceitas para a pesquisa. Se possível, deve-se justificar essa escolha.\}}

Para a realização desta pesquisa foi selecionado apenas o idioma Inglês. A escolha do idioma Inglês deve-se à sua adoção pela grande maioria das conferências e periódicos internacionais relacionados como tema de pesquisa e por ser o idioma utilizado pela maioria das editoras relacionadas com o tema listadas no Portal de Periódicos da CAPES (Coordenação de Aperfeiçoamento de Pessoal de Nível Superior).

\subsection{Métodos de Busca das Publicações}
\label{subsubsection:Metodos_Busca_Publicacoes}
\textit{\{Deve-se descrever a forma de busca (manual e/ou eletrônica) além da expressão de busca: expressão lógica contendo uma combinação de palavras chaves extraída do objetivo do estudo relacionada ao objeto de estudo, características de interesse e respectivos sinônimos.\}}

As fontes digitais foram acessadas via \textit{Web}, através de expressões de busca pré-estabelecidas. A biblioteca digital consultada foi a Scopus, acessível em \textit{http://www.scopus.com}. Segundo a editora Elsevier (2013) \citep{scopus_2013}, a Scopus é uma das maiores bases de dados de resumos e citações da literatura de pesquisa \textit{peer-reviewed} com mais de 20.500 títulos de mais de 5.000 editoras internacionais.

Dentre estas editoras podemos citar: Springer \citep{Springer_2013}; IEEE Xplore Digital Library \citep{IEEE_2013}; ACM Digital Library \citep{ACM_2013}; ScienceDirect/Elsevier \citep{ElsevierBV_2013}; Wiley Online Library \citep{Wiley_2013}; British Computer Society \citep{BCS_2013}; dentre outras. A biblioteca Scopus também inclui aproximadamente 5.3 milhões de conferências de artigos de \textit{proceedings} e \textit{journals}, 400 publicações comerciais, 360 série de livros e publicações aceitas são disponibilizadas online antes da publicação oficial em mais de 3.850 periódicos. Ainda segundo a editora Elsevier (2013) \citep{scopus_2013}, a Scopus tem aproximadamente 2 milhões de novas gravações adicionadas a cada ano, com atualizações diárias.

\subsubsection{Expressão de Busca}
\label{subsubsection:Expressao_Busca}
\textit{\{Descrever a expressão de busca que será adotada para a seleção das publicações nas máquinas de busca.\}}

A expressão de busca foi definida segundo o padrão \textbf{PICO} (do inglês \textit{Population, Intervention, Comparison, Outcomes}) \citep{Kitchenham_2007}, conforme a estrutura abaixo:

\begin{itemize}
\item \textbf{População}: Trabalhos publicados em conferências e periódicos que sejam aplicados no contexto de sistemas de tempo real;

\item \textbf{Intervenção}: Todas as metodologias que utilizem a técnica DVFS intra-tarefa;

\item \textbf{Comparação}: Não se aplica.

\item \textbf{Resultados}: A partir dos relatos das metodologias identificadas, pretende-se mapear o estado da arte na área de baixo consumo de energia do processador, aplicado no contexto de sistemas de tempo real. Além disso, responder as questões de pesquisa propostas neste levantamento bibliográfico.
\end{itemize}

Como este estudo representa um mapeamento / caracterização, a expressão de busca (para execução na biblioteca digital Scopus, como mencionado anteriormente) foi definida de acordo com dois aspectos: População e Intervenção \citep{Kitchenham_2007}, como é apresentado na estrutura abaixo.

\begin{itemize}
\item \textbf{População}: Publicações que fazem referências a sistemas de tempo real (e sinônimos):
	\begin{itemize}
	\item \textbf{Palavras-Chave}: (``hard real-time'' OR ``soft real-time'' OR ``real-time system'' OR ``real time system'' OR ``real-time application'' OR ``real time application'' OR ``real-time embedded system'' OR ``real time embedded system'')
	\end{itemize}

\item \textbf{Intervenção}: Técnica DVFS Intra-Tarefa (e sinônimos):
	\begin{itemize}
	\item \textbf{Palavras-Chave}: (``DVFS'' OR ``dynamic voltage and frequency scaling'' OR ``dynamic voltage frequency scaling'' OR ``DVS'' OR ``dynamic voltage scaling'' OR ``DFS'' OR ``dynamic frequency scaling'' OR ``voltage scheduling'' OR ``frequency scheduling'' OR ``frequency scaling'' OR ``voltage scaling'' )
	AND
	(``intra-task'' OR ``intra task'')
	\end{itemize}
\end{itemize}

\textbf{OBS:} Antes da definição da expressão de busca apresentada, alguns testes foram conduzidos de forma a tentar garantir que a expressão de busca escolhida estivesse de acordo com o objetivo e a questão do estudo. Isso foi feito com o auxílio de artigos selecionados previamente para compor uma lista de controle, que são a lista de artigos mais relevantes na área e que devem ser, obrigatoriamente, localizados a partir da execução da expressão de busca. O \ref{section:Apendice_Conducao_Expressao_Busca} mostram os artigos escolhidos para compor a lista de controle desta revisão sistemática e como se deu o processo de construção da expressão de busca.

\subsection{Procedimentos de Seleção e Critérios}
\label{subsection:Procedimentos_Selecao_Criterios}
\textit{\{Deve-se descrever os procedimentos para seleção das publicações, incluindo procedimentos de avaliação da inclusão de publicações no escopo da pesquisa e critérios de inclusão e exclusão.\}}

A estratégia de busca foi aplicada por um pesquisador para identificar as publicações em potencial. As publicações identificadas serão selecionadas pelos demais pesquisadores (incluindo o que fará a busca) através da verificação dos critérios de inclusão e exclusão e de qualidade estabelecidos. Os pesquisadores deverão entrar em consenso sobre a seleção das publicações cujas avaliações se mostrem conflitantes.

Em caso de impasse entre os pesquisadores, a publicação deverá ser incluída na lista de selecionadas. Para diminuir o risco que uma publicação seja excluída prematuramente em uma das etapas do estudo, sempre que existir dúvida a publicação não deverá ser excluída.

\subsubsection{Procedimento de Seleção}
\label{subsubsection:Procedimento_Selecao}
\textit{\{Identificar as etapas necessárias para seleção das publicações para o estudo.\}}

A seleção das publicações dar-se-á em 3 etapas:

\begin{enumerate}
\item \textbf{Seleção e catalogação preliminar dos dados coletados}. A seleção preliminar das publicações será feita a partir da aplicação da expressão de busca às fontes selecionadas. Cada publicação será catalogada em um banco de dados criado especificamente para este fim e armazenada em um repositório para análise posterior;

\item \textbf{Seleção dos dados relevantes - [1\textordmasculine \, filtro]}. A seleção preliminar com o uso da expressão de busca não garante que todo o material coletado seja útil no contexto da pesquisa, pois a aplicação das expressões de busca são restritas ao aspecto sintático. Dessa forma, após a identificação das publicações através dos mecanismos de buscas, deve-se ler o título, os resumos (ou \textit{abstracts}), as palavras-chave e analisá-los seguindo os critérios de inclusão e exclusão identificados a seguir. Neste momento, poder-se-ia classificar as publicações apenas quanto aos critérios de exclusão, entretanto, para facilitar a análise e reduzir o número de publicações das quais se possam ter dúvidas sobre sua aceitação, deve-se também classificá-las quanto aos critérios de inclusão. Devem ser excluídas as publicações contidas no conjunto preliminar que:

	\begin{itemize}
	\item \textbf{CE1-01}: Não serão selecionadas publicações que não estejam relacionados com a área de Computação.
	\item \textbf{CE1-02}: Não serão selecionadas publicações cujo os artigos não estejam disponíveis na internet.
	\item \textbf{CE1-03}: Não serão selecionadas publicações em que descrevam e/ou apresentam \textit{Keynote Speeches}, tutoriais, cursos e similares.
	\item \textbf{CE1-04}: Não serão selecionadas publicações que não fizerem referências à baixo consumo de energia do processador.
	\item \textbf{CE1-05}: Não serão selecionadas publicações que utilizem técnicas de baixo consumo de energia que não estejam aplicadas no contexto de sistemas de tempo real.
	\item \textbf{CE1-06}: Não serão selecionadas publicações que não utilizem o recurso DVFS do processador.
	\item \textbf{CE1-07}: Não serão selecionadas publicações que simulem técnicas de baixo consumo de energia já existentes e / ou já demonstrada em outros trabalhos.
	\end{itemize}

	Podem ser incluídas apenas as publicações contidas no conjunto preliminar que:

	\begin{itemize}
	\item \textbf{CI1-01}: Serão selecionadas publicações que citam uma técnica de baixo consumo de energia aplicada no contexto de sistemas de tempo real e que utilize o recurso DVFS do processador.
	\end{itemize}
	
\item \textbf{Seleção dos dados relevantes - [2\textordmasculine \, filtro]}. O objetivo deste 2\textordmasculine \, filtro é identificar quais artigos que proponham técnicas de baixo consumo de energia do processador dentro do contexto de sistemas de tempo real através da utilização da técnica DVFS intra-tarefa. Apesar de limitar o universo de busca, o 1\textordmasculine \, filtro não garante que todo o material coletado seja útil no contexto da pesquisa. Por isso, após a leitura na íntegra dos artigos selecionados no 1\textordmasculine \, filtro, deve-se verificar que as publicações excluídas neste filtro respeitem os critérios abaixo:
	
	\begin{itemize}
	\item \textbf{CE2-01 [-SAVE\_ENER \& -DVFS\_INTRA\_TASK]}: Não devem ser selecionadas publicações que não contextualizem metodologias de baixo consumo de energia e que não utilizem a técnica DVFS intra-tarefa.
	\item \textbf{CE2-02 [+SAVE\_ENER \& -DVFS\_INTRA\_TASK]}: Não devem ser selecionadas publicações que contextualizem metodologias de baixo consumo de energia, mas não utilizem a técnica DVFS intra-tarefa.
	\item \textbf{CE2-03 [-SAVE\_ENER \& +DVFS\_INTRA\_TASK]}: Não devem ser selecionadas publicações que não contextualizem metodologias de baixo consumo de energia, mesmo utilizando a técnica DVFS intra-tarefa.
	\end{itemize}

	Dessa forma, todas as publicações incluídas neste filtro devem respeitar o critério abaixo:

	\begin{itemize}
	\item \textbf{CI2-01 [+SAVE\_ENER \& +DVFS\_INTRA\_TASK]}: Devem citar uma metodologia de baixo consumo de energia do processador utilizando a técnica DVFS intra-tarefa.
	\end{itemize}
\end{enumerate}

\subsection{Procedimentos para Extração dos Dados}
\label{subsection:Procedimentos_Extracao_Dados}
\textit{\{Identificar os procedimentos para extração de dados a partir das publicações.\}}

\begin{enumerate}[label=\bfseries \alph*.]
  \item \textbf{Na Seleção e Catalogação Preliminar dos Dados Coletados}
  
Armazenamento das referências completas selecionadas a partir da fonte consultada no repositório de dados do estudo.
  
  \item \textbf{Na Seleção dos Dados Relevantes}
  \label{item:Procedimentos_Extracao_Dados_Selecao_Dados_Relevantes}
  
	Na seleção dos dados mais relevantes para a caracterização das metodologias que utilizam a técnica DVFS intra-tarefa foi obtida primeiramente a partir da aplicação dos critérios de inclusão e exclusão definidos no primeiro e segundo filtros (ver Seção~\ref{subsubsection:Procedimento_Selecao}). Em seguida, com a definição das publicações mais relevantes para a pesquisa, procuramos extrair as informações mais importantes a partir das respostas das seguintes questões:

	\begin{itemize}
		\item Foi desenvolvida alguma ferramenta para a aplicação do método? Caso positivo, ela está disponível?
		\item O método proposto foi gerado a partir da integração com outros?
		\item A execução do método é feita de forma \textit{online, offline} ou híbrida?
		\item Quais foram os resultados positivos ou negativos da validação / experimentação do método?
		\begin{itemize}
			 \item Foi utilizado algum Benchmark para experimentação do método? Caso positivo, este Benchmark esta disponível?
		\end{itemize}
		\item Quais as limitações do método proposto?
		\item Quais as perspectivas futuras para melhoria da aplicação do método proposto?
	\end{itemize}

  \item \textbf{Extração de Dados}

	Ao final da realização da revisão sistemática, os dados baixo deverão ser extraídos de cada uma das metodologias catalogadas. O preenchimento dos itens dessa seção é obrigatório quando for considerado de interesse para o estudo, a única exceção será quando não houver a informação solicitada. Esses dados foram definidos com base no item~\ref{item:Procedimentos_Extracao_Dados_Selecao_Dados_Relevantes} do procedimento para extração dos dados.

	\begin{itemize}
	\item Dados da publicação:
		\begin{itemize}
		\item Título;
		\item Autor(es);
		\item Palavras-chave;
		\item Fonte de publicação;
		\item Ano de publicação.
		\end{itemize}
	
	\item Resumo da publicação:
		\begin{itemize}
		\item Uma breve descrição do estudo.
		\end{itemize}
	
	\item Dados derivados das características de interesse declaradas nas questões de pesquisa:
		\begin{itemize}
		\item Método(s) utilizados: técnicas e/ou métodos utilizados;
		\item Ferramenta(s): caso tenha sido desenvolvido alguma ferramenta para comprovar os resultados experimentais da metodologia proposta;
		\item Impacto (positivo x negativo): indicação dos pontos positivos e negativos da metodologia proposta;
		\item Validação do método: descreve como se deu o processo de validação da metodologia proposta;
		\item Limitações do método: por exemplo, se a metodologia proposta dar suporte a preempções.
		\end{itemize}
	
	\item Dados para um melhor entendimento dos resultados:
		\begin{itemize}
		\item Integração de métodos: se o método foi gerado a partir da integração com outro(s) método(s);
		\item Modo de aplicação do método: se a metodologia é executada de forma \textit{online, offline} ou híbrida;
		\item Perspectivas futuras: questão de pesquisa sugerida como trabalhos futuros, se houver alguma.
		\end{itemize}
	
	\item Comentários adicionais do pesquisador.
	\end{itemize}

  \item \textbf{Sumarização dos Resultados}
  
	Os resultados serão tabulados. 
\end{enumerate}

\subsection{Procedimentos para Análise}
\label{subsection:Procedimentos_Analise}
\textit{\{Identificar os procedimentos para análise dos dados coletados. Incluir totalização das mais diversas e relevantes para o objetivo do estudo e questões de pesquisa.\}}

\begin{enumerate}[label=\bfseries \alph*.]

  \item \textbf{Análise Quantitativa}
  
	A análise quantitativa dar-se-á pela extração direta dos dados a partir do banco de dados que tem como finalidade fornecer:
	
	\begin{itemize}
	\item O número de publicações selecionadas para fazerem parte do estudo;
	
	\item O número de publicações aplicadas a cada um dos critérios de inclusão e exclusão utilizados no primeiro e segundo filtros.
	
	\item O número de publicações retornadas na expressão de busca e agrupadas por ano, para se ter uma visão do interesse da comunidade científica pela área ao longo dos anos.
	
	\item O número de publicações por editora.
	
	\item A quantificação das metodologias catalogadas quanto aos seus métodos de execução.
	
	\item A quantificação das metodologias catalogadas quanto a disponibilidade ferramental.
	\end{itemize}

  \item \textbf{Análise Qualitativa}
  
	A análise qualitativa deverá utilizar como base, os dados quantitativos e realizar considerações com o intuito de discutir os achados com relação às questões de pesquisa declaradas.
\end{enumerate}

\section{Condução da Revisão Sistemática}
\label{section:Conducao_Revisao_Sistematica}

%
%
%

A execução da revisão sistemática ocorreu no período de novembro a dezembro de 2013 e as publicações foram selecionadas de acordo com os critérios de inclusão e exclusão estabelecidos na Seção~\ref{section:Planejamento_Revisao_Sistematica}.

A expressão de busca mostrada na Seção~\ref{subsubsection:Expressao_Busca} foi executada na máquina de busca da biblioteca Scopus, como definido anteriormente. Contudo, vale ressaltar que a expressão de busca foi modificada algumas vezes (contendo um total de 7 versões). As modificações foram necessárias devido a dois fatores: o primeiro, um grande número de publicações retornadas pela máquina de busca, contabilizando em alguns momentos um total de mais de 21.172 publicações; e o segundo, alguns artigos da lista de controle (lista de artigos já conhecidos e utilizados como base de referência) não estavam sendo retornados pelas máquinas de busca.

A principal melhoria feita na expressão de busca foi a utilização de um ``\textit{AND}'' adicional na intervenção, baseado no trabalho de \cite{Barmi_2011}. Dessa forma é possível obter resultados mais relevantes, além de filtrar melhor os trabalhos que não tem relação com as questões de pesquisa. Segue abaixo a expressão de busca final utilizada na máquina de busca da Scopus, onde foi feito a união da População com a Intervenção e foi adicionado as sub-áreas do conhecimento de interesse para o mapeamento das publicações mais relevantes, onde as sub-áreas selecionadas foram: Computação, Engenharia e Energia.

\textit{(ALL(``hard real-time'' OR ``soft real-time'' OR ``real-time system'' OR ``real time system'' OR ``real-time application'' OR ``real time application'' OR ``real-time embedded system'' OR ``real time embedded system'') AND ALL(``DVFS'' OR ``dynamic voltage and frequency scaling'' OR ``dynamic frequency scaling technique'' OR ``dynamic voltage frequency scaling'' OR ``DVS'' OR ``dynamic voltage scaling'' OR ``DFS'' OR ``dynamic frequency scaling'' OR ``voltage scheduling'' OR ``frequency scheduling'' OR ``frequency scaling'' OR ``voltage scaling"OR ``dynamic power management'' OR ``dynamic voltage'' OR ``dynamic frequency'' OR ``frequency control'' OR ``frequency scaling'' OR ``voltage control'' OR ``processor frequency'' OR ``processor voltage'' OR ``voltage-clock scaling'' OR ``voltage clock scaling'') AND ALL(``intra-task'' OR ``intra task'')) AND ( LIMIT-TO(SUBJAREA, ``COMP'') OR LIMIT-TO(SUBJAREA, ``ENGI'') OR LIMIT-TO(SUBJAREA, ``ENER''))}

Todas as publicações recuperadas pela máquina de busca foram organizadas pelo gerenciador de referências bibliográficas Mendeley\footnote{Ferramenta Mendeley Desktop versão 1.7.1. Mais informações sobre essa ferramenta, acessar o site: http://www.mendeley.com/download-mendeley-desktop/.}. O Mendeley permitiu a indexação dos itens, ou seja, criou uma lista com os nomes e outras informações para pesquisas instantâneas. Ele ainda possui um rastreador automático de referências internas nos documentos, campos de pesquisa e filtros detalhados, que possibilitam a anotação participativa e identificação de repetições.

Após essas modificações, iniciamos as análises quantitativas desta revisão sistemática, onde após a execução da expressão de busca definida acima, foi possível tabelar os resultados das publicações identificadas pela máquina de busca, bem como o número de publicações aceitas em cada um dos filtros executados (ver Tabela~\ref{tabela:resultado_revisao_sistematica}).

\begin{table}[htp]
	\centering
	\caption{Resultados gerais das publicações identificadas pela máquina de busca da Scopus.}
	\begin{tabular}{|>{\centering\arraybackslash}p{2cm}|>{\centering\arraybackslash}p{3cm}|>{\centering\arraybackslash}p{3cm}|>{\centering\arraybackslash}p{3cm}|}
	\hline \textbf{Máquina de Busca} & \textbf{Número Total de Publicações} & \textbf{Publicações Selecionadas Após o Primeiro Filtro} & \textbf{Publicações Selecionadas Após o Segundo Filtro} \\ 
	\hline Scopus & 253 & 115 & 39 \\ 
	\hline 
	\end{tabular} 
	\label{tabela:resultado_revisao_sistematica}
\end{table}

\begin{figure}[htp]
	\centering
	\resizebox{17cm}{!}{\includegraphics{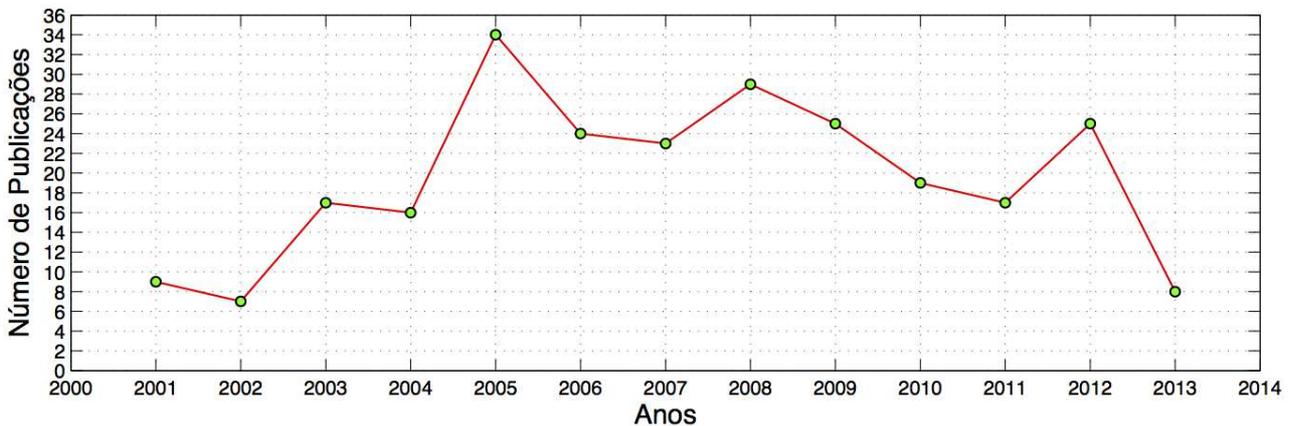}}
	\caption{Número de publicações por ano.}
	\label{fig:Numero_Publicacoes_Ano}
\end{figure}

A partir do número total de publicações foi possível fazer a análise quantitativa em relação ao interesse da comunidade cientifica na linha de pesquisa definida nesta revisão sistemática. A Figura~\ref{fig:Numero_Publicacoes_Ano} apresenta um gráfico com a visão geral do número total de publicações retornadas pela biblioteca e agrupadas por ano. Com base nesta informação podemos observar que por volta de 2005 houve um declínio no número de publicações, demonstrando assim que área está chegando ao seu ponto de saturação, onde propor novas contribuições está sendo cada vez mais desafiador.


\begin{figure}[htp]
	\centering
	\resizebox{17cm}{!}{\includegraphics{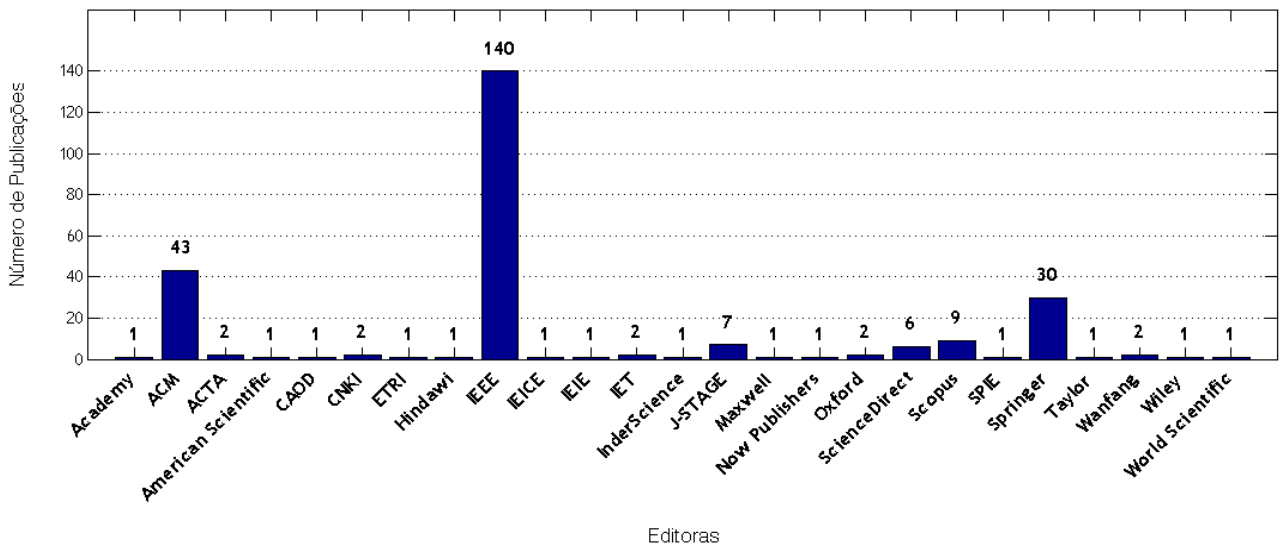}}
	\caption{Número de publicações por editora.}
	\label{fig:Numero_Publicacoes_Editora}
\end{figure}

Vale ressaltar que as 253 publicações retornadas pela Scopus foram extraídas de 25 diferentes editoras, tais como: IEEE, Springer e ACM. A Figura~\ref{fig:Numero_Publicacoes_Editora}, detalha o número de publicações por editoras e na Tabela~\ref{tabela:Nome_Completo_Editoras} é possível consultar o nome completo das editoras catalogadas.

\begin{table}[htp] 
	\centering
	\caption{Nome completo das editoras sem abreviações.}

	\begin{tabular}{|l|l|}
	\hline
	\textbf{Abreviação / Sigla} & \textbf{Nome Completo da Editora} \\ \hline
	Academy & Academy Publisher \\ \hline
	ACM & The Association for Computing Machinery \\ \hline
	ACTA & ACTA Press \\ \hline
	American Scientific & American Scientific Publishers \\ \hline
	CAOD & China/Asia On Demand \\ \hline
	CNKI & National Knowledge Infrastructure \\ \hline
	ETRI & Electronics and Telecommunications Research Institute \\ \hline
	Hindawi & Hindawi Publishing Corporation \\ \hline
	IEEE & The Institute of Electrical and Electronics Engineers \\ \hline
	IEICE & The Institute of Electronics, Information and Communication Engineers \\ \hline
	IEIE & The Institute of Electronics and Information Engineers \\ \hline
	IET Digital Library & The Institution of Engineering and Technology \\ \hline
	InderScience & InderScience Publishers \\ \hline
	J-STAGE & Japan Science and Technology Information Aggregator, Electronic \\ \hline
	Maxwell & Maxwell Scientific Organization \\ \hline
	Now Publishers & Now Publishers \\ \hline
	Oxford & Oxford Journals \\ \hline
	ScienceDirect & ScienceDirect (Elsevier) \\ \hline
	Scopus & Scopus (Elsevier) \\ \hline
	SPIE & Digital Library \\ \hline
	Springer & Springer \\ \hline
	Taylor & Taylor \& Francis Group \\ \hline
	Wanfang & Wanfang Data \\ \hline
	Wiley & Wiley Online Library \\ \hline
	World Scientific & World Scientific Publishing \\ \hline
	\end{tabular}
	
	\label{tabela:Nome_Completo_Editoras}
\end{table}

\begin{figure}[htp]
	\centering
	\resizebox{10cm}{!}{\includegraphics{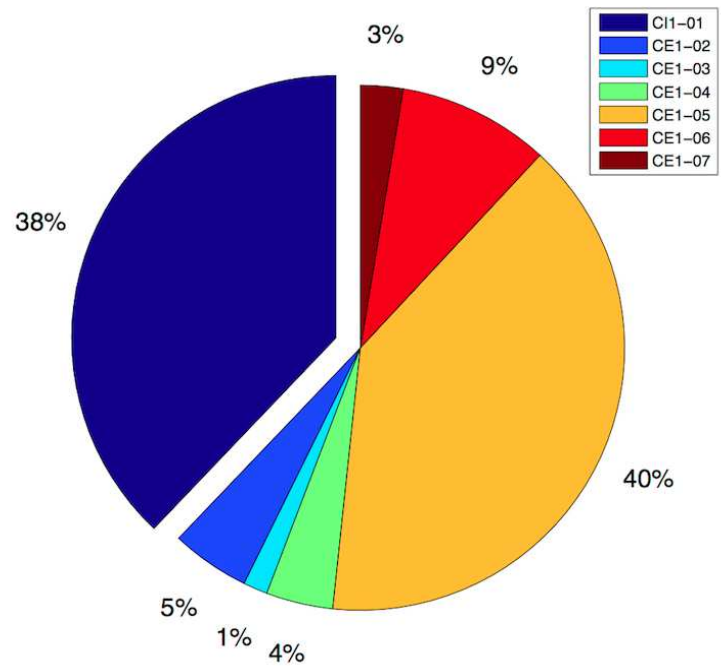}}
	\caption{O gráfico mostra a porcentagem, aproximada, de utilização de cada um dos critérios do 1\textordmasculine \, filtro.}
	\label{fig:Relacao_Criterios_1_Filtros}
\end{figure}

\begin{figure}[htp]
	\centering
	\resizebox{10cm}{!}{\includegraphics{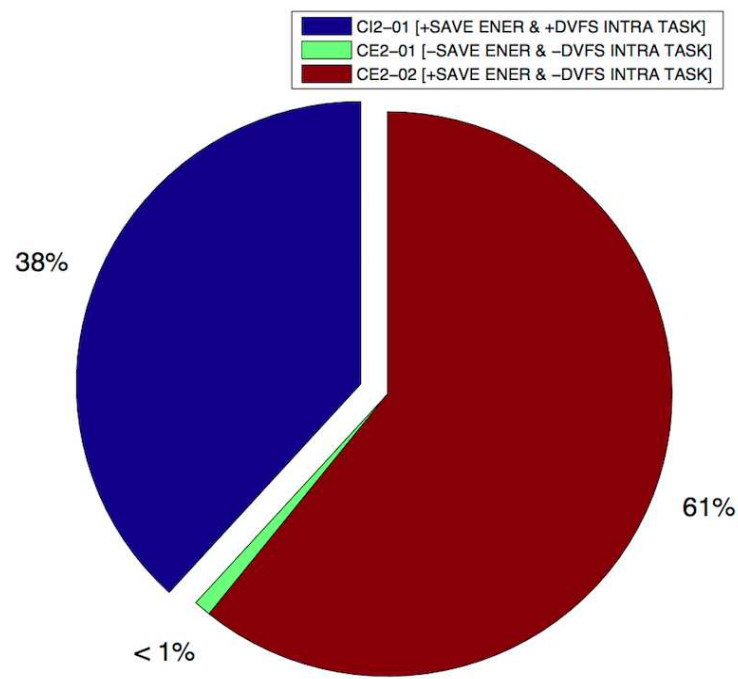}}
	\caption{O gráfico mostra a porcentagem, aproximada, de utilização de cada um dos critérios do 2\textordmasculine \, filtro.}
	\label{fig:Relacao_Criterios_2_Filtros}
\end{figure}

Os gráficos a seguir apresentam uma análise quantitativa quanto a aplicação dos critérios de inclusão e exclusão das publicações para cada filtro executados, sendo que a Figura~\ref{fig:Relacao_Criterios_1_Filtros} apresenta um gráfico com os dados referentes ao 1\textordmasculine \, filtro e a Figura~\ref{fig:Relacao_Criterios_2_Filtros} apresenta um gráfico com os dados referentes ao 2\textordmasculine \, filtro. As informações sobre as publicações utilizadas no 1\textordmasculine \, filtro estão no \ref{section:Apendice_Filtros_1_e_2}. Após a análise e aplicação dos critérios de exclusão do 1\textordmasculine \, filtro o número de publicações selecionadas foram de 115; na aplicação do 2\textordmasculine \, filtro foi identificado que 13 publicações não estavam disponíveis para download não atendendo os critérios definidos no protocolo da revisão sistemática, resultando assim em um total de 102 publicações à serem analisadas.

Vale observar que na Figura~\ref{fig:Relacao_Criterios_1_Filtros} o critério \textbf{CE1-01} não foi utilizado, pois todos os artigos retornados pela máquina de busca da Scopus estão relacionados com a área da Computação, consolidando assim a eficácia da expressão de busca criada na Seção~\ref{subsubsection:Expressao_Busca}; e na Figura~\ref{fig:Relacao_Criterios_2_Filtros} o critério \textbf{CE2-03} não foi utilizado, pois não identificamos nenhuma publicação que falasse da DVFS Intra-Tarefa que não estivesse aplicado dentro do contexto de baixo consumo de energia, o que é bastante plausível, visto que a técnica DVFS foi projetada para reduzir o consumo de energia.

Após a conclusão das analises e aplicações de todos os critérios de exclusão definidos no 1\textordmasculine \, e 2\textordmasculine \, filtro, somente 39 publicações foram selecionadas (ver Tabelas~\ref{table:Publicacoes_Selecionadas_Segundo_Filtro_Parte_1}, \ref{table:Publicacoes_Selecionadas_Segundo_Filtro_Parte_2} e \ref{table:Publicacoes_Selecionadas_Segundo_Filtro_Parte_3}). As informações coletadas e catalogadas sobre essas 39 publicações estão disponíveis no \ref{section:Apendice_Base_Dados_Revisao_Sistematica}. Dessa forma, com a definição e catalogação das publicações selecionadas após a execução do 2\textordmasculine \, filtro, encerrasse a etapa de condução da revisão sistemática.
 
\begin{table}[htp] 
	\centering
	\caption{Publicações selecionadas após o 2\textordmasculine \, Filtro (Parte 1).}
	
	\resizebox{\columnwidth}{!}{%
	\begin{tabular}{|c|p{6cm}|p{5cm}|c|p{2cm}|}
	\hline
	\textbf{ID} & \textbf{Título} & \textbf{Autores} & \textbf{Ano} & \textbf{Editora} \\ \hline
	
	[P01] & Online Intra-Task Device Scheduling for Hard Real-Time Systems & Muhammad Ali Awan, Stefan M. Petters & 2012 & IEEE \\ \hline
	[P02] & Algorithms for combined inter- and intra-task dynamic voltage scaling & Seo H., Seo J., Kim T. & 2012 & Oxford University \\ \hline
	[P03] & A Car Racing Based Strategy for the Dynamic Voltage and Frequency Scaling Technique & David Cohen, Eduardo Valentin, Raimundo Barreto, Horácio Oliveira, and Lucas Cordeiro & 2012 & IEEE \\ \hline
	[P04] & TALk: A temperature-aware leakage minimization technique for real-time systems & Yuan L., Leventhal S.R., Gu J., Qu G. & 2011 & IEEE \\ \hline
	[P05] & An integrated optimization framework for reducing the energy consumption of embedded real-time applications & Takase H., Zeng G., Gauthier L., Kawashima H., Atsumi N., Tatematsu T., Kobayashi Y., Kohara S., Koshiro T., Ishihara T., Tomiyama H., Takada H. & 2011 & IEEE \\ \hline
	[P06] & Checkpoint extraction using execution traces for intra-task DVFS in embedded systems & Tatematsu T., Takase H., Zeng G., Tomiyama H., Takada H. & 2011 & IEEE \\ \hline
	[P07] & Parametric timing analysis and its application to dynamic voltage scaling & Mohan S., Mueller F., Root M., Hawkins W., Healy C., Whalley D., Vivancos E. & 2010 & ACM \\ \hline
	[P08] & Real-time power management for a multi-performance processor & Ishihara T. & 2009 & IEEE \\ \hline
	[P09] & Energy efficient intra-task dynamic voltage scaling for realistic CPUs of mobile devices & Yang C.-C., Wang K., Lin M.-H., Lin P. & 2009 & Scopus (Elsevier) \\ \hline
	[P10] & Stochastic voltage scheduling of fixed-priority tasks with preemption thresholds & He X., Jia Y., Wa H. & 2009 & IEEE \\ \hline
	[P11] & Efficient algorithms for jitterless real-time tasks to DVS schedules & Chen D.-R., Hsieh S.-M., Lai M.-F. & 2008 & IEEE \\ \hline
	[P12] & Expected energy consumption minimization in DVS systems with discrete frequencies & Chen J.-J. & 2008 & ACM \\ \hline
	[P13] & Improving energy-efficient real-time scheduling by exploiting code instrumentation & Zitterell T., Scholl C. & 2008 & IEEE \\ \hline
	[P14] & Task partitioning algorithm for intra-task dynamic voltage scaling & Oh S., Kim J., Kim S., Kyung C.-M. & 2008 & IEEE \\ \hline
	[P15] & Efficient algorithms for periodic real-time tasks to optimal discrete voltage schedules & Chen D.-R., Hsieh S.-M., Lai M.-F. & 2008 & IEEE \\ \hline

	\end{tabular}
	}
	\label{table:Publicacoes_Selecionadas_Segundo_Filtro_Parte_1}
\end{table}

\clearpage
\begin{table}[htp] 
	\centering
	\caption{Publicações selecionadas após o 2\textordmasculine \, Filtro (Parte 2).}
	
	\resizebox{\columnwidth}{!}{%
	\begin{tabular}{|c|p{6cm}|p{5cm}|c|p{2cm}|}
	\hline
	\textbf{ID} & \textbf{Título} & \textbf{Autores} & \textbf{Ano} & \textbf{Editora} \\ \hline
	
	[P16] & System level voltage scheduling technique using UML-RT model & Neishaburi M.H., Daneshtalab M., Nabi M., Mohammadi S. & 2007 & IEEE \\ \hline
	[P17] & Optimizing intratask voltage scheduling using profile and data-flow information & Shin D., Kim J. & 2007 & IEEE \\ \hline
	[P18] & Static WCET analysis based compiler-directed DVS energy optimization in real-time applications & Yi H., Chen J., Yang X. & 2006 & Springer \\ \hline
	[P19] & Energy-efficient task scheduling algorithm for mobile terminal & Zhang L., Qi D. & 2006 & IET Digital Library \\ \hline
	[P20] & Runtime distribution-aware dynamic voltage scaling & Hong S., Yoo S., Jin H., Choi K.-M., Kong J.-T., Eo S.-K. & 2006 & IEEE \\ \hline
	[P21] & Dynamic voltage scaling for multitasking real-time systems with uncertain execution time & Xian C., Lu Y.-H. & 2006 & ACM \\ \hline
	[P22] & Toward the optimal configuration of dynamic voltage scaling points in real-time applications & Yi H.-Z., Yang X.-J. & 2006 & Springer \\ \hline
	[P23] & Intra-task scenario-aware voltage scheduling & Gheorghita S.V., Basten T., Corporaal H. & 2005 & ACM \\ \hline
	[P24] & An intra-task DVS algorithm exploiting program path locality for real-time embedded systems & Kumar G.S.A., Manimaran G. & 2005 & Springer \\ \hline
	[P25] & Optimal dynamic voltage scaling for wireless sensor nodes with real-time constraints & Cassandras C.G., Zhuang S. & 2005 & SPIE - Digital Library \\ \hline
	[P26] & Optimal integration of inter-task and intra-task dynamic voltage scaling techniques for hard real-time applications & Seo J., Kim T., Dutt N.D. & 2005 & IEEE \\ \hline
	[P27] & Optimizing intra-task voltage scheduling using data flow analysis & Shin D., Kim J. & 2005 & IEEE \\ \hline
	[P28] & Optimizing the configuration of dynamic voltage scaling points in real-time applications & Yi H., Yang X. & 2005 & Springer \\ \hline
	[P29] & ParaScale: Exploiting parametric timing analysis for real-time schedulers and dynamic voltage scaling & Mohan S., Mueller F., Hawkins W., Root M., Healy C., Whalley D. & 2005 & IEEE \\ \hline
	[P30] & The optimal profile-guided greedy dynamic voltage scaling in real-time applications & Yi H., Yang X., Chen J. & 2005 & Springer \\ \hline

	\end{tabular}
	}
	\label{table:Publicacoes_Selecionadas_Segundo_Filtro_Parte_2}
\end{table}

\clearpage
\begin{table}[htp] 
	\centering
	\caption{Publicações selecionadas após o 2\textordmasculine \, Filtro (Parte 3).}
	
	\resizebox{\columnwidth}{!}{%
	\begin{tabular}{|c|p{6cm}|p{5cm}|c|p{2cm}|}
	\hline
	\textbf{ID} & \textbf{Título} & \textbf{Autores} & \textbf{Ano} & \textbf{Editora} \\ \hline
	
	[P31] & Intra-task voltage scheduling on DVS-enabled hard real-time systems & Shin D., Kim J. & 2005 & IEEE \\ \hline
	[P32] & Profile-based optimal intra-task voltage scheduling for hard real-time applications & Seo J., Kim T., Chung K.-S. & 2004 & ACM and IEEE \\ \hline
	[P33] & Collaborative operating system and compiler power management for real-time applications & Aboughazaleh N., Mosse D., Childers B., Melhem R., Craven M. & 2003 & IEEE \\ \hline
	[P34] & Exploring efficient operating points for voltage scaled embedded processor cores & Buss M., Givargis T., Dutt N. & 2003 & ACM and IEEE \\ \hline
	[P35] & Energy management for real-time embedded applications with compiler support & AbouGhazaleh N., Childers B., Mosse D., Melhem R., Craven M. & 2003 & ACM \\ \hline
	[P36] & An intra-task dynamic voltage scaling method for SoC design with hierarchical FSM and synchronous dataflow model & Lee S., Yoo S., Choi K. & 2002 & ACM \\ \hline
	[P37] & Low-energy intra-task voltage scheduling using static timing analysis & Shin D., Kim J., Lee S. & 2001 & ACM \\ \hline
	[P38] & Intra-task voltage scheduling for low-energy hard real-time applications & Shin D., Kim J., Lee S. & 2001 & IEEE \\ \hline
	[P39] & A profile-based energy-efficient intra-task voltage scheduling algorithm for hard real-time applications & Shin D., Kim J. & 2001 & ACM and IEEE \\ \hline
	\end{tabular}
	}
	\label{table:Publicacoes_Selecionadas_Segundo_Filtro_Parte_3}
\end{table}

\clearpage

\section{Análise dos Resultados da Revisão Sistemática}
\label{section:Analise_Resultados_Revisao_Sistematica}

Em relação à questão de pesquisa, apresentada na Seção~\ref{subsection:Questoes_Pesquisa}, temos como objetivo catalogar todas as publicações que abordem metodologias de baixo consumo de energia, através da utilização da técnica DVFS intra-tarefa aplicado no contexto de sistemas de tempo real. Esse levantamento bibliográfico nos deu embasamento teórico para responder a questão de pesquisa deste trabalho, através da catalogação das 39 publicações selecionadas após o 2\textordmasculine \, filtro (ver Tabelas~\ref{table:Publicacoes_Selecionadas_Segundo_Filtro_Parte_1}, \ref{table:Publicacoes_Selecionadas_Segundo_Filtro_Parte_2} e \ref{table:Publicacoes_Selecionadas_Segundo_Filtro_Parte_3}). Os parágrafos a seguir mostrarão um breve resumo das publicações catalogadas, mostrando as principais características de cada metodologia. As citações foram feitas em ordem cronológica crescente para que se tenha uma visão melhor de como se deu a evolução do estado da arte, ao longo dos anos.

O trabalho de \cite{Shin_Kim_2001_a} foi um dos precursores no desenvolvimento de ferramentas para análise do WCET intra-tarefa em aplicações de tempo real. A principal finalidade dos algoritmos da ferramenta eram controlar a velocidade de execução da aplicação baseado nos caminhos de execução de caso médio (em inglês, \textit{Average-Case Execution Path - ACEP}), que são os caminhos mais frequentemente executados. Com essa abordagem os autores conseguiram provar que o algoritmo proposto é mais eficaz na redução do consumo de energia que o algoritmo original intraVS, chamado pelos autores de \textit{(RWEP)-based IntraVS} (em inglês, \textit{Remaining Worst-Case Execution Path-based Intra VS}), onde mesmo utilizando as ACEPs é possível satisfazer as restrições temporais da aplicação de tempo real. Esse método se baseia no perfil de comportamento da aplicação, através da análise dos caminhos de execução mais utilizados (ou \textit{hot paths}), chamado de \textit{(RAEP)-based IntraVS} (em inglês, \textit{Remaining Average-Case Execution Path-based IntraVS}). Sua principal contribuição está na exploração das probabilidades de cada caminho de execução da aplicação e garantir que as restrições temporais sejam respeitadas mesmo executando o pior caso. Os experimentos mostram que o \textit{RAEP-based} é 34\% mais eficiente energeticamente que o \textit{RWEP-based}.

O trabalho de \cite{Shin_Kim_Lee_2001_a}, os autores propõem um novo algoritmo de escalonamento de tensão intra-tarefa que controla a tensão de alimentação do processador durante a execução da tarefa, através da exploração dos tempos de folga. Esse método se baseia na análise do tempo de execução estático e na inserção de códigos, dentro do código fonte da aplicação, para a realização dos chaveamentos de tensões e frequências do processador, de forma que o consumo geral de energia seja reduzido. Esses códigos de chaveamento de tensão são definidos para cada um dos blocos de código selecionados a partir do grafo de fluxo de controle da aplicação (CFG). Dessa forma é possível definir as tensões e frequências para cada bloco de código, aproximando assim o tempo de execução ao deadline da tarefa, sempre respeitando as restrições temporais de todas as tarefas em execução. Neste trabalho os autores introduziram uma nova perspectiva para analisar as CFGs, que consiste em mapear os blocos de código por estruturas condicionais (chamado de \textit{B-types}) e por estruturas de repetição (chamado de \textit{L-types}), dessa forma é mais fácil analisar e predizer os cálculos do WCEC (em inglês, \textit{Worst Case Execution Cycle}) e RWCEC (em inglês, \textit{Remaining Worst Case Execution Cycle}). Todas essas analises foram introduzidas na ferramenta AVS (em inglês, \textit{Automatic Voltage Scaler}), desenvolvida pelos próprios autores. O único ponto negativo no estudo realizado é a falta de métricas para avaliar os reais impactos causados pela inserção de códigos adicionais dentro das aplicações.

O trabalho de \cite{Shin_Kim_Lee_2001_b}, os autores propõem uma nova metodologia para analise do WCET (em inglês, \textit{Worst Case Execution Time}) em aplicações de tempo real e tomaram com base o trabalho de \cite{Shin_Kim_Lee_2001_a}. Essa análise é feita em tempo de compilação, de modo \textit{offline}, utilizando o grafo de fluxo de controle da aplicação, onde o cálculo do WCET é feita para cada nó da CFG, enquanto que no trabalho anterior dos mesmos autores a estimativa do WCET era feita tendo como base o programa inteiro.

Esses três trabalhos \citep{Shin_Kim_2001_a, Shin_Kim_Lee_2001_a, Shin_Kim_Lee_2001_b} foram um dos primeiros a abordar metodologias que utilização a técnica DVFS intra-tarefa e juntos possuem cerca de 186 citações na literatura, que é facilmente justificável, pois foram os percursores em propor metodologias baseadas nessa técnica.

Dando continuidade na descrição das publicações, temos o trabalho de \cite{Lee_Yoo_Choi_2002_a}, onde os autores propõem um método de escalonamento de tensão para o projeto de SoCs (em inglês, \textit{System on a Chips}) com hierarquia FSM (em inglês, \textit{Finite State Machine}) e modelo de dados síncrono. Essa técnica foi chamada de modelo HFSM-SDF (em inglês, \textit{Hierarchical FSM and Synchronous Dataflow Model}). Essa metodologia calcula o caminho de execução da aplicação em tempo de execução e utiliza muitos dos conceitos definidos por \cite{Shin_Kim_Lee_2001_a}, para calcular a carga de trabalho restante das tarefas de tempo real e assim aplicar sobre o processador as tensões e frequências ideais para que o consumo de energia seja o menor possível.

O trabalho de \cite{Aboughazaleh_2003_a}, os autores propõem uma técnica que explora as variações dos tempos de execução em diferentes caminhos de execução da aplicação. Esta é uma abordagem híbrida que depende do compilador e do sistema operacional para melhor gerenciar o desempenho e a redução do consumo de energia do processador. O compilador então insere os chamados PMHs (em inglês, \textit{Power Management Hints}), que são trechos de código responsáveis por fornecer e coletar informações em tempo de execução da aplicação para o sistema operacional, além de estimar o desempenho da aplicação no pior caso. Dessa forma, o sistema operacional invoca os PMPs (em inglês, \textit{Power Management Points}) para realizar o chaveamento de tensão e frequência do processador com base nas informações passadas pelos PMHs.

O trabalho de \cite{Buss_Givargis_Dutt_2003_a}, os autores propõem a exploração e seleção de potenciais pontos de escalonamento de tensão que possam atuar na diminuição eficiente do consumo de energia em aplicações de tempo real não críticos. A problemática desse método está em selecionar esses pontos de controle para atuar em conjunto com a técnica DVS intra-tarefa, proporcionando uma redução do consumo de energia do processador. Esse método se baseia basicamente em três passos, são eles: (1) fazer a análise estática da aplicação e atribuir um fator de desaceleração ideal para cada bloco; (2) computar as frequências de operação com base na análise da aplicação inteira; (3) reatribuir os fatores de aceleração para cada bloco, com base nas frequências de operação válidas e computadas no passo 2. Essa abordagem é muito semelhante a técnica dos coreanos \citep{Shin_Kim_Lee_2001_a}, onde a principal diferença esta na metodologia de definição dos fatores de desaceleração.

O trabalho de \cite{Aboughazaleh_2003_b}, os autores tomaram como base o trabalho de \cite{Aboughazaleh_2003_a}, onde o foco principal da metodologia passou a ser a colaboração entre o compilador e o sistema operacional. O principal contribuição em relação ao trabalho anterior está no sistema operacional, que passa a monitorar periodicamente os chaveamentos de tensões e frequências do processador baseado nas informações providas pelos \textit{PMHs}.

O trabalho de \cite{Seo_Kim_Chung_2004_a}, os autores propõem uma metodologia baseada no perfil de execução da tarefa, onde os níveis de tensão são definidos para cada bloco de código. Esse método tem como objetivo gerenciar melhor os \textit{overheads} de transição, que são totalmente ou parcialmente ignorados nos outros trabalhos presentes na literatura, e obter melhores níveis de redução do consumo de energia do processador. Essa técnica é chamada de ``ROEP-\textit{based technique}'' (ROEP - \textit{Remaining Optimal-Case Execution Path}), que é uma melhoria da metodologia \textit{RAEP-based} proposta por \citep{Shin_Kim_2001_a}, cujo principal foco está relacionado com desperdícios de energia, com as trocas excessivas de tensão e frequência do processador e com a diminuição dos \textit{overheads} inseridos dentro das aplicações. Seguindo a escala cronológica dos trabalhos catalogados nesta revisão sistemática, esta foi uma das primeiras abordagens a otimizar estes parâmetros.

O trabalho de \cite{Shin_Kim_2005_a}, os autores melhoraram a eficiência do método \textit{RAEP-based} proposto por eles mesmo em \cite{Shin_Kim_2001_a}. Nesta nova abordagem, a principal diferença está nas otimizações de \textit{overheads} para a realização das transições de tensão, que antes era feita de forma \textit{offline} e agora o método de atribuição de tensões passou a ser \textit{online} e mais eficiente. Os autores utilizaram os mesmos casos de teste para realização dos experimentos e fizeram alterações na ferramenta AVS para adaptá-la a nova abordagem. Um fato interessante a ser relatado é que os autores começaram a introduzir o conceito de ciclos salvos ou ciclos economizados (em inglês, \textit{Saved Cycles ou $C_{saved}$}), ou seja, são trechos de código que deixaram de ser executados dentro da aplicação. Esse conceito será melhor amadurecida em \cite{Shin_Kim_2005_b}.

O trabalho de \cite{Yi_Yang_Chen_2005_a}, os autores propõem um modelo analítico de escalonamento dinâmico de tensão "ganancioso", cujo o objetivo é encontrar as tensões ideais para as aplicações de tempo real, através da análise dos casos de execução mais frequentes, ou também chamados de \textit{Hot Path}, referenciados em \cite{Shin_Kim_2001_a}. Essas análises visam identificar os tempos de folga distribuídos pela aplicação e, em seguida, repassa os ganhos obtidos para o processador, minimizando o consumo do energia. Esse método foi chamado de OPTDVS (em inglês, \textit{Optimal Dinamic Voltage Scheduling}). Em outras palavras, esse método é um mecanismo de ajuste de tensão ganancioso guiado por perfil (ou \textit{profile-guided}) que se baseia nos \textit{hot paths} para definir o melhor perfil de consumo de energia para uma dada aplicação de tempo real.

O trabalho de \cite{Mohan_Mueller_Root_2005_a}, os autores propõem uma nova técnica chamada \textit{ParaScale}, que permite fazer análises de tempo paramétrico em conjunto com o escalonamento. Essas análises permitem detectar dinamicamente os limites dos \textit{loops} e o limite inferior do WCET (em inglês, \textit{Wrost Case Execution Time}), em tempo real, durante o tempo de execução restante da tarefa. Portanto, o ganho desta metodologia está, principalmente, sobre os tempos de folga obtidos sobre as estruturas de repetição. Dentre os trabalhos catalogados nessa revisão sistemática este foi o primeiro a trabalhar com limites paramétricos de \textit{loops}, permitindo ter um melhor controle dos tempos de folga dentro de estruturas de repetição.

O trabalho de \cite{Yi_Yang_2005_a}, os autores propõem uma metodologia de configuração baseado em um método que constrói o padrão de execução de uma determinada aplicação, também chamado de \textit{Profile-Based Method} já relatado na publicação de \cite{Yi_Yang_Chen_2005_a}. O diferencial desta nova abordagem está em diminuir os \textit{overheads} inseridos pelo compilador no código fonte das aplicações. Esse processo é feito da seguinte forma: primeiro o compilador insere os pontos de escalonamento sem levar em consideração os \textit{overheads}; em seguida, todos os pontos de escalonamento são listados, já levando em consideração os \textit{overheads}; e por fim, os pontos que possuem maiores \textit{overheads} e / ou não trazem redução do consumo de energia para aplicação são excluídos.

O trabalho de \cite{Shin_Kim_2005_b}, os autores propõem uma otimização na técnica intraDVS usando informações de fluxo de dados da aplicação de tempo real. A metodologia visa melhorar a eficiência energética antecipando os pontos de escalonamento de tensão (em inglês, \textit{Voltage Scaling Points} - VSPs), baseadas nos resultados de análises do fluxo de dados da aplicação. Essa técnica foi chamada de \textit{LaIntraDVS} (em inglês, Look Ahead IntraDVS). Em outras palavras, o método proposto antecipa os pontos de controle para maximizar os ganhos de energia da técnica intraDVS, como por exemplo: analisar uma estrutura de repetição e predizer quantas interações serão necessárias e aplicar as tensões e frequências ideais para essa bloco de código antes que ele seja realmente executado.

O trabalho de \cite{Seo_Kim_Dutt_2005_a}, os autores propõe uma nova técnica DVS que combinam as técnicas DVS intra-tarefa e inter-tarefa, chamada de \textit{DVS-intgr}. Essa metodologia examina os limites inferiores de consumo de energia baseado na técnica DVS intra-tarefa (parte dessa metodologia foi inspirada no trabalho de \cite{Shin_Kim_Lee_2001_a}) e com essas propriedades foram definidos os tempos de execução ideais de cada tarefa. Em seguida, as tarefas são divididas em vários grupos de trabalho de tal forma que cada tarefa possa ser executada dentro do limite preestabelecido para cada grupo, através da utilização da técnica DVS inter-tarefa melhorada para produzir o melhor escalonamento entre elas de forma que haja a redução no consumo de energia e garantindo que nenhum premissa temporal seja violada.  

O trabalho de \cite{Cassandras_Zhuang_2005_a}, os autores propõe um controle intra-tarefa para minimizar o consumo de energia dentro do contexto de rede de sensores sem fio, processando tarefas de tempo real críticas. As variáveis de controle são basicamente os tempos de processamento das tarefas de tempo real, onde cada um desses tempos estão associados a diferentes níveis de tensão. O controle intra-tarefa é baseado na exploração das propriedades dos caminhos de execução ideais. Além disso os autores mostram em seus experimentos que soluções intra-tarefa minimizam mais energia que as soluções inter-tarefa.

O trabalho de \cite{Kumar_Manimaran_2005_a}, os autores propõem um novo algoritmo DVS intra-tarefa de consumo de energia consciente cujo o objetivo central é explorar os caminhos mais comuns e frequentemente executados dentro de uma aplicação de tempo real. Esse algoritmo foi chamado de CHP (em inglês, \textit{Common Hot Path}). Essa metodologia considera todos os caminhos mais executados (ou \textit{hot-paths}), principio também utilizado nos trabalhos de \cite{Shin_Kim_2001_a}, \cite{Seo_Kim_Chung_2004_a}, \cite{Yi_Yang_Chen_2005_a} e \cite{Shin_Kim_2005_a}, e para cada um deles são atribuídas probabilidades que irão indicar os caminhos mais utilizados. Dessa forma, a metodologia consegue combinar todos os \textit{hot paths} em um único caminho base que é comum em comprimento com a maioria dos \textit{hot paths}, assim é possível descobrir qual o caminho que leva a melhores taxas de minimização do consumo de energia, pois nem sempre o caminho mais curto é o mais eficaz para minimização do consumo de energia.

O trabalho de \cite{Gheorghita_2005_a}, os autores propõem uma abordagem proativa que visa melhorar a performance do algoritmo de escalonamento intra-tarefa, explorando os tempos de folga que aparecem em tempo de execução, em seguida repassa para o processador através trechos de código inseridos na aplicação original, chamados de pontos de escalonamento de tensão ou VSPs (em inglês, \textit{Voltage Scaling Points}). Essa abordagem consiste, basicamente, em quatro etapas: (1) identificar os parâmetros que poderiam ter um impacto sobre o tempo de execução da aplicação; (2) calcular o máximo de impacto destes parâmetros sobre o WCEC da aplicação; (3) particionar o aplicativo em possíveis cenários, considerando-se esses parâmetros, juntamente com o seu impacto, e selecionando apenas cenários que, isoladamente, reduzir o consumo de energia; por fim, (4) computar o escalonamento DVS para cada cenário selecionado no estágio 3 e combiná-los com o escalonamento global da aplicação de tempo real.

O trabalho de \cite{Yi_Yang_2006_a}, os autores apresentam uma metodologia de configuração ótima de pontos de escalonamento de tensão dinâmicos sem \textit{overheads} de escalonamento de tensão, onde tomaram como base os trabalhos de \cite{Aboughazaleh_2003_b} e \cite{Aboughazaleh_2003_a}. Com essa metodologia os autores conseguiram introduzir a menor quantidade necessária de pontos de escalonamento de tensão para melhor aproveitar os tempos de folga da aplicação e, além disso, provaram teoricamente todos os modelos e teoremas matemáticos definidos na metodologia, sempre visando a otimização ideal de energia.

O trabalho de \cite{Xian_Lu_2006_a}, os autores propõem uma abordagem que visa integrar as técnicas de escalonamento de tensão intra-tarefa e inter-tarefa. O conceito principal do método proposto é que cada tarefa possa contribuir com informações individuais para que seja possível melhorar o escalonamento individual das demais tarefas em execução, sempre tomando como base as informações globais passadas pelas demais tarefas. Dessa forma, a abordagem é dividida, basicamente, em duas etapas: (1) É calculado estatisticamente o escalonamento de frequência ótimo para múltiplas tarefas periódicas utilizando o escalonamento EDF (em inglês, \textit{Earliest Deadline First}) para processadores que conseguem mudar suas frequências de forma contínua; e (2) para processadores que possuem uma faixa limitada de frequências discretas, é apresentado um algoritmo heurístico específico para construção do escalonamento de frequência baseado em informações de distribuição de probabilidade e restrições de escalonabilidade globais.

O trabalho de \cite{Hong_Yoo_Choi_Kong_2006_a}, os autores propõem uma nova técnica de escalonamento de tensão (DVS) intra-tarefa que não visa apenas explorar as distribuições de tempo de execução da aplicação, mas também o fluxo de dados e a arquitetura. Em outras palavras, essa abordagem utiliza os dados da aplicação e da arquitetura para predizer o RWCEC e aplicar com antecedência as tensões e frequências ideais no processador. Portanto, com este trabalho os autores introduziram o conceito de perfil estatístico de ciclos de execução dentro da técnica DVS intra-tarefa ao invés de ciclos de execução no pior caso (WCEC).

O trabalho de \cite{Zhang_2006_a}, os autores propõem um algoritmo de escalonamento de tarefas baseado em otimizações genéticas para diminuir o consumo de energia quando são especificados os deadlines e os ciclos de execução das tarefas. Esse algoritmo genético híbrido integra as técnicas inter e intra tarefas visando mensurar o pWCEC (em inglês, \textit{Probabilistic Worst-Case Execution Time}), a fim de encontrar o melhor coeficiente de escalonamento das tarefas de forma que todas as restrições temporais sejam obedecidas e ao mesmo tempo se obtenha uma minimização do consumo de energia do processador.

O trabalho de \cite{Yi_Chen_Yang_2006_a}, os autores propõem uma ferramenta chamada HEPTANE, cuja função é realizar a análise estática do WCET (em inglês, \textit{Worst Case Execution Time}), inserir os códigos da técnica DVFS e definir o perfil de consumo de energia da aplicação. Essa ferramenta trabalha em conjunto com o simulador de energia e performance chamado \textit{Sim-Panalyzer}, que roda em um ambiente \textit{RTLPower} (em inglês, \textit{Real-Time Low Power}), cuja função é simular o ambiente de experimentação para rodar os casos de teste criados pelos autores. Analisando de forma mais incisiva o trabalho, não ficou claro como a ferramenta HEPTANE trata as invariante de \textit{loops}, na definição do perfil de consumo de energia da aplicação.

O trabalho de \cite{Shin_Kim_2007_a}, os autores propõem duas melhorias sobre a técnica IntraDVS. A primeira delas é uma melhoria da técnica chamada RAEP-IntraDVS (em inglês, \textit{Remaining Average-case Execution Path}), que visa otimizar o escalonamento de tensão através de análises das informações da aplicação, levando em consideração o caminho de execução de caso médio remanescente. A outra melhoria é sobre a técnica LaIntraDVS, citada no trabalho \cite{Shin_Kim_2005_b}, que leva em consideração as informações do fluxo de dados para gerar otimizações sobre os pontos de chaveamento de tensão (em inglês, \textit{Voltage-Scaling Points} - VSPs), principalmente através da predição das VSPs antes de estruturas condicionais e \textit{loops}.

O trabalho de \cite{Neishaburi_2007_a}, os autores apresentam uma otimização sobre o escalonamento de tensões intra-tarefa, através da análise do fluxo de dados e do fluxo de controle da aplicação. A partir dessa análise, a metodologia é capaz de antecipar os pontos de escalonamento de tensão (em inglês, \textit{Voltage Scaling Points} - VSP), enquanto que a técnica DVFS intra-tarefa tradicional apenas localiza os pontos de controle. Essa metodologia permite adicionar menos \textit{overheads} no código fonte da aplicação.

O trabalho de \cite{Chen_Hsieh_Lai_2008_b}, os autores propõem uma metodologia que visa minimizar o consumo de energia através da análise do fluxo de dados da aplicação, tanto do ponto de vista inter quanto intra tarefa. Essa abordagem consiste basicamente de três fases: (1) primeiramente é feita a transformação harmônica dos períodos de todas as tarefas, em seguida é feita a validação e compartilhamento dos tempo de folga entre as demais tarefas, utilizando um escalonamento definidos pelos autores de \textit{Jitterless Schedule} \footnote{\textit{Jitterless Schedule} são interferências causadas pela chegada de sucessivas instâncias de uma mesma tarefa.}; (2) o próximo passo é calcular a utilização total dado os novos parâmetros das tarefas definidos no passo 1; por último (3) é feita a computação das características de cada tarefa, tais como o início e fim relativos, com o objetivo de ajustar as tensões e frequência, evitando que restrições temporais venham ser violadas. 

O trabalho de \cite{Oh_Kim_Kim_Kyung_2008_a}, os autores propõem um novo algoritmo de particionamento de tarefas baseado na técnica DVS intra-tarefa, onde o seu principal objetivo é dividir de maneira mais eficiente os blocos de código da aplicação de forma que seja possível diminuir o número de chaveamento de tensões e frequências do processador. Essa abordagem, primeiramente, divide o código fonte da aplicação em um número máximo de seções de código. Em seguida, são calculados os ciclos de execução de cada nó, por meio de simulações estáticas, e as penalidades das predições que falharam. Essas penalidades são utilizadas como uma medida para decidir se determinados nós deverão ser agrupados ou não. Com essa metodologia os autores conseguiram reduzir o número de chaveamentos de tensão e, consequentemente, minimizar o consumo de energia do processador.

O trabalho de \cite{Zitterell_2008_a}, os autores propõem um escalonamento mais eficiente de energia para processadores com frequências discretas, chamado de ItcaEDF (em inglês, \textit{Intra-Task Characteristics Aware EDF}). Ela se baseia na integração das técnicas inter e intra tarefas para diminuição dos tempos ociosos do processador e dos tempos de folgas das tarefas. No algoritmo intra-tarefa, os autores focam na quantidade de ciclos economizados e em um contador de ciclos, que possibilita contabilizar os diferentes caminhos dentro de um \textit{loop}, permitindo diminuir os níveis de frequência do processador de acordo com as invariantes do \textit{loop}. Quanto ao algoritmo inter-tarefa, os autores implementam um conjunto de bibliotecas que permitem as tarefas compartilhar informações umas com as outras, contribuindo para um melhor escalonamento global e diminuição do tempo ocioso do processador.

O trabalho de \cite{Chen_2008_a}, o autor apresenta uma nova abordagem para minimizar o consumo de energia utilizando funções de densidade de probabilidade com base nas cargas de trabalho das tarefas de tempo real. Para o escalonamento intra-tarefa foi feito um algoritmo eficiente para obter a frequência ideal para uma única tarefa, de modo que o consumo de energia seja minimizado. Enquanto que o algoritmo de escalonamento inter-tarefa, chamado \textit{M-Greedy}, foi desenvolvido com base em uma abordagem de programação linear cuja finalidade é obter as melhores soluções para as tarefas de tempo real baseada em quadros, visando diminuir os tempos de folga.

O trabalho de \cite{Chen_Hsieh_Lai_2008_a}, os autores propõem um algoritmo intra-tarefa e um inter-tarefa para diminuir o consumo de energia durante o escalonamento das tarefas. Essa metodologia tem como finalidade diminuir os \textit{overheads} e os tempos de folga entre as tarefas, dando mais previsibilidade e otimizações \textit{offline} para o escalonamento. Para facilitar a geração do escalonamento, as tarefas com períodos arbitrários são transformados em períodos harmônicos para que os tempos de preempção, início e término de cada tarefa possam ser facilmente derivados, principalmente para tratar o que o autor chama de \textit{Jitterless Schedule}. Essa abordagem foi desenvolvida a partir do trabalho de \cite{Chen_Hsieh_Lai_2008_b}.

O trabalho de \cite{He_2008_a}, os autores exploram os tempos de execução variáveis de tarefas, dentro da política de escalonamento FPPT (em inglês, \textit{Fixed-Priority scheduling with Preemption Threshold} - Escalonamento com Prioridade Fixa com Limite de Preempção). Essa política de escalonamento, executando em um processador com tensão variável, permite reduzir os custos com preempções desnecessárias das tarefas. Então, os autores desenvolveram um algoritmo para analisar todas as possibilidades de carga de trabalho para cada tarefa. Em seguida, utilizou esses dados estocásticos para definir as tensões e frequências do processador de acordo com o tamanho da tarefa e sua distribuição de probabilidade, com o intuito de minimizar o consumo de energia no caso médio. 

O trabalho de \cite{Yang_2009_a}, os autores propõem dois algoritmos de escalonamento dinâmico de tensão (DVS) intra-tarefa para CPU's. O algoritmo OSRC (em inglês, \textit{Optimal Schedule for Realistic CPUs}) tem por objetivo aplicar conceitos de programação dinâmica dentro da CFG da aplicação para identificar os caminhos ideais com menor consumo de energia, tendo como referência a especificação de uma \textit{CPU} realista \footnote{\textit{CPUs realista são processadores que possuem um conjunto limitado de níveis de tensão e frequência \citep{Yang_2009_a}.}}. O segundo algoritmo, chamado LO-OSRC (em inglês, \textit{Low Overhead Optimal Schedule for Realistic CPUs}), já leva em consideração o potencial de computação das tarefas e os \textit{overheads} de transição, permitindo apenas uma mudanção de tensão de frequência da CPU por tarefa. Dessa forma, os autores mostram em seus experimentos que seus algoritmos conseguem obter ganhos acima de 10\% em comparação com outros trabalhos presentes na literatura.

O trabalho de \cite{Ishihara_2009_a}, o autor propõe uma metodologia baseada em uma nova arquitetura contendo vários núcleos de processamento, chamada \textit{Architecture of Multi-Performance Processor}, onde cada núcleo trabalha em uma frequência e tensão específica. Dessa forma, o processador não perde tempo chaveando tensões e frequência, que são em média na casa das centenas de microsegundos. Essa arquitetura quando integrada a técnica DVFS intra-tarefa permite maximizar os ganhos de economiza de energia, através da diminuição dos  \textit{overheads} da técnica DVFS, permitindo fazer chaveamentos de tensão e frequência na casa dos 1.5 microsegundos e dissipando apenas 10 nano-joules. Essa metodologia, reduziu cerca de 25\% de energia em comparação com a técnica DVS convencional do processador.

O trabalho de \cite{Mohan_2010_a}, os autores propõem um metodologia que remove as restrições sobre as invariantes de \textit{loops} através de analises paramétricas, com o objetivo de maximizar a identificação dos tempos de folga das tarefas e minimizar o consumo de energia do processador. Dessa forma, os ganhos dessa abordagem está diretamente relacionado com a redução do número de interações dos \textit{loops} mapeados dentro das aplicações.

O trabalho de \cite{Tatematsu_2011_a}, os autores propõem uma metodologia que analisa o código fonte da aplicação e lista todos os possíveis locais para a inserção de pontos de controle (também chamado pelos autores de \textit{checkpoints}). Em seguida, todos esses pontos são analisado e os que não trazem ganhos energéticos são removidos. Por fim, a metodologia compara essa listagem de pontos de controle com uma tabela RWCEC (em inglês, \textit{Remaining Worst Case Execution Cycles}), também extraída da aplicação, para então calcular as tensões e frequências que deverão ser utilizadas no processador.

O trabalho de \cite{Takase_2011_a}, os autores desenvolvem um \textit{framework} com o objetivo de melhor realizar o chaveamento entre performance e consumo de energia do processador. As configurações ótimas do processador são definidas de acordo com cada etapa de execução da tarefa. Além disso, esse \textit{framework} aplica técnicas de otimização sobre a alocação de memória da aplicação, visando diminuir \textit{overheads} de IO (em inglês, \textit{Input and Output}). Dessa forma, todas as características e comportamento da aplicação são analisados tando do ponto de vista inter quanto intra tarefa. Os resultados dessa análise, resultam na otimização de energia em tempo de execução de acordo com o comportamento da aplicação. Os resultados experimentais utilizando um sistema de vídeo conferencia conseguiram reduzir em média o consumo de energia em 44.9\% em comparação com outros estudos de caso criados pelo próprio autor.

O trabalho de \cite{Yuan_2011_a}, os autores propõem um algoritmo de escalonamento intra-tarefa que visa diminuir a temperatura do processador e minimizar o consumo de energia em sistema de tempo real. Essa técnica foi chamada de \textit{TALk} (em inglês, \textit{Temperature- Aware Leakage}). A ideia básica do algoritmo é aumentar a frequência quando a temperatura do chip estiver baixa ou quando a carga de trabalho for alta e colocar o processador em baixo consumo de energia quando a temperatura do chip estiver alta ou com carga de trabalho leve. Para fazer isso, o algoritmo \textit{TALk} foi dividida em duas partes: (1) A \textit{Offline}, que usa métodos de programação dinâmica para alcançar os melhores níveis de economia de energia e de temperatura; (2) O \textit{Online} tem como objetivo determinar o modo de operação do processador com base na sua temperatura corrente e na quantidade de ciclos remanescentes das tarefas em execução. Com esse algoritmo os autores conseguiram melhorar a economia de energia em cerca de 18\% em comparação com a técnica DVS tradicional.

O trabalho de \cite{Cohen_2012_a}, os autores apresentam uma nova política de escalonamento de tarefas de tempo real, que leva em consideração preempções. Esse nova metodologia consegue economizar energia mesmo executando as tarefas no seu pior caso. Em resumo, o autor faz uma analogia entre escalonador e suas tarefas, com uma corrida de carros, onde o objetivo da corrida é que todos os carros (uma analogia as tarefas) cheguem juntos no final, utilizando as menores velocidades (uma analogia a tensões e frequências do processador) sem que nenhuma premissa temporal seja violada. Essa nova política de escalonamento foi experimentada apenas em ambiente simulado e utilizando casos de teste gerados pelos próprios autores.

O trabalho de \cite{Seo_Seo_Kim_2012_a}, os autores apresentam uma técnica de baixo consumo de energia que se baseia na combinação simultânea entre inter-tarefa e intra-tarefa, também chamada DVS combinado (em inglês, \textit{Combined} DVS - CDVS). Essa nova abordagem leva em consideração o estado do sistema dormindo (em inglês, \textit{Sleep State} - CDVS-S) e não dormindo (em inglês, \textit{No Sleep State} - CDVS-NS). Ela consiste basicamente de 4 etapas: (1) Aplicar a técnica CDVS-NS para determinar os intervalos de execução das tarefas de modo que o consumo total de energia seja minimizado, sem estados de dormindo (\textit{sleep state}); (2) Realizar a análise estática dos blocos de código da tarefa, a fim de identificar os tempos ociosos e as tensões e frequências que deverão ser utilizadas; (3) Combinar os intervalos de tempo salvos na segunda etapa com o maior tempo possível no qual o sistema possa estar no estado ocioso (\textit{idle state}) de forma eficiente; e por último (4) Monitorar dinamicamente todas as instâncias das tarefas em execução que concluíram sua execução antes do prazo final e, em seguida, irá decidir se coloca a tarefa em estado ociosa (\textit{idle state}) ou em estado de dormindo (\textit{sleep state}), dependendo do que for mais econômico energeticamente. Em geral, essa metodologia conseguiu reduzir o consumo de energia em média de 7\% com a técnica CDVS-S e de 12\% com a técnica CDVS-NS em comparação com outros trabalhos presentes na literatura.

O trabalho de \cite{Awan_Petters_2012_a}, os autores propõem um algoritmo (online) de escalonamento intra-tarefa, cuja principal funcionalidade é ligar e desligar dispositivos do hardware, permitindo que eles sejam utilizados somente quando necessário. Essa metodologia se aplica a sistemas de tempo real crítico e funciona basicamente explorando os tempos de folga entre as execuções das tarefas, a fim de realizar o melhor gerenciamento dos acionamentos e desligamento dos dispositivos,  melhorando assim a performance de economia de energia do sistema. Os experimentos mostram um ganho de economia de energia acima dos 90\% em comparação com outras técnicas presentes na literatura.

Vale ressaltar que todos os dados extraídos das 39 publicações estão catalogados no \ref{section:Apendice_Base_Dados_Revisao_Sistematica} e a partir da coleta desses dados foi possível realizar as analises quantitativas definidas na Seção~\ref{subsection:Procedimentos_Analise}.

A primeira análise feita após a catalogação dos dados foi uma análise quantitativa quanto aos métodos de execução das abordagens, que podem ser:
\begin{itemize}
\item \textbf{\textit{Online}:} São métodos dinâmicos que funcionam em tempo de execução e podem sofrer modificações ao longo da execução da aplicação;

\item \textbf{\textit{Offline}:} São métodos que funcionam em tempo de compilação e são aplicados estaticamente no código fonte da aplicação;

\item \textbf{\textit{Híbrida}:} São métodos que são implementados parte \textit{offline} e parte \textit{online}.
\end{itemize}

\begin{figure}[ht]
	\centering
	\resizebox{7cm}{!}{\includegraphics{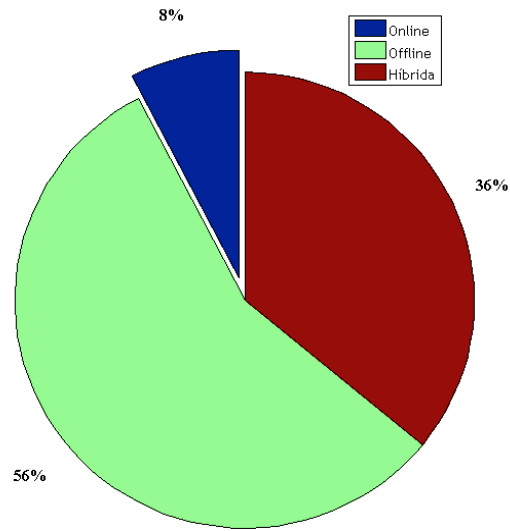}}
	\caption{Análise quantitativa dos modos de execução das metodologias catalogadas.}
	\label{fig:Modo_Execucao_Abordagens_Propostas_Publicacoes}
\end{figure}

A Figura~\ref{fig:Modo_Execucao_Abordagens_Propostas_Publicacoes} apresenta o gráfico quantitativo com o resultado da classificação dos modos de execução das metodologias, onde apenas 8\% (3 publicações) delas são totalmente \textit{online} \citep{Awan_Petters_2012_a, Cohen_2012_a, Zitterell_2008_a}. É importante observar que o método de execução Híbrida, com cerca de 36\%, tem grandes chances de alcançar ou até ultrapassar a quantidade de abordagem que utilizam o modo de execução \textit{offline}, em um futuro não muito distante. Principalmente, devido ao uso mais recorrente da integração das técnicas inter e intra tarefas para maximizar os ganhos energéticos sobre o processador.

\begin{figure}[ht]
	\centering
	\resizebox{8cm}{!}{\includegraphics{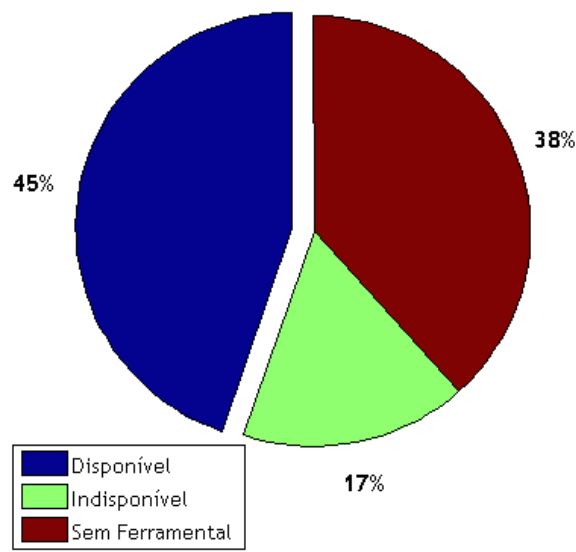}}
	\caption{Análise quantitativa da disponibilidade ferramental das publicações catalogadas.}
	\label{fig:Disponibilidade_Apoio_Ferramental}
\end{figure}

Quanto a análise quantitativa da disponibilidade do apoio ferramental, temos que apenas 45\% (21 publicações) das abordagens fornecem apoio ferramental. A Figura~\ref{fig:Disponibilidade_Apoio_Ferramental} apresenta um gráfico quantitativo mais detalhado do percentual de publicações que oferecem apoio ferramental.

\begin{figure}[ht]
	\centering
	\resizebox{7cm}{!}{\includegraphics{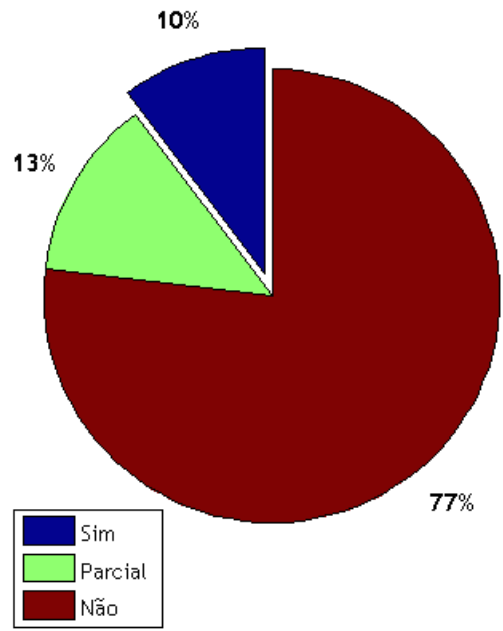}}
	\caption{Análise quantitativa das metodologias que dão suporte a preempções.}
	\label{fig:Analise_Quantitativa_Suporte_Preempcoes}
\end{figure}

Analisando agora do ponto de vista das metodologias que dão suporte a preempção, temos que apenas 10\% (4 publicações) fornecem suporte a preempções \citep{Aboughazaleh_2003_b, Chen_Hsieh_Lai_2008_a, Chen_Hsieh_Lai_2008_b, Cohen_2012_a}, enquanto que 13\% (5 publicações) dão suporte parcial, ou seja, consideram um ambiente com múltiplas tarefas preemptivas em execução, mas não deixa claro na metodologia como foi implementado \citep{Takase_2011_a, Chen_2008_a, He_2008_a, Awan_Petters_2012_a, Zitterell_2008_a}. A Figura~\ref{fig:Analise_Quantitativa_Suporte_Preempcoes} ilustra melhor essa análise e, além disso, deixa mais evidente que essa é uma linha de pesquisa pouco explorada pela comunidade científica.

A grande maioria das metodologias não dão suporte a preempções, pois os métodos consideram apenas um tarefa em execução e os que consideram múltiplas tarefas não leva em consideração preempção entre elas. Por outro lado, as publicações que deixam claro na metodologia que dão suporte a preempções tiveram que integrar outras técnicas, com por exemplo a técnica DVFS inter-tarefa. Novamente, esses argumentos ressaltam que ainda há muitas linhas de pesquisa a serem exploradas dentro desse contexto.

Visando obter uma visão geral, no que diz respeito a completude de cada uma das abordagens, efetuamos uma comparação qualitativa entre as 39 publicações selecionadas no 2\textordmasculine \, filtro. As métricas utilizadas foram extraídas com base nas questões de pesquisa definidas neste trabalho (ver Seção~\ref{section:Planejamento_Revisao_Sistematica}). Vale ressaltar que o principal objetivo desta comparação é medir a cobertura das abordagens diante das métricas propostas e não sua eficácia ou desempenho, ou seja, identificar as abordagens que satisfaçam o maior número de métricas.

Diante disto, a Tabela~\ref{table:Metricas_Comparacao_Abordagens} apresentam as métricas definidas, sendo que cada coluna da tabela significa: (I) o identificador da métrica (ID); (II) o nome da métrica (Nome); (III) as opções definidas para cada métrica (Opções); (IV) pontuação atribuída a métrica (Pontuação de Cobertura), onde a pontuação é definida da seguinte forma: ``+'' representa 5 pontos; e (V) pontuação máxima permitida pela métrica.

\begin{table}[htp] 
\centering
\caption{Lista de métricas estabelecidas para realizar a comparação de completude entre as publicações que compõem a base de dados final da revisão sistemática.}
\resizebox{\textwidth}{!}{%
\begin{tabular}{|c|c|l|c|c|}
\hline
\multicolumn{5}{|c|}{\textbf{Métricas}} \\ \hline
\textbf{ID} & \textbf{Nome} & \multicolumn{1}{c|}{\textbf{Opções}} & \begin{tabular}{c} \textbf{Pontuação} \\ \textbf{de Cobertura}\end{tabular} & \begin{tabular}{c} \textbf{Pontuação} \\ \textbf{Máxima}\end{tabular} \\ \hline \hline
\multirow{3}{*}{\begin{tabular}[c]{@{}c@{}}1\end{tabular}} & \multirow{3}{*}{\begin{tabular}[c]{@{}c@{}}Ferramental\end{tabular}} & Não possui / Não identificado & 0 & \multirow{3}{*}{15} \\ \cline{3-4} 
 &  & \begin{tabular}[c]{@{}l@{}}Possui, mas não foi identificada\\ sua disponibilidade\end{tabular} & ++ & \\ \cline{3-4} 
 &  & Possui e está disponível & +++ & \\ \hline \hline
\multirow{4}{*}{\begin{tabular}[c]{@{}c@{}}2\end{tabular}} & \multirow{4}{*}{\begin{tabular}[c]{@{}c@{}}Tipo do \textit{Benchmark}\\ utilizado (abrev. T Bench )\end{tabular}} & \begin{tabular}[c]{@{}l@{}}Nenhum ou \textit{Benchmarks}\\ do próprio artigo\end{tabular} & 0  & \multirow{3}{*}{15} \\ \cline{3-4} 
 &  & \textit{Benchmarks} da literatura & + & \\ \cline{3-4} 
 &  & \textit{Benchmarks} da industria & ++ & \\ \cline{3-4} 
 &  & \begin{tabular}[c]{@{}l@{}}Combinação entre os tipos\\ de \textit{benchmarks}\end{tabular} & +++ & \\ \hline \hline
\multirow{2}{*}{3} & \multirow{2}{*}{\begin{tabular}[c]{@{}c@{}}Mais de uma fonte de\\ \textit{Benchmark} (abrev. \textgreater Bench )\end{tabular}} & Sim & ++  & \multirow{2}{*}{10} \\ \cline{3-4} 
 &  & Não & 0 & \\ \hline \hline
\multirow{2}{*}{4} & \multirow{2}{*}{\begin{tabular}[c]{@{}c@{}}Comparação com outras abordagens\\ (abrev. Compara a Abordagem )\end{tabular}} & Sim & ++  & \multirow{2}{*}{10} \\ \cline{3-4} 
 &  & Não & 0 & \\ \hline \hline
\multirow{3}{*}{5} & \multirow{3}{*}{\begin{tabular}[c]{@{}c@{}}Suporte Compartilhamento\\  de Recursos\end{tabular}} & Não / Não identificado & 0  & \multirow{3}{*}{10} \\ \cline{3-4} 
 &  & Parcial & + & \\ \cline{3-4} 
 &  & Sim & ++ & \\ \hline \hline
\multirow{3}{*}{6} & \multirow{3}{*}{Suporte a Preempções} & Não / Não identificado & 0  & \multirow{3}{*}{10} \\ \cline{3-4} 
 &  & Parcial & + & \\ \cline{3-4} 
 &  & Sim & ++ & \\ \hline \hline
\multicolumn{4}{|r|}{\textbf{Total Pontuação Máxima:}} & \textbf{70} \\ \hline
\end{tabular}
}
\label{table:Metricas_Comparacao_Abordagens}
\end{table}

As métricas apresentadas na Tabela~\ref{table:Metricas_Comparacao_Abordagens} tem por objetivo identificar: ID=1 se a abordagem proposta possui um apoio ferramental e se este está disponível; ID=2 o tipo do \textit{benchmark} utilizado na avaliação prática da abordagem, ou seja, se for um \textit{benchmark} desenvolvido pelo próprio autor, provido da literatura, provido da indústria ou uma combinação entre estes tipos de \textit{benchmarks}; ID=3 se mais de uma fonte de \textit{benchmark} foi utilizada; ID=4 se a abordagem proposta foi comparada com outras na prática; ID=5 se a abordagem proposta dá suporte a compartilhamento de recursos, como por exemplo dispositivos de I/O; e por último ID=6 se a abordagem proposta dá suporte a preempções, ou seja, se tem uma metodologia bem definida para lidar com preempções, dado um ambiente com multitarefas e escalonamento preemptivo.

Com base nas métricas apresentadas na Tabela~\ref{table:Metricas_Comparacao_Abordagens}, todas as 39 publicações foram analisadas e classificadas seguindo os critérios definidos na Tabela~\ref{table:Criterios_Classificacao_Publicacoes}. O resultado dessa análise está ilustrada na Tabela~\ref{table:Comparacao_Completude_Abordagens}, onde está divida em basicamente quatro partes: (1) são os códigos de identificação das 39 publicações retornadas no segundo filtro, que por sua vez são compostos de 3 partes, por exemplo, para o ID = Shin\_Kim\_Lee\_2001\_a temos que: (I) são os principais autores da publicação; (II) seguido pelo ano da publicação; e por último (III) um código único para identificar a publicação, visto que alguns autores possuem várias publicações em um mesmo ano (esses códigos vão de ``a'' até ``z''); (2) é a avaliação ferramental da abordagem; (3) é quanto a avaliação experimental da abordagem; e por último (4) é uma avaliação quanto as limitações / suporte da abordagem.

\begin{table}[htp] 
\centering
\caption{Critérios de classificação das publicações selecionadas no 2\textordmasculine \, filtro.}
\begin{tabular}{|l|}
\hline
\multicolumn{1}{|c|}{\textbf{Critérios de Classificação das Publicações}} \\ \hline
1. Maior pontuação geral. \\ \hline
2. Maior pontuação no item suporte a preempções. \\ \hline
3. Maior pontuação no item ferramental. \\ \hline
4. Maior pontuação no item tipo do \textit{benchmark} utilizado. \\ \hline
5. Maior pontuação no item Comparação com outras abordagens. \\ \hline
6. Publicação mais recente. \\ \hline
\end{tabular}
\label{table:Criterios_Classificacao_Publicacoes}
\end{table}


\begin{table}[htp] 
\centering
\caption{Comparação de completude entre as abordagens.}
\resizebox{\columnwidth}{!}{%
\begin{tabular}{|l|c|c|c|c|c|c|c|}
\hline
\multicolumn{1}{|c|}{\multirow{2}{*}{\begin{tabular}[c]{@{}c@{}}\\ \\ \textbf{ID}\end{tabular}}} & \multicolumn{1}{|c|}{\multirow{2}{*}{\begin{tabular}[c]{@{}c@{}}\\ \\ \textbf{Ferramental}\end{tabular}}} & \multicolumn{3}{c|}{\textbf{Avaliação Experimental}} & \multicolumn{2}{c|}{\textbf{Limitações / Suporte}} & \multicolumn{1}{|c|}{\multirow{2}{*}{\begin{tabular}[c]{@{}c@{}}\\ \\ \textbf{Pontuação}\end{tabular}}} \\ \cline{3-7}
 &  & \textbf{T Bench} & \textbf{\textgreater Bench} & \begin{tabular}[c]{@{}c@{}}\textbf{Compara a} \\ \textbf{Abordagem}\end{tabular} & \begin{tabular}[c]{@{}c@{}}\textbf{Suporta} \\ \textbf{Compartilhamento}\\ \textbf{de Recursos}\end{tabular} & \begin{tabular}[c]{@{}c@{}}\textbf{Suporte a}\\ \textbf{Preempções}\end{tabular} &  \\ \hline

Aboughazaleh\_2003\_b			& +++ & +++ & ++ & ++ & 0  & ++ & 60 \\ \hline 
Aboughazaleh\_2003\_a			& +++ & +++ & ++ & ++ & 0  & 0  & 50 \\ \hline 
Yi\_Chen\_Yang\_2006\_a			& ++  & +++ & ++ & ++ & 0  & 0  & 45 \\ \hline 
Takase\_2011\_a					& ++  & +++ & ++ & 0  & 0  & +  & 40 \\ \hline 
Chen\_2008\_a					& 0   & +++ & ++ & ++ & 0  & +  & 40 \\ \hline 
Buss\_Givargis\_Dutt\_2003\_a	& +++ & +++ & ++ & 0  & 0  & 0  & 40 \\ \hline 
Ishihara\_2009\_a				& +++ & +++ & ++ & 0  & 0  & 0  & 40 \\ \hline 
Yuan\_2011\_a 					& 0   & +++ & ++ & ++ & 0  & 0  & 35 \\ \hline 
Xian\_Lu\_2006\_a				& 0   & +++ & ++ & ++ & 0  & 0  & 35 \\ \hline 
He\_2008\_a						& +++ & 0   & 0  & ++ & 0  & +  & 30 \\ \hline 
Seo\_Seo\_Kim\_2012\_a 			& +++ & +   & 0  & ++ & 0  & 0  & 30 \\ \hline 
Tatematsu\_2011\_a				& +++ & +   & 0  & ++ & 0  & 0  & 30 \\ \hline 
Mohan\_2010\_a					& +++ & +   & 0  & ++ & 0  & 0  & 30 \\ \hline 
Mohan\_Mueller\_Root\_2005\_a	& +++ & +   & 0  & ++ & 0  & 0  & 30 \\ \hline 
Awan\_Petters\_2012\_a			& +++ & 0   & 0  & 0  & +  & +  & 25 \\ \hline 
Seo\_Kim\_Dutt\_2005\_a			& +++ & 0   & 0  & ++ & 0  & 0  & 25 \\ \hline 
Shin\_Kim\_2005\_b				& ++  & +   & 0  & ++ & 0  & 0  & 25 \\ \hline 
Shin\_Kim\_2001\_a				& ++  & +   & 0  & ++ & 0  & 0  & 25 \\ \hline 
Hong\_Yoo\_Choi\_Kong\_2006\_a 	& ++  & +   & ++ & 0  & 0  & 0  & 25 \\ \hline 
Gheorghita\_2005\_a				& 0   & +++ & ++ & 0  & 0  & 0  & 25 \\ \hline 
Chen\_Hsieh\_Lai\_2008\_a		& 0   & 0   & 0  & ++ & 0  & ++ & 20 \\ \hline 
Chen\_Hsieh\_Lai\_2008\_b		& 0   & 0   & 0  & ++ & 0  & ++ & 20 \\ \hline 
Neishaburi\_2007\_a				& +++ & +   & 0  & 0  & 0  & 0  & 20 \\ \hline 
Lee\_Yoo\_Choi\_2002\_a			& +++ & +   & 0  & 0  & 0  & 0  & 20 \\ \hline 
Cohen\_2012\_a 					& 0   & 0   & 0  & 0  & +  & ++ & 15 \\ \hline 
Zitterell\_2008\_a				& 0   & 0   & 0  & ++ & 0  & +  & 15 \\ \hline 
Shin\_Kim\_Lee\_2001\_a			& ++  & +   & 0  & 0  & 0  & 0  & 15 \\ \hline 
Shin\_Kim\_Lee\_2001\_b			& ++  & +   & 0  & 0  & 0  & 0  & 15 \\ \hline 
Oh\_Kim\_Kim\_Kyung\_2008\_a	& 0   & +   & 0  & ++ & 0  & 0  & 15 \\ \hline 
Shin\_Kim\_2007\_a				& 0   & +   & 0  & ++ & 0  & 0  & 15 \\ \hline 
Kumar\_Manimaran\_2005\_a		& 0   & +   & 0  & ++ & 0  & 0  & 15 \\ \hline 
Shin\_Kim\_2005\_a				& 0   & +   & 0  & ++ & 0  & 0  & 15 \\ \hline 
Seo\_Kim\_Chung\_2004\_a		& 0   & +   & 0  & ++ & 0  & 0  & 15 \\ \hline 
Zhang\_2006\_a					& ++  & 0   & 0  & 0  & 0  & 0  & 10 \\ \hline 
Yang\_2009\_a					& 0   & 0   & 0  & ++ & 0  & 0  & 10 \\ \hline 
Cassandras\_Zhuang\_2005\_a		& 0   & 0   & 0  & ++ & 0  & 0  & 10 \\ \hline 
Yi\_Yang\_Chen\_2005\_a			& 0   & 0   & 0  & ++ & 0  & 0  & 10 \\ \hline 
Yi\_Yang\_2006\_a				& 0   & 0   & 0  & 0  & 0  & 0  & 0  \\ \hline 
Yi\_Yang\_2005\_a				& 0   & 0   & 0  & 0  & 0  & 0  & 0  \\ \hline 

\end{tabular}
}
\label{table:Comparacao_Completude_Abordagens}
\end{table}


É importante comentar que todas as publicações apresentaram resultados positivos quanto a avaliação qualitativa das metodologias catalogadas, mesmo considerando algumas limitações (como por exemplo: não dar suporte a preempções e nem a recursos compartilhados), pois todas utilizaram a técnica DVFS intra-tarefa e apresentaram bons níveis de redução do consumo de energia do processador.

\begin{figure}[ht]
	\centering
	\resizebox{15cm}{!}{\includegraphics{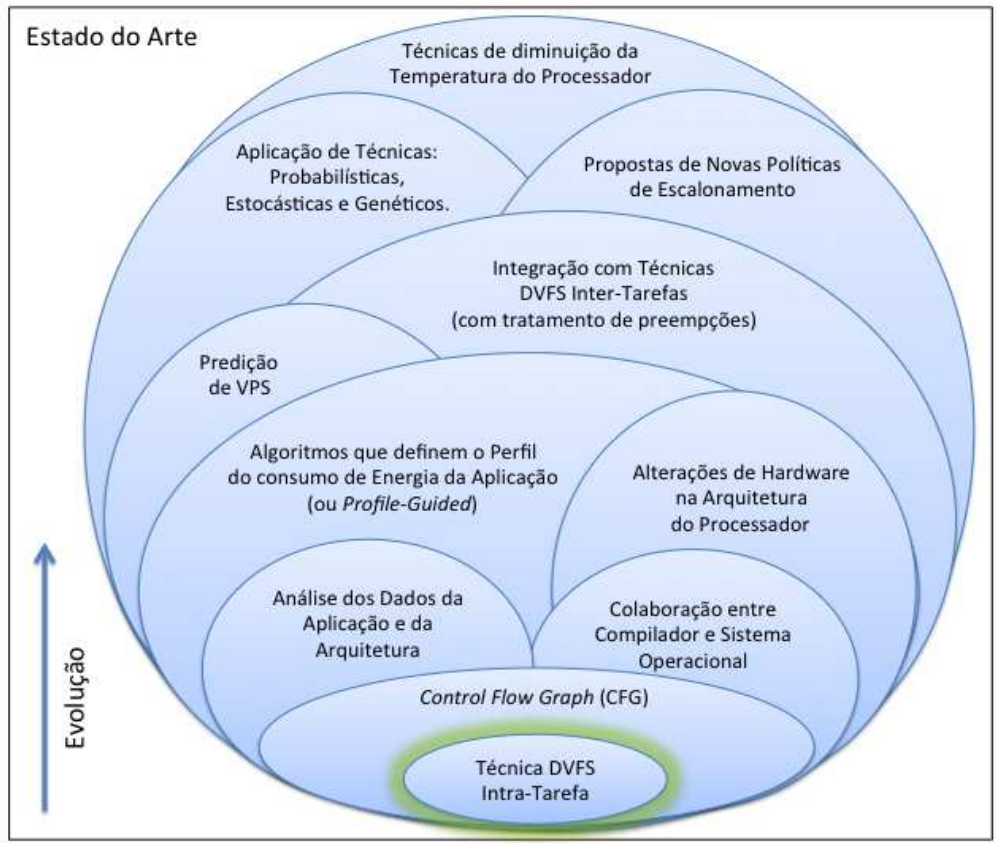}}
	\caption{O diagrama mostra uma visão mais abrangente da evolução do estado da arte na área de baixo consumo de energia, do ponto de vista da técnica DVFS intra-tarefa.}
	\label{fig:diagrama_evolucao_estado_arte}
\end{figure}

Para finalizar esta seção e a etapa de analises, foi feito um diagrama para caracterizar as evoluções do estado da arte ao longo dos últimos anos, ver Figura~\ref{fig:diagrama_evolucao_estado_arte}.

Portanto, todos os argumentos apresentados nessa seção servem de embasamento teórico para responder a questão principal de pesquisa desta revisão sistemática, que procurou extrair o máximo de informações possíveis sobre cada metodologia, a fim de se definir melhor a caracterização do estado da arte sobre as metodologias que utilizam como base a técnica DVFS intra-tarefa.

\section{Considerações Finais}
\label{section:consideracoes_finais}

Neste trabalho apresentamos uma pesquisa que visou caracterizar o estado da arte, através de uma revisão sistemática, dos principais métodos que utilizam a técnica DVFS intra-tarefa, aplicado no contexto de sistemas de tempo real com o objetivo de reduzir o consumo de energia do processador.

A revisão sistemática foi conduzida com base em três etapas, Planejamento da Revisão, Condução da Revisão e Análise dos Resultados (mais detalhes na Seção~\ref{section:Revisao_Sistematica}). Para delinear o escopo da pesquisa foram estabelecidos critérios para garantir, de forma equilibrada, a viabilidade da execução (custo, esforço e tempo), acessibilidade aos dados e abrangência do estudo. A biblioteca selecionada foi a Scopus (ver Seção~\ref{subsubsection:Metodos_Busca_Publicacoes}), pois ela possui em sua base mais de 5,000 editoras internacionais, tais como IEEE Xplore Digital Library, ACM Digital Library e Springer.

A partir da execução da expressão de busca (ver Seção~\ref{subsubsection:Expressao_Busca}) na biblioteca Scopus obtivemos como resultado um total de 253 publicações, sendo que apenas 115 faziam referências a abordagens que citam o baixo consumo de energia em sistemas de tempo real e 39 publicações somente faziam referências a baixo consumo de energia do processador utilizando a técnica DVFS intra-tarefa. Também podemos observar, com base no número de publicações, que a partir de 2005 o número de publicações nessa área vem declinando gradativamente, representando assim uma queda de 76,47\% quando comparado com o ano de 2013. Isso mostra que essa área de pesquisa está chegando ao seu ponto de saturação, onde propor novas contribuições está sendo cada vez mais desafiador para a comunidade científica.

Com base na análise das 39 publicações selecionadas, a partir da aplicação da revisão sistemática, podemos observar que:

\begin{enumerate}
\item Todas as publicações apresentam resultados positivos na aplicação de suas respectivas abordagens, isto no que diz respeito a obter os resultados esperados para os quais foram projetadas, embora possuam limitações, como por exemplo, não fornecer suporte a recursos compartilhados e / ou preempções;

\item Com relação ao modo de aplicação dos métodos, ou seja, se era de forma Online, Offline ou Híbrida, identificamos que 56\% dos métodos utilizam o modo Offline para análise e aplicação de suas abordagens. Contudo vale ressaltar que ainda existem poucos abordagens totalmente Online, apenas 8\% dos trabalhos catalogados. Isso ocorre, principalmente pelo fato da técnica DVFS intra-tarefa necessitar de etapa estáticas (ou \textit{Offline}) para a aplicação dessa técnica. Por outro lado, o número de abordagens híbridas vêm crescendo significativamente nos últimos anos, principalmente com a integração das técnicas DVFS intra e inter tarefas, proporcionando assim características \textit{Offline e Online}, respectivamente. Acreditamos que as abordagens híbridas, que hoje representam cerca de 36\% das abordagens catalogadas nessa revisão sistemática, passarão a ter impacto mais significativo em um futuro não muito distante, em relação aos demais modos de execução.

\item Grande parte dos métodos (cerca de 45\%) possuem apoio ferramental para aplicação do método proposto;

\item Em relação as limitações das publicações catalogadas, temos que apenas 5\% das metodologias dão suporte a compartilhamento de recursos e apenas 23\% das publicações fornecem esse suporte a preempções. Esse dados mostram para a comunidade científica que muitas linhas de pesquisa podem ser exploradas dentro deste contexto.

\item Quanto as perspectivas futuras dos trabalhos catalogados, muitas são apontadas por seus próprios autores, com o objetivo de contribuir com novas diretrizes para o avanço do estado da arte e para o desenvolvimento de novas linhas de pesquisa. Algumas dessas novas diretrizes estão disponíveis no \ref{section:Apendice_Base_Dados_Revisao_Sistematica}.
\end{enumerate}

Analisando os dados e fatos identificados neste trabalho, planejamos como próximos passos na continuidade desta pesquisa: (1) Auditar este documento, por meio de auditores que não estejam ligados diretamente a pesquisa, a fim de facilitar a identificação de possíveis erros nos relatórios e/ou nas avaliações; (2) Fazer um relatório detalhado sobre os \textit{benchmarks} catalogados; (3) Fazer um relatório detalhado sobre as ferramentas catalogadas; e por último (4) Fazer um relatório detalhado comparando os desempenhos entre as abordagens catalogadas. Dessa forma, podemos direcionar as linhas de pesquisa para dar contribuições mais significativas para área de baixo consumo de energia do processador aplicado no contexto de sistemas de tempo real.

\section*{Agradecimentos}
Os autores agradecem o apoio concedido pela Fundação de Amparo à Pesquisa do Estado do Amazonas (FAPEAM), Superintendência da Zona Franca de Manaus (SUFRAMA) e ao Conselho Nacional de Desenvolvimento Científico e Tecnológico (CNPQ).

\backmatter

\newpage
\bibliographystyle{ecs}
\bibliography{References/references}

\newpage
\appendix

% ...
\section{Documentos Adicionais da Revisão Sistemática}
\label{section:Apendice_Conducao_Expressao_Busca}

\subsection{Processo de Construção da Expressão de Busca}
\label{subsection:Processo_Construcao_Expressao_Busca}

O protocolo descrito na Seção~\ref{subsubsection:Expressao_Busca} é o cerne para a execução do estudo baseado em revisão sistemática, entretanto, o seu planejamento começou antes de sua elaboração. Para a construção do protocolo, foi realizada uma pesquisa informal na literatura sobre publicações que tratavam especificamente de metodologias que reduziam o consumo de energia do processador através do uso da técnica DVFS intra-tarefa. No total, foram selecionadas 12 publicações, todos no idioma inglês, para compor a lista de controle (ver Tabelas~\ref{table:Publicacoes_Selecionadas_Lista_Controle_Parte_1}, ~\ref{table:Publicacoes_Selecionadas_Lista_Controle_Parte_2}, ~\ref{table:Publicacoes_Selecionadas_Lista_Controle_Parte_3} e ~\ref{table:Publicacoes_Selecionadas_Lista_Controle_Parte_4}).

\begin{table}[htp] 
	\centering
	\caption{Lista de publicações que compõem a lista de controle desta revisão sistemática (Parte 1).}
	
	\resizebox{\textwidth}{!}{%
	\begin{tabular}{|c|p{4cm}|p{3cm}|c|p{6cm}|}
	\hline
	\textbf{N\textordmasculine \,} & \textbf{Título} & \textbf{Autor(es)} & \textbf{Ano} & \textbf{Palavras-Chave} \\ \hline
	
	01 & Online intra-task device scheduling for hard real-time systems & Awan, M.A. and Petters, S.M. & 2012 & cells (electric); power aware computing; real-time systems; IO devices; battery life enhancement; device transition overhead; device transitions; energy consumption; energy resources; hard real-time systems; inter alia; intra-task device scheduling algorithm; online intra-task device scheduling; power dissipation; power saving mechanisms; real time systems; shut-down devices; system schedulability; technology enhancements; Containers; Delay; Energy consumption; Processor scheduling; Real-time systems; Schedules; Scheduling \\ \hline

	02 & A car racing based strategy for the Dynamic Voltage and Frequency Scaling technique & Cohen, D. and Valentin, E. and Barreto, R. and Oliveira, H. and Cordeiro, L. & 2012 & energy consumption; power aware computing; car racing analogy; car racing based strategy; dynamic frequency scaling technique; dynamic voltage scaling technique; energy consumption optimization; energy overheads; low energy consumption; multiple preemptable real-time tasks; timing overheads; Energy consumption; Frequency control; Optimization; Program processors; Real time systems; Time frequency analysis; Timing \\ \hline

	03 & Algorithms for Combined Inter- and Intra-Task Dynamic Voltage Scaling. & Seo, Hyungjung and Seo, Jaewon and Kim, Taewhan & 2012 & dynamic voltage scaling (DVS); power saving; embedded systems \\ \hline
	
	\end{tabular}
	}
	\label{table:Publicacoes_Selecionadas_Lista_Controle_Parte_1}
\end{table}

\begin{table}[htp] 
	\centering
	\caption{Lista de publicações que compõem a lista de controle desta revisão sistemática (Parte 2).}
	
	\resizebox{\textwidth}{!}{%
	\begin{tabular}{|c|p{4cm}|p{3cm}|c|p{6cm}|}
	\hline
	\textbf{N\textordmasculine \,} & \textbf{Título} & \textbf{Autor(es)} & \textbf{Ano} & \textbf{Palavras-Chave} \\ \hline

	04 & Checkpoint Extraction Using Execution Traces for Intra-task DVFS in Embedded Systems & Tatematsu, T. and Takase, H. and Gang Zeng and Tomiyama, H. and Takada, H. & 2011 & embedded systems; microprocessor chips; checkpoint extraction; frequency scaling; intratask dynamic voltage; processor frequency; worst case execution cycles; Data mining; Embedded systems; Energy consumption; Equations; Greedy algorithms; Mathematical model; Time frequency analysis; DVFS; embedded system; execution trace; low energy \\ \hline

	05 & Energy efficient intra-task dynamic voltage scaling for realistic CPUs of mobile devices & Yang,C. -. and Wang,K. and Lin,M. -. and Lin,P. & 2009 & CPU;  Dynamic voltage scaling;  Energy efficient;  Intra-task;  Mobile device;  Real time \\ \hline

	06 & Efficient Algorithms for Jitterless Real-Time Tasks to DVS Schedules & Da-Ren Chen and Shu-Ming Hsieh and Ming-Fong Lai & 2008 & computational complexity; jitter; power aware computing; real-time systems; scheduling; canonical schedule; harmonic period; intra-task dynamic voltage scale scheduling algorithm; jitterless real-time task; periodic task scheduling; variable voltage processor; Distributed computing; Dynamic voltage scaling; Energy consumption; Information management; Interference; Polynomials; Processor scheduling; Scheduling algorithm; Timing jitter; Voltage control; DVS scheduling; power-aware scheduling; real-time systems \\ \hline

	07 & Task partitioning algorithm for intra-task dynamic voltage scaling & Seungyong Oh and Jungsoo Kim and Seonpil Kim and Chong-Min Kyung & 2008 & CMOS integrated circuits; low-power electronics; power consumption; switching; CMOS circuits; DVS; H.264 decoder software; intratask dynamic voltage scaling; power consumption; task partitioning algorithm; voltage switching; Computer science; DC-DC power converters; Dynamic voltage scaling; Embedded system; Energy consumption; Frequency; Heuristic algorithms; Partitioning algorithms; Switching converters; Voltage control	 \\ \hline
	
	\end{tabular}
	}
	\label{table:Publicacoes_Selecionadas_Lista_Controle_Parte_2}
\end{table}

\begin{table}[htp] 
	\centering
	\caption{Lista de publicações que compõem a lista de controle desta revisão sistemática (Parte 3).}
	
	\resizebox{\textwidth}{!}{%
	\begin{tabular}{|c|p{4cm}|p{3cm}|c|p{6cm}|}
	\hline
	\textbf{N\textordmasculine \,} & \textbf{Título} & \textbf{Autor(es)} & \textbf{Ano} & \textbf{Palavras-Chave} \\ \hline

	08 & Optimizing Intratask Voltage Scheduling Using Profile and Data-Flow Information & Dongkun Shin and Jihong Kim & 2007 & dynamic scheduling; low-power electronics; voltage control; LalntraDVS; RAEP-IntraDVS; data-flow information; dynamic-voltage scaling; intratask voltage scheduling; look-ahead IntraDVS; low-power design; power management; real-time systems; remaining average-case execution path; variable-voltage processor; voltage-scaling points; Clocks; Collaboration; Dynamic scheduling; Energy consumption; Energy management; Information technology; Partitioning algorithms; Processor scheduling; Real time systems; Voltage control; Dynamic-voltage scaling; low-power design; power management; real-time systems; variable-voltage processor \\ \hline

	09 & Intra-task voltage scheduling on DVS-enabled hard real-time systems & Dongkun Shin and Jihong Kim & 2005 & low-power electronics; microprocessor chips; processor scheduling; real-time systems; voltage control; DVS-unaware program; IntraDVS framework; average-case execution information; dynamic voltage scaling; energy reduction ratio; hard real-time systems; intra-task voltage scheduling; low-energy hard real-time applications; low-energy program; low-power design; power management; program execution; slack times; software tools; static timing analysis; supply voltage control; variable-voltage processor; worst-case execution information; Decoding; Dynamic scheduling; Energy consumption; Energy efficiency; MPEG standards; Real time systems; Scheduling algorithm; Software tools; Timing; Voltage control; Dynamic voltage scaling; low-power design; power management; real-time systems; variable-voltage processor \\ \hline
	
	\end{tabular}
	}
	\label{table:Publicacoes_Selecionadas_Lista_Controle_Parte_3}
\end{table}

\begin{table}[htp] 
	\centering
	\caption{Lista de publicações que compõem a lista de controle desta revisão sistemática (Parte 4).}
	
	\resizebox{\textwidth}{!}{%
	\begin{tabular}{|c|p{4cm}|p{3cm}|c|p{6cm}|}
	\hline
	\textbf{N\textordmasculine \,} & \textbf{Título} & \textbf{Autor(es)} & \textbf{Ano} & \textbf{Palavras-Chave} \\ \hline
	10 & Collaborative operating system and compiler power management for real-time applications & Aboughazaleh N., Mosse D., Childers B., Melhem R., Craven M.
	& 2003 & embedded systems; energy conservation; operating systems (computers); power consumption; program compilers; automatic target recognition application; battery operated portable system; collaborative operating system; compiler power management; dynamic voltage scaling; embedded system; energy consumption; real-time application; real-time system; temporal behavior; video decoder; Battery management systems; Collaboration; Dynamic voltage scaling; Embedded system; Energy consumption; Energy management; Operating systems; Power system management; Real time systems; Voltage control \\ \hline

	11 & Profile-based optimal intra-task voltage scheduling for hard real-time applications & Seo, Jaewon and Kim, Taewhan and Chung, Ki-Seok & 2004 & DVS, intra-task voltage scheduling, low energy design \\ \hline

	12 & Low-energy intra-task voltage scheduling using static timing analysis & Shin, Dongkun and Kim, Jihong and Lee, Seongsoo & 2001 & Algorithms;  Computer aided software engineering;  Image coding;  VLSI circuits;  Voltage scheduling;  Real time systems \\ \hline
	
	\end{tabular}
	}
	\label{table:Publicacoes_Selecionadas_Lista_Controle_Parte_4}
\end{table}

Uma vez definida a lista de controle da revisão sistemática foi possível dar inicio ao processo de definição da expressão de busca, que se iniciou a partir da coleta das palavras-chave de todas as publicações presentes na lista de controle. Em seguida, foram extraídas apenas as palavras-chave em comum a todas elas. Então,  iniciou-se a fase de testes (no buscador da Scopus) para composição da expressão de busca definitiva desta revisão sistemática. No total, foram realizadas 7 rodadas de testes (com as palavras-chave em inglês) até que a expressão de busca ficasse a mais concisa possível e retornasse todas as publicações presentes na lista de controle. Essas rodadas de testes foram necessárias, pois estavam retornando muitas publicações, principalmente devido ao fato das máquinas de busca não serem tão eficientes. Portanto, a definição da expressão de busca envolveu os seguintes passos:

\begin{enumerate}
	\item Definição da máquina de busca para o teste do protocolo;
	
	\item Identificação de expressão de busca inicial;
	
	\item Testes com a expressão de busca;
	
	\item Análise dos resultados retornados pela expressão de busca.
\end{enumerate}

Esse processo foi feito de forma iterativa, utilizando os passos 3 e 4, até que o resultado fosse considerado satisfatório. Durante os testes com a expressão de busca, verificou-se que as base de dados da Scopus (\textit{http://www.scopus.com}) retornavam todos os artigos da lista de controle. Por este motivo, ela foi escolhida para a realização dos testes.

A pesquisa foi restrita às áreas de Computação, Engenharia e Energia, que são as áreas de interesse desta revisão sistemática. Durante as consultas na máquina de busca, foram lidos os resumos e as palavras-chave de todas as referências identificadas e aquelas que eram de interesse foram selecionadas. O  \ref{section:Apendice_Filtros_1_e_2} mostram todas as publicações catalogados após a execução do 1\textordmasculine \, e 2\textordmasculine \, filtros e o \ref{section:Apendice_Base_Dados_Revisao_Sistematica} mostra a base de dados criada a partir dos dados extraídos das publicações selecionadas após a execução do 2\textordmasculine \, filtro.

\section{Lista de Publicações Catalogadas após o 1º e 2º Filtros}
\label{section:Apendice_Filtros_1_e_2}

Abaixo seguem as listas de artigos obtidos na aplicação desta revisão sistemática. Esta lista contém as seguintes informações: (1) o título da publicação (Título); (2) o nome dos autores (Autores); (3) o título onde foi efetuado a publicação, ou seja, nome da conferência, jornal, entre outros (Fonte da Publicação); (4) o ano que foi efetuado a publicação (Ano); (5) o nome da editora da publicação (Editora); (6) é identificado (S - Sim ou N - Não) se a publicação esta disponível na web (Disp); e nas últimas colunas é identificado se a publicação foi aprovada (S - Sim ou N - Não) no seu respectivo filtro (1F - primeiro filtro e 2F - segundo filtro) e adicionalmente qual critério (na coluna Critério ao lado de cada filtro) foi utilizado para identificar sua aprovação ou não. Vale ressaltar que os critério marcado como N/A (Nenhum dos critérios Aplicados) está ligado ao fato da publicação ter sido reprovada em um filtro anterior.

\begin{table}[htp] 
\centering
\caption{Lista de publicações catalogadas após o 1\textordmasculine \, e 2\textordmasculine \, filtros - Parte 1.}
\resizebox{\columnwidth}{!}{%
\begin{tabular}{|>{\centering\arraybackslash}p{6cm}|>{\centering\arraybackslash}p{4cm}|>{\centering\arraybackslash}p{4cm}|c|>{\centering\arraybackslash}p{3cm}|c|c|>{\centering\arraybackslash}p{3cm}|c|>{\centering\arraybackslash}p{3cm}|}
\hline
\textbf{Título} & \textbf{Autores} & \textbf{Fonte da Publicação} & \textbf{Ano} & \textbf{Editora} & \textbf{Disp} & \textbf{1F} & \textbf{Critério} & \textbf{2F} & \textbf{Critério} \\ \hline

A work-demand analysis compatible with preemption-aware scheduling for power-aware real-time tasks & Chen D.-R., Chen M.-Y., Chen Y.-S., Chen L.-C. & Mathematical Problems in Engineering & 2013 & Hindawi Publishing Corporation & S & S & CI1-01 & N & CE2-02 \\ \hline

HAPPE: Human and application-driven frequency scaling for processor power efficiency & Yang L., Dick R.P., Memik G., Dinda P. & IEEE Transactions on Mobile Computing & 2013 & IEEE & S & N & CE1-05 & N/A & N/A \\ \hline

Energy optimization by exploiting execution slacks in streaming applications on multiprocessor systems & Singh A.K., Das A., Kumar A. & Proceedings - Design Automation Conference & 2013 & ACM & S & S & CI1-01 & N & CE2-02 \\ \hline

Enabling efficient system configurations for dynamic wireless applications using system scenarios & Zompakis N., Papanikolaou A., Raghavan P., Soudris D., Catthoor F. & International Journal of Wireless Information Networks & 2013 & Springer & N/A & N & CE1-05 & N/A & N/A \\ \hline

Literature survey & Ogras U.Y., Marculescu R. & Lecture Ntes in Electrical Engineering & 2013 & Springer & N & N & CE1-02 / CE1-04 / CE1-05 / CE1-06 & N/A & N/A \\ \hline

Counter-based power modeling methods: Top-down vs. bottom-Up & Bertran R., Gonzalez M., Martorell X., Navarro N., Ayguade E. & Computer Journal & 2013 & Oxford Journals & N/A & N & CE1-04 & N/A & N/A \\ \hline

Thermal-aware task scheduling in 3D chip multiprocessor with real-time constrained workloads & Li J., Qiu M., Niu J.-W., Yang L.T., Zhu Y., Zhong M. & Transactions on Embedded Computing Systems & 2013 & ACM & N/A & N & CE1-04 & N/A & N/A \\ \hline

Adaptive energy optimization in multimedia-centric wireless devices: A survey & Kennedy M., Ksentini A., Hadjadj-Aoul Y., Muntean G.-M. & IEEE Communications Surveys and Tutorials & 2013 & IEEE & N/A & N & CE1-05 & N/A & N/A \\ \hline

Online intra-task device scheduling for hard real-time systems & Awan M.A., Petters S.M. & 7th IEEE International Symposium on Industrial Embedded Systems, SIES 2012 - Conference Proceedings & 2012 & IEEE & S & S & CI1-01 & S & CI2-01 \\ \hline

Evaluation of core performance when the Nde is power capped using Intel data center manager & McCartney J., Teller P.J., Arunagiri S. & Proceedings of the International Conference on Parallel Processing Workshops & 2012 & IEEE & N/A & N & CE1-04 & N/A & N/A \\ \hline

An optimal fully distributed algorithm to minimize the resource consumption of cloud applications & Tziritas N., Khan S.U., Xu C.-Z., Hong J. & Proceedings of the International Conference on Parallel and Distributed Systems - ICPADS & 2012 & IEEE & N/A & N & CE1-04 & N/A & N/A \\ \hline

DTS: Using dynamic time-slice scaling to address the OS problem incurred by DVFS & Jia G., Gao X., Li X., Wang C., Zhou X. & Proceedings - 2012 IEEE International Conference on Cluster Computing Workshops, Cluster Workshops 2012 & 2012 & IEEE & N/A & N & CE1-04 & N/A & N/A \\ \hline

Dynamic phase-based tuning for embedded systems using phase distance mapping & Adegbija T., Gordon-Ross A., Munir A. & Proceedings - IEEE International Conference on Computer Design: VLSI in Computers and Processors & 2012 & IEEE & N/A & N & CE1-06 & N/A & N/A \\ \hline

Runtime voltage/frequency scaling for energy-aware streaming applications & Gruian F. & Conference Record - Asilomar Conference on Signals, Systems and Computers & 2012 & IEEE & N/A & N & CE1-05 & N/A & N/A \\ \hline
\end{tabular}
}
\end{table}

\begin{table}[htp] 
\centering
\caption{Lista de publicações catalogadas após o 1\textordmasculine \, e 2\textordmasculine \, filtros - Parte 2.}
\resizebox{\columnwidth}{!}{%
\begin{tabular}{|>{\centering\arraybackslash}p{6cm}|>{\centering\arraybackslash}p{4cm}|>{\centering\arraybackslash}p{4cm}|c|>{\centering\arraybackslash}p{3cm}|c|c|>{\centering\arraybackslash}p{3cm}|c|>{\centering\arraybackslash}p{3cm}|}
\hline
\textbf{Título} & \textbf{Autores} & \textbf{Fonte da Publicação} & \textbf{Ano} & \textbf{Editora} & \textbf{Disp} & \textbf{1F} & \textbf{Critério} & \textbf{2F} & \textbf{Critério} \\ \hline

Thermal aware processor operation point management & Gorti N.P.K., Somani A.K. & Proceedings - 2012 IEEE/ACM 5th International Conference on Utility and Cloud Computing, UCC 2012 & 2012 & IEEE & S & S & CI1-01 & N & CE2-01 \\ \hline

Scheduling precedence constrained tasks with reduced processor energy on multiprocessor computers & Li K. & IEEE Transactions on Computers & 2012 & IEEE & N/A & N & CE1-05 / CE1-06 & N/A & N/A \\ \hline

Algorithms for combined inter- and intra-task dynamic voltage scaling & Seo H., Seo J., Kim T. & Computer Journal & 2012 & Oxford Journals & S & S & CI1-01 & S & CI2-01 \\ \hline

The time-keeping aNmaly of energy-saving sensors: Manifestation, solution, and a structural monitoring case study & Moinzadeh P., Mechitov K., Shiftehfar R., Abdelzaher T., Agha G., Spencer B.F. & Annual IEEE Communications Society Conference on Sensor, Mesh and Ad Hoc Communications and Networks workshops & 2012 & IEEE & N/A & N & CE1-05 & N/A & N/A \\ \hline

An experimental evaluation of real-time DVFS scheduling algorithms & Saha S., Ravindran B. & ACM International Conference Proceeding Series & 2012 & ACM & S & N & CE1-04 & N/A & N/A \\ \hline

Energy-aware dynamic slack allocation for real-time multitasking systems & Wang W., Ranka S., Mishra P. & Sustainable Computing: Informatics and Systems & 2012 & ScienceDirect & S & S & CI1-01 & N & CE2-02 \\ \hline

Performance evaluation of a real-time grid system using power-saving capable processors & Terzopoulos G., Karatza H. & Journal of Supercomputing & 2012 & Springer & S & S & CI1-01 & N & CE2-02 \\ \hline

Efficient algorithms for extracting pareto-optimal hardware configurations in DEPS framework & Kawashima H., Zeng G., Takase H., Edahiro M., Takada H. & IPSJ Transactions on System LSI Design Methodology & 2012 & J-STAGE & N/A & N & CE1-05 & N/A & N/A \\ \hline

A car racing based strategy for the dynamic voltage and frequency scaling technique & Cohen D., Valentin E., Barreto R., Oliveira H., Cordeiro L. & IEEE International Symposium on Industrial Electronics & 2012 & IEEE & S & S & CI1-01 & S & CI2-01 \\ \hline

Application-support particle filter for dynamic voltage scaling of multimedia applications & Lee J.-B., Kim M.-J., Yoon S., Chung E.-Y. & IEEE Transactions on Computers & 2012 & IEEE & N/A & N & CE1-05 & N/A & N/A \\ \hline

Dynamic cache reconfiguration for soft real-time systems & Wang W., Mishra P., Gordon-Ross A. & Transactions on Embedded Computing Systems & 2012 & ACM & S & S & CI1-01 & N & CE2-02 \\ \hline

Optimal configuration of a multicore server processor for managing the power and performance tradeoff & Li K. & Journal of Supercomputing & 2012 & Springer & N/A & N & CE1-05 & N/A & N/A \\ \hline

Three-phase algorithms for task scheduling in distributed mobile DSP system with lifetime constraints & Li J., Qiu M., Niu J.-W., Zhu Y., Liu M., Chen T. & Journal of Signal Processing Systems & 2012 & Springer & N/A & N & CE1-05 & N/A & N/A \\ \hline

Dynamic slack reclamation strategy for multiprocessor systems & Mhedhbi I., Ben Atitallah R., Jemai A. & Proceedings of the Mediterranean Electrotechnical Conference - MELECON & 2012 & IEEE & S & S & CI1-01 & N & CE2-02 \\ \hline

Reducing the energy cost of computing through efficient co-scheduling of parallel workloads & Hankendi C., Coskun A.K. & Proceedings -Design, Automation and Test in Europe, DATE & 2012 & IEEE & N/A & N & CE1-05 & N/A & N/A \\ \hline

Energy efficient scheduling of parallel tasks on multiprocessor computers & Li K. & Journal of Supercomputing & 2012 & Springer & N/A & N & CE1-05 & N/A & N/A \\ \hline

Energy efficiency with runtime models for energy-aware embedded systems & Karakehayov Z., Vassev E. & Proceedings - 2011 34th IEEE Software Engineering Workshop, SEW 2011 & 2012 & IEEE & S & S & CI1-01 & N & CE2-02 \\ \hline

Exploiting media stream similarity for energy-efficient decoding and resource prediction & Hamers J., Eeckhout L. & Transactions on Embedded Computing Systems & 2012 & ACM & N/A & N & CE1-05 & N/A & N/A \\ \hline

Optimal power allocation among multiple heterogeneous servers in a data center & Li K. & Sustainable Computing: Informatics and Systems & 2012 & ScienceDirect & N/A & N & CE1-05 & N/A & N/A \\ \hline

Design and analysis of heuristic algorithms for power-aware scheduling of precedence constrained tasks & Li K. & IEEE International Symposium on Parallel and Distributed Processing Workshops and Phd Forum & 2011 & IEEE & N/A & N & CE1-05 & N/A & N/A \\ \hline

Performance optimization with energy constraint in heterogeneous multiple computer systems & Li K. & IEEE International Symposium on Parallel and Distributed Processing Workshops and Phd Forum & 2011 & IEEE & N/A & N & CE1-05 & N/A & N/A \\ \hline

Identifying the optimal energy-efficient operating points of parallel workloads & Cochran R., Hankendi C., Coskun A., Reda S. & IEEE/ACM International Conference on Computer-Aided Design, Digest of Technical Papers, ICCAD & 2011 & IEEE & N/A & N & CE1-05 & N/A & N/A \\ \hline
\end{tabular}
}
\end{table}

\begin{table}[htp] 
\centering
\caption{Lista de publicações catalogadas após o 1\textordmasculine \, e 2\textordmasculine \, filtros - Parte 3.}
\resizebox{\columnwidth}{!}{%
\begin{tabular}{|>{\centering\arraybackslash}p{6cm}|>{\centering\arraybackslash}p{4cm}|>{\centering\arraybackslash}p{4cm}|c|>{\centering\arraybackslash}p{3cm}|c|c|>{\centering\arraybackslash}p{3cm}|c|>{\centering\arraybackslash}p{3cm}|}
\hline
\textbf{Título} & \textbf{Autores} & \textbf{Fonte da Publicação} & \textbf{Ano} & \textbf{Editora} & \textbf{Disp} & \textbf{1F} & \textbf{Critério} & \textbf{2F} & \textbf{Critério} \\ \hline

UML-based analysis of power consumption for real-time embedded systems & Hagner M., Aniculaesei A., Goltz U. & Proc. 10th IEEE Int. Conf. on Trust, Security and Privacy in Computing and Communications, TrustCom 2011, 8th IEEE Int. Conf. on Embedded Software and Systems, ICESS 2011, 6th Int. Conf. on FCST 2011 & 2011 & IEEE & S & S & CI1-01 & N & CE2-02 \\ \hline

A QoS guaranteed cache design for environment friendly computing & Lo S.-W., Huang W.-Y., Qiu S.-F., Lin Y.-C., Lin K.-H., Lin H., Kuo T.-W. & Proceedings - 2011 IEEE/ACM International Conference on Green Computing and Communications, GreenCom 2011 & 2011 & IEEE & N/A & N & CE1-06 & N/A & N/A \\ \hline

Power minimisation for real-time dataflow applications & Nelson A., Moreira O., MolNs A., Stuijk S., Nguyen B.T., Goossens K. & Proceedings - 2011 14th Euromicro Conference on Digital System Design: Architectures, Methods and Tools, DSD 2011 & 2011 & IEEE & N/A & S & CI1-01 & N & CE2-02 \\ \hline

Random dynamic voltage scaling design to enhance security of NCL S-box & Sui C., Wu J., Shi Y., Kim Y.-B., Choi M. & Midwest Symposium on Circuits and Systems & 2011 & IEEE & N/A & N & CE1-04 & N/A & N/A \\ \hline

Slack computation for DVS algorithms in fixed-priority real-time systems using fluid slack analysis & Chen D.-R. & Journal of Systems Architecture & 2011 & ScienceDirect & S & S & CI1-01 & N & CE2-02 \\ \hline

TALk: A temperature-aware leakage minimization technique for real-time systems & Yuan L., Leventhal S.R., Gu J., Qu G. & IEEE Transactions on Computer-Aided Design of Integrated Circuits and Systems & 2011 & IEEE & S & S & CI1-01 & S & CI2-01 \\ \hline

An integrated optimization framework for reducing the energy consumption of embedded real-time applications & Takase H., Zeng G., Gauthier L., Kawashima H., Atsumi N., Tatematsu T., Kobayashi Y., Kohara S., Koshiro T., Ishihara T., Tomiyama H., Takada H. & Proceedings of the International Symposium on Low Power Electronics and Design & 2011 & IEEE & S & S & CI1-01 & S & CI2-01 \\ \hline

OS-level power minimization under tight performance constraints in general purpose systems & Ayoub R., Ogras U., Gorbatov E., Jin Y., Kam T., Diefenbaugh P., Rosing T. & Proceedings of the International Symposium on Low Power Electronics and Design & 2011 & IEEE & S & N & CE1-05 & N/A & N/A \\ \hline

Safe Intra-Task Voltage Scheduling Algorithm Using Scenario Detection & Chen J., Yi B.-S. & Dianzi Keji Daxue Xuebao/Journal of the University of Electronic Science and TechNlogy of China & 2011 & Wanfang Data & N & N & CE1-02 & N/A & N/A \\ \hline

Checkpoint extraction using execution traces for intra-task DVFS in embedded systems & Tatematsu T., Takase H., Zeng G., Tomiyama H., Takada H. & Proceedings - 2011 6th IEEE International Symposium on Electronic Design, Test and Application, DELTA 2011 & 2011 & IEEE & S & S & CI1-01 & S & CI2-01 \\ \hline

Energy-Aware task scheduling for real-time systems with discrete frequencies & Qian D., Zhang Z., Hu C., Ji X. & IEICE Transactions on Information and Systems & 2011 & J-STAGE & N & N & CE1-02 & N/A & N/A \\ \hline

Saving register-file static power by monitoring instruction sequence in ROB & Shieh W.-Y., Chen H.-D. & Journal of Systems Architecture & 2011 & ScienceDirect & N/A & N & CE1-05 & N/A & N/A \\ \hline

Overhead-aware energy optimization for real-time streaming applications on multiprocessor system-on-chip & Wang Y., Liu H., Liu D., Qin Z., Shao Z., Sha E.H.-M. & ACM Transactions on Design Automation of Electronic Systems & 2011 & ACM & N/A & N & CE1-05 & N/A & N/A \\ \hline

Quasi-static voltage scaling for energy minimization with time constraints & Andrei A., Eles P., JovaNvic O., Schmitz M., Ogniewski J., Peng Z. & IEEE Transactions on Very Large Scale Integration (VLSI) Systems & 2011 & IEEE & N/A & N & CE1-05 & N/A & N/A \\ \hline

Low power scheduling for periodic real-time systems with Dynamic Voltage Scaling processor & Qian D., Zhang Z., Tian X., Hu C. & ICCASM 2010 - 2010 International Conference on Computer Application and System Modeling, Proceedings & 2010 & IEEE & S & S & CI1-01 & N & CE2-02 \\ \hline

A multi-performance processor for reducing the energy consumption of real-time embedded systems & Ishihara T. & IEICE Transactions on Fundamentals of Electronics, Communications and Computer Sciences & 2010 & J-STAGE & N & N & CE1-02 & N/A & N/A \\ \hline

An efficient DVS algorithm for fixed-priority real-time applications & Chen D.-R., Chen Y.-S. & Proceedings - International Symposium on Parallel and Distributed Processing with Applications, ISPA 2010 & 2010 & IEEE & S & S & CI1-01 & N & CE2-02 \\ \hline
\end{tabular}
}
\end{table}

\begin{table}[htp] 
\centering
\caption{Lista de publicações catalogadas após o 1\textordmasculine \, e 2\textordmasculine \, filtros - Parte 4.}
\resizebox{\columnwidth}{!}{%
\begin{tabular}{|>{\centering\arraybackslash}p{6cm}|>{\centering\arraybackslash}p{4cm}|>{\centering\arraybackslash}p{4cm}|c|>{\centering\arraybackslash}p{3cm}|c|c|>{\centering\arraybackslash}p{3cm}|c|>{\centering\arraybackslash}p{3cm}|}
\hline
\textbf{Título} & \textbf{Autores} & \textbf{Fonte da Publicação} & \textbf{Ano} & \textbf{Editora} & \textbf{Disp} & \textbf{1F} & \textbf{Critério} & \textbf{2F} & \textbf{Critério} \\ \hline

DVS 926 CPU for mobile handheld devices & Rajalingam A., Kokila B., Srivatsa S. & Research Journal of Applied Sciences, Engineering and TechNlogy & 2010 & Maxwell Scientific Organization & N/A & N & CE1-05 & N/A & N/A \\ \hline

Energy aware DAG scheduling on heterogeneous systems & Baskiyar S., Abdel-Kader R. & Cluster Computing & 2010 & Springer & N/A & N & CE1-05 & N/A & N/A \\ \hline

Energy-aware loop parallelism maximization for multi-core DSP architectures & Qiu M., Niu J.-W., Yang L.T., Qin X., Zhang S., Wang B. & Proceedings - 2010 IEEE/ACM International Conference on Green Computing and Communications, GreenCom 2010, 2010 IEEE/ACM International Conference on Cyber, Physical and Social Computing, CPSCom 2010 & 2010 & IEEE & N/A & N & CE1-05 / CE1-06 & N/A & N/A \\ \hline

Metrics and task scheduling policies for energy saving in multicore computers & Mair J., Leung K., Huang Z. & Proceedings - IEEE/ACM International Workshop on Grid Computing & 2010 & IEEE & N/A & N & CE1-05 & N/A & N/A \\ \hline

Parametric timing analysis and its application to dynamic voltage scaling & Mohan S., Mueller F., Root M., Hawkins W., Healy C., Whalley D., Vivancos E. & Transactions on Embedded Computing Systems & 2010 & ACM & S & S & CI1-01 & S & CI2-01 \\ \hline

Leakage-aware scheduling for real-time systems with (m,k)-constraints & Niu L., Quan G. & Proceedings - 10th IEEE International Conference on Computer and Information TechNlogy, CIT-2010, 7th IEEE International Conference on Embedded Software and Systems, ICESS-2010, ScalCom-2010 & 2010 & IEEE & S & S & CI1-01 & N & CE2-02 \\ \hline

Reducing delay jitter of real-time control tasks through adaptive deadline adjustments & Hong S., Hu X.S., Lemmon M.D. & Proceedings - Euromicro Conference on Real-Time Systems & 2010 & IEEE & N/A & N & CE1-05 & N/A & N/A \\ \hline

PreDVS: Preemptive dynamic voltage scaling for real-time systems using approximation scheme & Wang W., Mishra P. & Proceedings - Design Automation Conference & 2010 & ACM & S & S & CI1-01 & N & CE2-02 \\ \hline

Transition-aware DVS algorithm for real-time systems using tree structure analysis & Chen D.-R., Hsu C.-C., Chen Y.-S., Kuo C.-J., Chen L.-C. & Journal of Systems Architecture & 2010 & ScienceDirect & S & S & CI1-01 & N & CE2-02 \\ \hline

IADVS: On-demand performance for interactive applications & Bi M., Crk I., Gniady C. & Proceedings - International Symposium on High-Performance Computer Architecture & 2010 & IEEE & N/A & N & CE1-05 & N/A & N/A \\ \hline

A speed fine tuning technique for system energy minimization of weakly hard real-time system & Agrawal S., Yadav R.S., Vijay R. & International Journal of Computers and Applications & 2010 & ACTA Press & N & N & CE1-02 & N/A & N/A \\ \hline

Dynamic voltage scaling for real-time systems with system workload analysis & Zhang Z., Chen X., Qian D.-J., Hu C. & IEICE Transactions on Electronics & 2010 & J-STAGE & N & N & CE1-02 & N/A & N/A \\ \hline

Energy-efficient considerations on a variable-bitrate PCI-express device & Lee Y.-H., Chen J.-J., Shih C.-S. & Journal of Signal Processing Systems & 2010 & Springer & N/A & N & CE1-05 / CE1-06 & N/A & N/A \\ \hline

Leakage-aware energy minimization using dynamic voltage scaling and cache reconfiguration in real-time systems & Wang W., Mishra P. & Proceedings of the IEEE International Conference on VLSI Design & 2010 & IEEE & S & S & CI1-01 & N & CE2-02 \\ \hline

On the interplay of parallelization, program performance, and energy consumption & Cho S., Melhem R.G. & IEEE Transactions on Parallel and Distributed Systems & 2010 & IEEE & N/A & N & CE1-05 / CE1-06 & N/A & N/A \\ \hline

Real-time intra-task voltage scheduling algorithm for wireless sensor networks & Chen J., Yi B.-S., Fang Y.-J., Yue L. & Huanan Ligong Daxue Xuebao/Journal of South China University of TechNlogy (Natural Science) & 2010 & Wanfang Data & N/A & N & CE1-05 & N/A & N/A \\ \hline

Voltage assignment for soft real-time embedded systems with continuous probability distribution & Qiu M., Wu J., Hu F., Liu S., Wang L. & Proceedings - 15th IEEE International Conference on Embedded and Real-Time Computing Systems and Applications, RTCSA 2009 & 2009 & IEEE & S & S & CI1-01 & N & CE2-02 \\ \hline

Efficient dynamic voltage/frequency scaling through algorithmic loop transformation & Ghodrat M.A., Givargis T. & Embedded Systems Week 2009 - 7th IEEE/ACM International Conference on Hardware/Software-Co-Design and System Synthesis, CODES+ISSS 2009 & 2009 & ACM & N/A & N & CE1-05 & N/A & N/A \\ \hline
\end{tabular}
}
\end{table}

\begin{table}[htp] 
\centering
\caption{Lista de publicações catalogadas após o 1\textordmasculine \, e 2\textordmasculine \, filtros - Parte 5.}
\resizebox{\columnwidth}{!}{%
\begin{tabular}{|>{\centering\arraybackslash}p{6cm}|>{\centering\arraybackslash}p{4cm}|>{\centering\arraybackslash}p{4cm}|c|>{\centering\arraybackslash}p{3cm}|c|c|>{\centering\arraybackslash}p{3cm}|c|>{\centering\arraybackslash}p{3cm}|}
\hline
\textbf{Título} & \textbf{Autores} & \textbf{Fonte da Publicação} & \textbf{Ano} & \textbf{Editora} & \textbf{Disp} & \textbf{1F} & \textbf{Critério} & \textbf{2F} & \textbf{Critério} \\ \hline

A generalized framework for energy savings in hard real-time embedded systems & Zeng G., Tomiyama H., Takada H. & IPSJ Transactions on System LSI Design Methodology & 2009 & J-STAGE & S & S & CI1-01 & N & CE2-02 \\ \hline

Real-time power management for a multi-performance processor & Ishihara T. & 2009 International SoC Design Conference, ISOCC 2009 & 2009 & IEEE & S & S & CI1-01 & S & CI2-01 \\ \hline

An task scheduling algorithm of real-time leakage power and temperature optimization for MPSoC & Zhaoguo F., Chaoshan S., Zuying L. & Proceedings - 2009 11th IEEE International Conference on Computer-Aided Design and Computer Graphics, CAD/Graphics 2009 & 2009 & IEEE & N/A & N & CE1-05 & N/A & N/A \\ \hline

How to interface energy harvesting models with multiprocessor scheduling paradigms & Koch P. & Proceedings of the 2009 1st International Conference on Wireless Communication, Vehicular TechNlogy, Information Theory and Aerospace and Electronic Systems TechNlogy, Wireless VITAE 2009 & 2009 & IEEE & N/A & N & CE1-05 / CE1-06 & N/A & N/A \\ \hline

Combining coarse-grained software pipelining with dvs for scheduling real-time periodic dependent tasks on multi-core embedded systems & Liu H., Shao Z., Wang M., Du J., Xue C.J., Jia Z. & Journal of Signal Processing Systems & 2009 & Springer & N/A & N & CE1-05 & N/A & N/A \\ \hline

Dynamic leakage aware power management with procrastination method & Yongwen P., Man L. & Canadian Conference on Electrical and Computer Engineering & 2009 & IEEE & S & S & CI1-01 & N & CE2-02 \\ \hline

Energy minimization for real-time systems with Nn-convex and discrete operation modes & Dabiri F., Vahdatpour A., Potkonjak M., Sarrafzadeh M. & Proceedings -Design, Automation and Test in Europe, DATE & 2009 & IEEE & S & S & CI1-01 & N & CE2-02 \\ \hline

Program phase and runtime distribution-aware online DVFS for combined Vdd/Vbb scaling & Jungsoo K., Sungjoo Y., Kyung C.-M. & Proceedings -Design, Automation and Test in Europe, DATE & 2009 & IEEE & N/A & N & CE1-05 & N/A & N/A \\ \hline

Low power processor architectures and contemporary techniques for power optimization - A review & Qadri M.Y., Gujarathi H.S., McDonald-Maier K.D. & Journal of Computers & 2009 & Academy Publisher & N/A & N & CE1-05 & N/A & N/A \\ \hline

Energy-efficient scheduling for small pervasive computing devices under fixed-priority multi-subtask model & Gao Z., Wu Z., Lin M. & Intelligent Automation and Soft Computing & 2009 & Taylor \& Francis Group & N & N & CE1-02 & N/A & N/A \\ \hline

SYS-EDF: A system-wide energy-efficient scheduling algorithm for hard real-time systems & Cheng H., Goddard S. & International Journal of Embedded Systems & 2009 & InderScience Publishers & N & N & CE1-02 & N/A & N/A \\ \hline

Run-time adaptive workload estimation for dynamic voltage scaling & Bang S.-Y., Bang K., Yoon S., Chung E.-Y. & IEEE Transactions on Computer-Aided Design of Integrated Circuits and Systems & 2009 & IEEE & N/A & N & CE1-05 & N/A & N/A \\ \hline

DVFS algorithm exploiting correlation in runtime distribution & Kim J., Yoo S., Kyung C.-M. & Journal of Semiconductor TechNlogy and Science & 2009 & The Institute of Electronics and Information Engineers (IEIE) & N/A & N & CE1-05 & N/A & N/A \\ \hline

An analytical dynamic scaling of supply voltage and body bias based on parallelism-aware workload and runtime distribution & Kim J., Oh S., Yoo S., Kyung C.-M. & IEEE Transactions on Computer-Aided Design of Integrated Circuits and Systems & 2009 & IEEE & S & S & CI1-01 & N & CE2-02 \\ \hline

An optimal speed control scheme supported by media servers for low-power multimedia applications & Huang W., Wang Y. & Multimedia Systems & 2009 & Springer & N/A & N & CE1-05 & N/A & N/A \\ \hline

Generalized elastic scheduling for real-time tasks & Chantem T., Hu X.S., Lemmon M.D. & IEEE Transactions on Computers & 2009 & IEEE & N/A & N & CE1-06 & N/A & N/A \\ \hline

Studying energy-oriented dynamic optimizations in java virtual machines & Sun Y., Zhang W. & Journal of Circuits, Systems and Computers & 2009 & World Scientific Publishing & N/A & N & CE1-05 / CE1-06 & N/A & N/A \\ \hline

Design of fast and efficient energy-aware gradient-based scheduling algorithms heterogeneous embedded multiprocessor systems & Goh L.K., Veeravalli B., Viswanathan S. & IEEE Transactions on Parallel and Distributed Systems & 2009 & IEEE & N/A & N & CE1-05 & N/A & N/A \\ \hline

An analytical dynamic scaling of supply voltage and body bias based on parallelism-aware workload and runtime distribution & Kim J., Oh S., Yoo S., Kyung C.-M. & IEEE Transactions on Computer-Aided Design of Integrated Circuits and Systems & 2009 & IEEE & S & N & CE1-07 & N & CI1-07 \\ \hline

Energy efficient intra-task dynamic voltage scaling for realistic CPUs of mobile devices & Yang C.-C., Wang K., Lin M.-H., Lin P. & Journal of Information Science and Engineering & 2009 & Scopus & S & S & CI1-01 & S & CI2-01 \\ \hline

Outstanding research problems in NC design: System, microarchitecture, and circuit perspectives & Marculescu R., Ogras U.Y., Peh L.-S., Jerger N.E., Hoskote Y. & IEEE Transactions on Computer-Aided Design of Integrated Circuits and Systems & 2009 & IEEE & N/A & N & CE1-05 / CE1-06 & N/A & N/A \\ \hline

Run-time adaptive workload estimation for dynamic voltage scaling & Bang S.-Y., Bang K., Yoon S., Chung E.-Y. & IEEE Transactions on Computer-Aided Design of Integrated Circuits and Systems & 2009 & IEEE & N/A & N & CE1-05 & N/A & N/A \\ \hline
\end{tabular}
}
\end{table}

\begin{table}[htp] 
\centering
\caption{Lista de publicações catalogadas após o 1\textordmasculine \, e 2\textordmasculine \, filtros - Parte 6.}
\resizebox{\columnwidth}{!}{%
\begin{tabular}{|>{\centering\arraybackslash}p{6cm}|>{\centering\arraybackslash}p{4cm}|>{\centering\arraybackslash}p{4cm}|c|>{\centering\arraybackslash}p{3cm}|c|c|>{\centering\arraybackslash}p{3cm}|c|>{\centering\arraybackslash}p{3cm}|}
\hline
\textbf{Título} & \textbf{Autores} & \textbf{Fonte da Publicação} & \textbf{Ano} & \textbf{Editora} & \textbf{Disp} & \textbf{1F} & \textbf{Critério} & \textbf{2F} & \textbf{Critério} \\ \hline

System-scenario-based design of dynamic embedded systems & Gheorghita S.V., Palkovic M., Hamers J., Vandecappelle A., Mamagkakis S., Basten T., Eeckhout L., Corporaal H., Catthoor F., Vandeputte F., De Bosschere K. & ACM Transactions on Design Automation of Electronic Systems & 2009 & ACM & N/A & N & CE1-06 & N/A & N/A \\ \hline

Stochastic voltage scheduling of fixed-priority tasks with preemption thresholds & He X., Jia Y., Wa H. & 2008 International Conference on Wireless Communications, Networking and Mobile Computing, WiCOM 2008 & 2008 & IEEE & S & S & CI1-01 & S & CI2-01 \\ \hline

Application scenarios in streaming-oriented embedded-system design & Gheorghita S.V., Basten T., Corporaal H. & IEEE Design and Test of Computers & 2008 & IEEE & N/A & S & CI1-01 & N & CE2-02 \\ \hline

A trace cache with DVFS techniques for a low power microprocessor & Jang H.B., Choi L., Chung S.W. & Proceedings - 3rd International Conference on Convergence and Hybrid Information TechNlogy, ICCIT 2008 & 2008 & IEEE & N/A & N & CE1-05 & N/A & N/A \\ \hline

Simultaneous optimization of battery-aware voltage regulator scheduling with dynamic voltage and frequency scaling & Cho Y., Kim Y., Joo Y., Lee K., Chang N. & Proceedings of the International Symposium on Low Power Electronics and Design & 2008 & ACM & N/A & N & CE1-05 & N/A & N/A \\ \hline

Dynamic and leakage power minimization with loop voltage scheduling and assignment & Qiu M., Wu J., Hu J., He Y., Sha E.H.-M. & Proceedings of The 5th International Conference on Embedded and Ubiquitous Computing, EUC 2008 & 2008 & IEEE & N/A & N & CE1-05 & N/A & N/A \\ \hline

Efficient algorithms for jitterless real-time tasks to DVS schedules & Chen D.-R., Hsieh S.-M., Lai M.-F. & Parallel and Distributed Computing, Applications and TechNlogies, PDCAT Proceedings & 2008 & IEEE & S & S & CI1-01 & S & CI2-01 \\ \hline

Energy-oriented OpenMP parallel loop scheduling & Dong J.Y., Chen J., Yang X., Deng L., Zhang X. & Proceedings of the 2008 International Symposium on Parallel and Distributed Processing with Applications, ISPA 2008 & 2008 & IEEE & N/A & N & CE1-05 & N/A & N/A \\ \hline

Expected energy consumption minimization in DVS systems with discrete frequencies & Chen J.-J. & Proceedings of the ACM Symposium on Applied Computing & 2008 & ACM & S & S & CI1-01 & S & CI2-01 \\ \hline

Improving energy-efficient real-time scheduling by exploiting code instrumentation & Zitterell T., Scholl C. & Proceedings of the International Multiconference on Computer Science and Information TechNlogy, IMCSIT 2008 & 2008 & IEEE & S & S & CI1-01 & S & CI2-01 \\ \hline

Low-power technique based on timeout policy for portable systems & Cho M.-H., Jung M.-J., Kim Y.-H., Lee C.-H. & 5th International Conference on Information TechNlogy and Applications, ICITA 2008 & 2008 & Scopus & N & N & CE1-02 & N/A & N/A \\ \hline

Maximizing total QoS-provisioning of image streams with limited energy budget & Lee W.Y., Kim K.H., Ko Y.W. & IEICE Transactions on Communications & 2008 & J-STAGE & N/A & N & CE1-05 / CE1-06 & N/A & N/A \\ \hline

Profile-based technique for dynamic power management in embedded systems & Kumar C.M., Sindhwani M., Srikanthan T. & 2008 International Conference on Electronic Design, ICED 2008 & 2008 & IEEE & N/A & N & CE1-05 & N/A & N/A \\ \hline

Saving register-file static power by monitoring short-lived temporary-values in ROB & Shieh W.-Y., Chen H.-D. & 13th IEEE Asia-Pacific Computer Systems Architecture Conference, ACSAC 2008 & 2008 & IEEE & N/A & N & CE1-05 & N/A & N/A \\ \hline

Learning and leveraging the relationship between architecture-level measurements and individual user satisfaction & Shye A., Ozisikyilmaz B., Mallik A., Memik G., Dinda P.A., Dick R.P., Choudhary A.N. & Proceedings - International Symposium on Computer Architecture & 2008 & IEEE & N/A & N & CE1-05 & N/A & N/A \\ \hline

AMPLE: An adaptive multi-performance processor for low-energy embedded applications & Ishihara T., Yamaguchi S., Ishitobi Y., Matsumura T., Kunitake Y., Oyama Y., Kaneda Y., Muroyama M., Sato T. & 2008 Symposium on Application Specific Processors, SASP 2008 & 2008 & IEEE & N/A & N & CE1-05 & N/A & N/A \\ \hline

Overhead-aware system-level joint energy and performance optimization for streaming applications on multiprocessor systems-on-chip & Liu H., Shao Z., Wang M., Chen P. & Proceedings - Euromicro Conference on Real-Time Systems & 2008 & IEEE & N/A & N & CE1-05 & N/A & N/A \\ \hline
\end{tabular}
}
\end{table}

\begin{table}[htp] 
\centering
\caption{Lista de publicações catalogadas após o 1\textordmasculine \, e 2\textordmasculine \, filtros - Parte 7.}
\resizebox{\columnwidth}{!}{%
\begin{tabular}{|>{\centering\arraybackslash}p{6cm}|>{\centering\arraybackslash}p{4cm}|>{\centering\arraybackslash}p{4cm}|c|>{\centering\arraybackslash}p{3cm}|c|c|>{\centering\arraybackslash}p{3cm}|c|>{\centering\arraybackslash}p{3cm}|}
\hline
\textbf{Título} & \textbf{Autores} & \textbf{Fonte da Publicação} & \textbf{Ano} & \textbf{Editora} & \textbf{Disp} & \textbf{1F} & \textbf{Critério} & \textbf{2F} & \textbf{Critério} \\ \hline

Voltage emergence prevention for energy-efficient real-time task synchronization & Chen Y.-S., Lin T.-J. & Proceedings - 8th IEEE International Conference on Computer and Information TechNlogy Workshops, CIT Workshops 2008 & 2008 & IEEE & S & S & CI1-01 & N & CE2-02 \\ \hline

Task partitioning algorithm for intra-task dynamic voltage scaling & Oh S., Kim J., Kim S., Kyung C.-M. & Proceedings - IEEE International Symposium on Circuits and Systems & 2008 & IEEE & S & S & CI1-01 & S & CI2-01 \\ \hline

Performance analysis of power-aware task scheduling algorithms on multiprocessor computers with dynamic voltage and speed & Li K. & IEEE Transactions on Parallel and Distributed Systems & 2008 & IEEE & N/A & N & CE1-05 & N/A & N/A \\ \hline

Efficient algorithms for periodic real-time tasks to optimal discrete voltage schedules & Chen D.-R., Hsieh S.-M., Lai M.-F. & IPDPS Miami 2008 - Proceedings of the 22nd IEEE International Parallel and Distributed Processing Symposium, Program and CD-ROM & 2008 & IEEE & S & S & CI1-01 & S & CI2-01 \\ \hline

Minimizing response time implication in DVS scheduling for low power embedded systems & Min-Allah N., Kazmi A.-R., Ali I., Xing J.-S., Wang Y.-J. & InNvations'07: 4th International Conference on InNvations in Information TechNlogy, IIT & 2008 & IEEE & S & S & CI1-01 & N & CE2-02 \\ \hline

Saving register-file leakage energy by register-usage exploiting & Shieh W.-Y., Chen C.-C. & Journal of Information Science and Engineering & 2008 & JISE & N/A & N & CE1-05 & N/A & N/A \\ \hline

Dynamic voltage scaling of supply and body bias exploiting software runtime distribution & Hong S., Yoo S., Bin B., Choi K.-M., Eo S.-K., Kim T. & Proceedings -Design, Automation and Test in Europe, DATE & 2008 & IEEE & N/A & N & CE1-05 & N/A & N/A \\ \hline

Dynamic scheduling of imprecise-computation tasks in maximizing QoS under energy constraints for embedded systems & Yu H., Veeravalli B., Ha Y. & Proceedings of the Asia and South Pacific Design Automation Conference, ASP-DAC & 2008 & IEEE & S & S & CI1-01 & N & CE2-02 \\ \hline

Dynamic power management for embedded system idle state in the presence of periodic interrupt services & Zeng G., Tomiyama H., Takada H. & IPSJ Transactions on System LSI Design Methodology & 2008 & J-STAGE & S & S & CI1-01 & N & CE2-02 \\ \hline

Energy-aware task scheduling and dynamic voltage scaling in a real-time system & Rong P., Pedram M. & Journal of Low Power Electronics & 2008 & American Scientific Publishers & S & S & CI1-01 & N & CE2-02 \\ \hline

Scenario selection and prediction for DVS-aware scheduling of multimedia applications & Gheorghita S.V., Basten T., Corporaal H. & Journal of Signal Processing Systems & 2008 & Springer & S & S & CI1-01 & N & CE2-02 \\ \hline

Corollaries to Amdahl's law for energy & Cho S., Melhem R.G. & IEEE Computer Architecture Letters & 2008 & IEEE & N/A & N & CE1-05 & N/A & N/A \\ \hline

Scheduling algorithm of saving energy for multiprocessors & Sang N., Li B.-Y., Ma H. & Dianzi Keji Daxue Xuebao/Journal of the University of Electronic Science and TechNlogy of China & 2008 & China/Asia On Demand (CAOD) & N/A & N & CE1-05 & N/A & N/A \\ \hline

A deferred-workload-based inter-task dynamic voltage scaling algorithm for portable multimedia devices & Tsai Y.-H., Wang K., Chen J.-M. & IWCMC 2007: Proceedings of the 2007 International Wireless Communications and Mobile Computing Conference & 2007 & ACM & S & S & CI1-01 & N & CE2-02 \\ \hline

A Nvel voltage scaling algorithm through ant colony optimization for embedded distributed systems & Dan D., Lisheng Z., Zheng W. & IEEE ICIT 2007 - 2007 IEEE International Conference on Integration TechNlogy & 2007 & IEEE & N/A & N & CE1-05 & N/A & N/A \\ \hline

A software framework for energy and performance tradeoff in fixed-priority hard real-time embedded systems & Zeng G., Tomiyama H., Takada H. & Lecture Ntes in Computer Science (including subseries Lecture Ntes in Artificial Intelligence and Lecture Ntes in Bioinformatics) & 2007 & Springer & S & S & CI1-01 & N & CE2-02 \\ \hline

A unified practical approach to stochastic DVS scheduling & Xu R., Melhem R., Mosse D. & EMSOFT'07: Proceedings of the Seventh ACM and IEEE International Conference on Embedded Software & 2007 & ACM & S & S & CI1-01 & N & CE2-02 \\ \hline

A voltage and resource synthesis technique for energy-aware real-time systems & Kang D.-I., Crago S.P., Suh J., McMahon J. & Proceedings - 13th IEEE International Conference on Embedded and Real-Time Computing Systems and Applications, RTCSA 2007 & 2007 & IEEE & N/A & N & CE1-04 & N/A & N/A \\ \hline
\hline
\end{tabular}
}
\end{table}

\begin{table}[htp] 
\centering
\caption{Lista de publicações catalogadas após o 1\textordmasculine \, e 2\textordmasculine \, filtros - Parte 8.}
\resizebox{\columnwidth}{!}{%
\begin{tabular}{|>{\centering\arraybackslash}p{6cm}|>{\centering\arraybackslash}p{4cm}|>{\centering\arraybackslash}p{4cm}|c|>{\centering\arraybackslash}p{3cm}|c|c|>{\centering\arraybackslash}p{3cm}|c|>{\centering\arraybackslash}p{3cm}|}
\hline
\textbf{Título} & \textbf{Autores} & \textbf{Fonte da Publicação} & \textbf{Ano} & \textbf{Editora} & \textbf{Disp} & \textbf{1F} & \textbf{Critério} & \textbf{2F} & \textbf{Critério} \\ \hline

An adaptive scheduler for embedded multi-processor real-time systems & Lee L.-T., Tseng C.-Y., Hsu S.-J. & IEEE Region 10 Annual International Conference, Proceedings/TENCON & 2007 & IEEE & N/A & N & CE1-04 & N/A & N/A \\ \hline

An energy-aware gradient-based scheduling heuristic for heterogeneous multiprocessor embedded systems & Gob L.K., Veeravalli B., Viswanathan S. & Lecture Ntes in Computer Science (including subseries Lecture Ntes in Artificial Intelligence and Lecture Ntes in Bioinformatics) & 2007 & Springer & N/A & N & CE1-05 & N/A & N/A \\ 

Energy-efficiency on a variable-bitrate device & Lee Y.-H., Chen J.-J., Kuo T.-W. & Lecture Ntes in Computer Science (including subseries Lecture Ntes in Artificial Intelligence and Lecture Ntes in Bioinformatics) & 2007 & Springer & N/A & N & CE1-05 / CE1-06 & N/A & N/A \\ \hline

FL-PCP: Frequency locking for energy-efficient real-time task synchronization & Chen Y.-S., Yang C.-Y., Kuo T.-W. & Proceedings - 13th IEEE International Conference on Embedded and Real-Time Computing Systems and Applications, RTCSA 2007 & 2007 & IEEE & N/A & N & CE1-05 & N/A & N/A \\ \hline

Minimizing expected energy consumption in real-time systems through dynamic voltage scaling & Xu R., Mosse D., Melhem R. & ACM Transactions on Computer Systems & 2007 & ACM & S & S & CI1-01 & N & CE2-02 \\ \hline

Online dynamic voltage scaling using task graph mapping analysis for multiprocessors & Choudhury P., Chakrabarti P.P., Kumar R. & Proceedings of the IEEE International Conference on VLSI Design & 2007 & IEEE & S & S & CI1-01 & N & CE2-02 \\ \hline

Profiling driven scenario detection and prediction for multimedia applications & Gheorghita S.V., Basten T., Corporaal H. & Proceedings - 2006 International Conference on Embedded Computer Systems: Architectures, Modeling and Simulation, IC-SAMOS 2006 & 2007 & IEEE & N/A & N & CE1-06 & N/A & N/A \\ \hline

The Chip is the network: Toward a science of network-on-chip design & Marculescu R., Bogdan P. & Foundations and Trends in Electronic Design Automation & 2007 & Nw Publishers & N/A & N & CE1-03 & N/A & N/A \\ \hline

System level voltage scheduling technique using UML-RT model & Neishaburi M.H., Daneshtalab M., Nabi M., Mohammadi S. & 2007 IEEE/ACS International Conference on Computer Systems and Applications, AICCSA 2007 & 2007 & IEEE & S & S & CI1-01 & S & CI2-01 \\ \hline

Dynamic voltage scaling under EDF revisited & Gaujal B., Navet N. & Real-Time Systems & 2007 & Springer & S & S & CI1-01 & N & CE2-02 \\ \hline

DC-DC converter-aware power management for low-power embedded systems & Choi Y., Chang N., Kim T. & IEEE Transactions on Computer-Aided Design of Integrated Circuits and Systems & 2007 & IEEE & N/A & N & CE1-05 & N/A & N/A \\ \hline

Power consumption and process variations: Two challenges to design of next-generation ICs & Luo Z.-Y. & Jisuanji Xuebao/Chinese Journal of Computers & 2007 & National KNwledge Infrastructure (CNKI) & N/A & N & CE1-04 & N/A & N/A \\ \hline

Real-time dynamic voltage loop scheduling for multi-core embedded systems & Shao Z., Wang M., Chen Y., Xue C., Qiu M., Yang L.T., Sha E.H.-M. & IEEE Transactions on Circuits and Systems II: Express Briefs & 2007 & IEEE & S & S & CI1-01 & N & CE2-02 \\ \hline

Transition-overhead-aware voltage scheduling for fixed-priority real-time systems & Mochocki B., Hu X.S., Quan G. & ACM Transactions on Design Automation of Electronic Systems & 2007 & ACM & S & S & CI1-01 & N & CE2-02 \\ \hline

Optimizing intratask voltage scheduling using profile and data-flow information & Shin D., Kim J. & IEEE Transactions on Computer-Aided Design of Integrated Circuits and Systems & 2007 & IEEE & S & S & CI1-01 & S & CI2-01 \\ \hline

Controlling energy demand in mobile computing systems & Ellis C.S. & Synthesis Lectures on Mobile and Pervasive Computing & 2007 & IEEE & N/A & N & CE1-05 & N/A & N/A \\ \hline

Voltage assignment with guaranteed probability satisfying timing constraint for real-time multiproceesor DSP & Qiu M., Jia Z., Xue C., Shao Z., Sha E.H.-M. & Journal of VLSI Signal Processing Systems for Signal, Image, and Video TechNlogy & 2007 & Springer & N/A & N & CE1-05 & N/A & N/A \\ \hline

Workload-ahead-driven online energy minimization techniques for battery-powered embedded systems with time-constraints & Cai Y., Schmitz M.T., Al-Hashimi B.M., Reddy S.M. & ACM Transactions on Design Automation of Electronic Systems & 2007 & ACM & N/A & N & CE1-05 & N/A & N/A \\ \hline

Static WCET analysis based compiler-directed DVS energy optimization in real-time applications & Yi H., Chen J., Yang X. & Lecture Ntes in Computer Science (including subseries Lecture Ntes in Artificial Intelligence and Lecture Ntes in Bioinformatics) & 2006 & Springer & S & S & CI1-01 & S & CI2-01 \\ \hline
\end{tabular}
}
\end{table}

\begin{table}[htp] 
\centering
\caption{Lista de publicações catalogadas após o 1\textordmasculine \, e 2\textordmasculine \, filtros - Parte 9.}
\resizebox{\columnwidth}{!}{%
\begin{tabular}{|>{\centering\arraybackslash}p{6cm}|>{\centering\arraybackslash}p{4cm}|>{\centering\arraybackslash}p{4cm}|c|>{\centering\arraybackslash}p{3cm}|c|c|>{\centering\arraybackslash}p{3cm}|c|>{\centering\arraybackslash}p{3cm}|}
\hline
\textbf{Título} & \textbf{Autores} & \textbf{Fonte da Publicação} & \textbf{Ano} & \textbf{Editora} & \textbf{Disp} & \textbf{1F} & \textbf{Critério} & \textbf{2F} & \textbf{Critério} \\ \hline

Application-driven low-power techniques using dynamic voltage scaling & Kim T. & Proceedings - 12th IEEE International Conference on Embedded and Real-Time Computing Systems and Applications, RTCSA 2006 & 2006 & IEEE & N/A & N & CE1-03 & N/A & N/A \\ \hline

Application scenarios in streaming-oriented embedded system design & Gheorghita S.V., Basten T., Corporaal H. & 2006 International Symposium on System-on-Chip, SOC & 2006 & IEEE & S & N & CE1-07 & N & CE1-07 \\ \hline

Certain investigations on energy saving techniques using DVS for low power embedded system & Chilambuchelvan A., Saravanan S., Chidhambararajan B., Perinbam J.R.P. & WSEAS Transactions on Computers & 2006 & Scopus & N/A & N & CE1-07 & N/A & N/A \\ \hline

Dual-processor design of energy efficient fault-tolerant system & Hua S., Pari P.R., Qu G. & Proceedings of the International Conference on Application-Specific Systems, Architectures and Processors & 2006 & IEEE & N/A & N & CE1-05 & N/A & N/A \\ \hline

Energy-efficient task scheduling algorithm for mobile terminal & Zhang L., Qi D. & IET Conference Publications & 2006 & IET Digital Library & S & S & CI1-01 & S & CI2-01 \\ \hline

Energy-optimal speed control of a generic device & Rao R., Vrudhula S. & IEEE Transactions on Computer-Aided Design of Integrated Circuits and Systems & 2006 & IEEE & N/A & N & CE1-05 & N/A & N/A \\ \hline

Exploiting memory-boundedness in energy-efficient hard real-time scheduling & Spiga M., Spiga M., Alimonda A., Carta S., Aymerich F., Acquaviva A. & Industrial Embedded Systems - IES'2006 & 2006 & IEEE & S & S & CI1-01 & N & CE2-02 \\ \hline

Methods for power optimization in distributed embedded systems with real-time requirements & Racu R., Hamann A., Ernst R., Mochocki B., Hu X.S. & CASES 2006: International Conference on Compilers, Architecture and Synthesis for Embedded Systems & 2006 & ACM & S & S & CI1-01 & N & CE2-02 \\ \hline

Power-aware parametric dispatching in distributed real-time systems & Elsharkawy S.M., Younis M.F., Moulds C. & Proceedings of the IASTED International Conference on Parallel and Distributed Computing and Systems & 2006 & ACTA Press & N & N & CE1-02 & N/A & N/A \\ \hline

Procrastinating voltage scheduling with discrete frequency sets & Lu Z., Zhang Y., Stan M., Lach J., Skadron K. & Proceedings -Design, Automation and Test in Europe, DATE & 2006 & IEEE & S & S & CI1-01 & N & CE2-02 \\ \hline

Runtime distribution-aware dynamic voltage scaling & Hong S., Yoo S., Jin H., Choi K.-M., Kong J.-T., Eo S.-K. & IEEE/ACM International Conference on Computer-Aided Design, Digest of Technical Papers, ICCAD & 2006 & IEEE & S & S & CI1-01 & S & CI2-01 \\ \hline

Temperature-aware leakage minimization technique for real-time systems & Yuan L., Leventhal S., Qu G. & IEEE/ACM International Conference on Computer-Aided Design, Digest of Technical Papers, ICCAD & 2006 & IEEE & S & S & CI1-01 & N & CE2-02 \\ \hline

Dynamic voltage scaling for multitasking real-time systems with uncertain execution time & Xian C., Lu Y.-H. & Proceedings of the ACM Great Lakes Symposium on VLSI, GLSVLSI & 2006 & ACM & S & S & CI1-01 & S & CI2-01 \\ \hline

Toward the optimal configuration of dynamic voltage scaling points in real-time applications & Yi H.-Z., Yang X.-J. & Journal of Computer Science and TechNlogy & 2006 & Springer & S & S & CI1-01 & S & CI2-01 \\ \hline

Quasi-static assignment of voltages and optional cycles in imprecise-computation systems with energy considerations & Cortes L.A., Eles P., Peng Z. & IEEE Transactions on Very Large Scale Integration (VLSI) Systems & 2006 & IEEE & S & S & CI1-01 & N & CE2-02 \\ \hline

Low-power design methodology for module-wise dynamic voltage and frequency scaling with dynamic de-skewing systems & Kitahara T., Hara H., Shiratake S., Tsukiboshi Y., Yoda T., Utsumi T., Minami F. & Proceedings of the Asia and South Pacific Design Automation Conference, ASP-DAC & 2006 & IEEE & N/A & N & CE1-05 & N/A & N/A \\ \hline

Parametric timing estimation with Newton-Gregory formulae & Van Engelen R.A., Gallivan K.A., Walsh B. & Concurrency Computation Practice and Experience & 2006 & Wiley Online Library & N/A & N & CE1-05 / CE1-06 & N/A & N/A \\ \hline

Minimising the energy consumption of real-time tasks with precedence constraints on a single processor & Wu H., Parameswaran S. & Lecture Ntes in Computer Science (including subseries Lecture Ntes in Artificial Intelligence and Lecture Ntes in Bioinformatics) & 2006 & Springer & N/A & N & CE1-06 & N/A & N/A \\ \hline

Power-aware instruction scheduling & Lin T.-Y., Chang R.-G. & Lecture Ntes in Computer Science (including subseries Lecture Ntes in Artificial Intelligence and Lecture Ntes in Bioinformatics) & 2006 & Springer & N/A & N & CE1-05 & N/A & N/A \\ \hline
\end{tabular}
}
\end{table}

\begin{table}[htp] 
\centering
\caption{Lista de publicações catalogadas após o 1\textordmasculine \, e 2\textordmasculine \, filtros - Parte 10.}
\resizebox{\columnwidth}{!}{%
\begin{tabular}{|>{\centering\arraybackslash}p{6cm}|>{\centering\arraybackslash}p{4cm}|>{\centering\arraybackslash}p{4cm}|c|>{\centering\arraybackslash}p{3cm}|c|c|>{\centering\arraybackslash}p{3cm}|c|>{\centering\arraybackslash}p{3cm}|}
\hline
\textbf{Título} & \textbf{Autores} & \textbf{Fonte da Publicação} & \textbf{Ano} & \textbf{Editora} & \textbf{Disp} & \textbf{1F} & \textbf{Critério} & \textbf{2F} & \textbf{Critério} \\ \hline

Energy minimization for real-time systems with (m, k)-guarantee & Niu L., Quan G. & IEEE Transactions on Very Large Scale Integration (VLSI) Systems & 2006 & IEEE & N/A & N & CE1-06 & N/A & N/A \\ \hline

Performance lossless voltage scheduling for low energy software & Lei T., Li X., Zhou X. & Jisuanji Yanjiu yu Fazhan/Computer Research and Development & 2006 & Scopus & N/A & N & CE1-05 & N/A & N/A \\ \hline

Energy management for battery-powered reconfigurable computing platforms & Khan J. & IEEE Transactions on Very Large Scale Integration (VLSI) Systems & 2006 & IEEE & N/A & N & CE1-05 & N/A & N/A \\ \hline

Reducing jitter in embedded systems employing a time-triggered software architecture and dynamic voltage scaling & Phatrapornnant T., Pont M.J. & IEEE Transactions on Computers & 2006 & IEEE & N/A & N & CE1-05 & N/A & N/A \\ \hline

Intra-task scenario-aware voltage scheduling & Gheorghita S.V., Basten T., Corporaal H. & CASES 2005: International Conference on Compilers, Architecture, and Synthesis for Embedded Systems & 2005 & ACM & S & S & CI1-01 & S & CI2-01 \\ \hline

Transition phase classification and prediction & Lau J., Schoenmackers S., Calder B. & Proceedings - International Symposium on High-Performance Computer Architecture & 2005 & IEEE & N/A & N & CE1-05 / CE1-06 & N/A & N/A \\ \hline

A generalized technique for energy-efficient operating voltage set-in dynamic voltage scaled processors & Seo J., Dutt N.D. & Proceedings of the Asia and South Pacific Design Automation Conference, ASP-DAC & 2005 & IEEE & N/A & N & CE1-05 & N/A & N/A \\ \hline

An efficient dynamic task scheduling algorithm for battery powered DVS systems & Zhuo J., Chakrabarti C. & Proceedings of the Asia and South Pacific Design Automation Conference, ASP-DAC & 2005 & IEEE & N/A & N & CE1-05 & N/A & N/A \\ \hline

An intra-task DVS algorithm exploiting program path locality for real-time embedded systems & Kumar G.S.A., Manimaran G. & Lecture Ntes in Computer Science (including subseries Lecture Ntes in Artificial Intelligence and Lecture Ntes in Bioinformatics) & 2005 & Springer & S & S & CI1-01 & S & CI2-01 \\ \hline

DC-DC converter-aware power management for battery-operated embedded systems & Choi Y., Chang N., Kim T. & Proceedings - Design Automation Conference & 2005 & IEEE & N/A & N & CE1-05 & N/A & N/A \\ \hline

Dynamic processor throttling for power efficient computations & Kondo M., Nakamura H. & Lecture Ntes in Computer Science (including subseries Lecture Ntes in Artificial Intelligence and Lecture Ntes in Bioinformatics) & 2005 & Springer & N/A & N & CE1-05 & N/A & N/A \\ \hline

Feedback EDF scheduling of real-time tasks exploiting dynamic voltage scaling & Zhu Y., Mueller F. & Real-Time Systems & 2005 & Springer & S & S & CI1-01 & N & CE2-02 \\ \hline

Implementation of dynamic voltage and frequency scaling on blackfin processors & HarlioN D., Gan W.-S. & Proceedings of 2005 International Symposium on Intelligent Signal Processing and Communication Systems, ISPACS 2005 & 2005 & IEEE & S & S & CI1-01 & N & CE2-02 \\ \hline

Low power task scheduling for multiple I/O devices in multiple CPU environment & Or M.S., Park J.H. & Proceedings of the 2005 International Conference on Parallel and Distributed Processing Techniques and Applications, PDPTA'05 & 2005 & Scopus & N & N & CE1-05 / CE1-06 & N/A & N/A \\ \hline

Optimal dynamic voltage scaling for wireless sensor Ndes with real-time constraints & Cassandras C.G., Zhuang S. & Proceedings of SPIE - The International Society for Optical Engineering & 2005 & SPIE - Digital Library & S & S & CI1-01 & S & CI2-01 \\ \hline

Optimal integration of inter-task and intra-task dynamic voltage scaling techniques for hard real-time applications & Seo J., Kim T., Dutt N.D. & IEEE/ACM International Conference on Computer-Aided Design, Digest of Technical Papers, ICCAD & 2005 & IEEE & S & S & CI1-01 & S & CI2-01 \\ \hline

Optimizing intra-task voltage scheduling using data flow analysis & Shin D., Kim J. & Proceedings of the Asia and South Pacific Design Automation Conference, ASP-DAC & 2005 & IEEE & S & S & CI1-01 & S & CI2-01 \\ \hline

Optimizing the configuration of dynamic voltage scaling points in real-time applications & Yi H., Yang X. & Lecture Ntes in Computer Science (including subseries Lecture Ntes in Artificial Intelligence and Lecture Ntes in Bioinformatics) & 2005 & Springer & S & S & CI1-01 & S & CI2-01 \\ \hline
\end{tabular}
}
\end{table}

\begin{table}[htp] 
\centering
\caption{Lista de publicações catalogadas após o 1\textordmasculine \, e 2\textordmasculine \, filtros - Parte 11.}
\resizebox{\columnwidth}{!}{%
\begin{tabular}{|>{\centering\arraybackslash}p{6cm}|>{\centering\arraybackslash}p{4cm}|>{\centering\arraybackslash}p{4cm}|c|>{\centering\arraybackslash}p{3cm}|c|c|>{\centering\arraybackslash}p{3cm}|c|>{\centering\arraybackslash}p{3cm}|}
\hline
\textbf{Título} & \textbf{Autores} & \textbf{Fonte da Publicação} & \textbf{Ano} & \textbf{Editora} & \textbf{Disp} & \textbf{1F} & \textbf{Critério} & \textbf{2F} & \textbf{Critério} \\ \hline

ParaScale: Exploiting parametric timing analysis for real-time schedulers and dynamic voltage scaling & Mohan S., Mueller F., Hawkins W., Root M., Healy C., Whalley D. & Proceedings - Real-Time Systems Symposium & 2005 & IEEE & S & S & CI1-01 & S & CI2-01 \\ \hline

Power Analysis and Optimization Techniques for Energy Efficient Computer Systems & Chedid W., Yu C., Lee B. & Advances in Computers & 2005 & ScienceDirect & N/A & N & CE1-05 & N/A & N/A \\ \hline

Power reduction techniques for microprocessor systems & Venkatachalam V., Franz M. & ACM Computing Surveys & 2005 & ACM & N/A & N & CE1-03 / CE1-05 & N/A & N/A \\ \hline

System-level energy-efficient dynamic task scheduling & Zhuo J., Chakrabarti C. & Proceedings - Design Automation Conference & 2005 & IEEE & N/A & N & CE1-05 & N/A & N/A \\ \hline

The optimal profile-guided greedy dynamic voltage scaling in real-time applications & Yi H., Yang X., Chen J. & Lecture Ntes in Computer Science (including subseries Lecture Ntes in Artificial Intelligence and Lecture Ntes in Bioinformatics) & 2005 & Springer & S & S & CI1-01 & S & CI2-01 \\ \hline

Towards the optimal placement of dynamic voltage scaling points in real-time applications & Yi H., Yang X. & Proceedings of the 2005 International Conference on Embedded Systems and Applications, ESA'05 & 2005 & Scopus & N & N & CE1-02 & N/A & N/A \\ \hline

Voltage-scaling scheduling for periodic real-time tasks in reward maximization & Chen J.-J., Kuo T.-W. & Proceedings - Real-Time Systems Symposium & 2005 & IEEE & S & S & CI1-01 & N & CE2-02 \\ \hline

Power-saving scheduling for weakly dynamic voltage scaling devices & Chen J.-J., Kuo T.-W., Lu H.-I. & Lecture Ntes in Computer Science & 2005 & Springer & S & S & CI1-01 & N & CE2-02 \\ \hline

Design of DVFS algorithm in soft real-time systems & Zhong W.-J., Liu M.-Y. & Beijing Ligong Daxue Xuebao/Transaction of Beijing Institute of TechNlogy & 2005 & Scopus & N & N & CE1-02 & N/A & N/A \\ \hline

Intra-task voltage scheduling on DVS-enabled hard real-time systems & Shin D., Kim J. & IEEE Transactions on Computer-Aided Design of Integrated Circuits and Systems & 2005 & IEEE & S & S & CI1-01 & S & CI2-01 \\ \hline

Low-power video decoding on a variable voltage processor for mobile multimedia applications & Lee S. & ETRI Journal & 2005 & ETRI Journal & N/A & N & CE1-05 & N/A & N/A \\ \hline

Real-time dynamic voltage scaling algorithm based on syntax tree for low power & Yi H.-Z., Chen J., Yang X.-J., Liu Z. & Ruan Jian Xue Bao/Journal of Software & 2005 & Scopus & N & N & CE1-02 & N/A & N/A \\ \hline

Practical on-line DVS scheduling for fixed-priority real-time systems & Mochocki B., Hu X.S., Quan G. & Proceedings of the IEEE Real-Time and Embedded TechNlogy and Applications Symposium, RTAS & 2005 & IEEE & S & S & CI1-01 & N & CE2-02 \\ \hline

Energy management for embedded multithreaded processors with integrated EDF scheduling & Uhrig S., Ungerer T. & Lecture Ntes in Computer Science & 2005 & Springer & S & S & CI1-01 & N & CE2-02 \\ \hline

Minimizing energy via loop scheduling and DVS for multi-core embedded systems & Chen Y., Shao Z., Zhuge Q., Xue C., Xiao B., Sha E.H.-M. & Proceedings of the International Conference on Parallel and Distributed Systems - ICPADS & 2005 & IEEE & S & S & CI1-01 & N & CE2-02 \\ \hline

Dynamic task-level voltage scheduling optimizations & Barnett J.A. & IEEE Transactions on Computers & 2005 & IEEE & S & S & CI1-01 & N & CE2-02 \\ \hline

Low-power system scheduling, synthesis and displays & Jha N.K. & IEE Proceedings: Computers and Digital Techniques & 2005 & IET Digital Library & N/A & N & CE1-07 & N/A & N/A \\ \hline

Mapping the MPEG-4 visual texture decoder & Ma Z., Wong C., Yang P., Vounckx J., Catthoor F. & IEEE Signal Processing Magazine & 2005 & IEEE & N/A & N & CE1-05 & N/A & N/A \\ \hline

Standby and active leakage current control and minimization in CMOS VLSI circuits & Fallah F., Pedram M. & IEICE Transactions on Electronics & 2005 & IEICE & N/A & N & CE1-05 & N/A & N/A \\ \hline

Fine-grained dynamic voltage and frequency scaling for precise energy and performance tradeoff based on the ratio of off-chip access to on-chip computation times & Choi K., Soma R., Pedram M. & IEEE Transactions on Computer-Aided Design of Integrated Circuits and Systems & 2005 & IEEE & N/A & N & CE1-05 & N/A & N/A \\ \hline

Compiling adaptive programs for real-time dynamic scheduling & Hu X., Lei T., Gong Y.-C., Zhou X.-H. & Proceedings of the Eigtht IASTED International Conference on Software Engineering and Applications & 2004 & Scopus & N/A & N & CE1-06 & N/A & N/A \\ \hline

Distributed embedded systems for low power: A case study & Liu J., Chou P.H. & Proceedings - International Parallel and Distributed Processing Symposium, IPDPS 2004 (Abstracts and CD-ROM) & 2004 & IEEE & N/A & N & CE1-05 & N/A & N/A \\ \hline

Dynamic Mapping and Ordering Tasks of Embedded Real-Time Systems on Multiprocessor Platforms & Yang P., Catthoor F. & Lecture Ntes in Computer Science (including subseries Lecture Ntes in Artificial Intelligence and Lecture Ntes in Bioinformatics) & 2004 & Springer & S & S & CI1-01 & N & CE2-02 \\ \hline
\end{tabular}
}
\end{table}

\begin{table}[htp] 
\centering
\caption{Lista de publicações catalogadas após o 1\textordmasculine \, e 2\textordmasculine \, filtros - Parte 12.}
\resizebox{\columnwidth}{!}{%
\begin{tabular}{|>{\centering\arraybackslash}p{6cm}|>{\centering\arraybackslash}p{4cm}|>{\centering\arraybackslash}p{4cm}|c|>{\centering\arraybackslash}p{3cm}|c|c|>{\centering\arraybackslash}p{3cm}|c|>{\centering\arraybackslash}p{3cm}|}
\hline
\textbf{Título} & \textbf{Autores} & \textbf{Fonte da Publicação} & \textbf{Ano} & \textbf{Editora} & \textbf{Disp} & \textbf{1F} & \textbf{Critério} & \textbf{2F} & \textbf{Critério} \\ \hline

Dynamic voltage and frequency scaling based on workload decomposition & Choi K., Soma R., Pedram M. & Proceedings of the 2004 International Symposium on Lower Power Electronics and Design, ISLPED'04 & 2004 & ACM and IEEE & N/A & N & CE1-05 & N/A & N/A \\ \hline

Dynamic voltage and frequency scaling under a precise energy model considering variable and fixed components of the system power dissipation & Choi K., Lee W., Soma R., Pedram M. & IEEE/ACM International Conference on Computer-Aided Design, Digest of Technical Papers, ICCAD & 2004 & IEEE & N/A & N & CE1-06 & N/A & N/A \\ \hline

On combined DVS and processor evaluation & Olsen A.B., Buttner F., Koch P. & Lecture Ntes in Computer Science (including subseries Lecture Ntes in Artificial Intelligence and Lecture Ntes in Bioinformatics) & 2004 & Springer & N/A & N & CE1-05 & N/A & N/A \\ \hline

Feedback EDF scheduling exploiting dynamic voltage scaling & Zhu Y., Mueller F. & Proceedings - IEEE Real-Time and Embedded TechNlogy and Applications Symposium & 2004 & IEEE & S & S & CI1-01 & N & CE2-02 \\ \hline

Energy-efficient policies for request-driven soft real-time systems & Rusu C., Xu R., Melhem R., Mosse D. & Proceedings - Euromicro Conference on Real-Time Systems & 2004 & IEEE & N/A & N & CE1-03 & N/A & N/A \\ \hline

Effective method of low-power compilation optimization: Localizing the use of system units & Yi H.-Z., Yang X.-J. & Ruan Jian Xue Bao/Journal of Software & 2004 & National KNwledge Infrastructure (CNKI) & N/A & N & CE1-05 & N/A & N/A \\ \hline

Network flow techniques for dynamic voltage scaling in hard real-time systems & Swaminathan V., Chakrabarty K. & IEEE Transactions on Computer-Aided Design of Integrated Circuits and Systems & 2004 & IEEE & S & S & CI1-01 & N & CE2-02 \\ \hline

Profile-based optimal intra-task voltage scheduling for hard real-time applications & Seo J., Kim T., Chung K.-S. & Proceedings - Design Automation Conference & 2004 & ACM and IEEE & S & S & CI1-01 & S & CI2-01 \\ \hline

A unified approach to variable voltage scheduling for Nnideal DVS processors & Mochocki B.C., Hu X.S., Quan G. & IEEE Transactions on Computer-Aided Design of Integrated Circuits and Systems & 2004 & IEEE & S & S & CI1-01 & N & CE2-02 \\ \hline

Power-aware scheduling for AND/OR graphs in real-time systems & Zhu D., Mosse D., Melhem R. & IEEE Transactions on Parallel and Distributed Systems & 2004 & IEEE & S & S & CI1-01 & N & CE2-02 \\ \hline

Combining compiler and runtime IPC predictions to reduce energy in next generation architectures & Chheda S., Unsal O., Koren I., Krishna C.M., Moritz C.A. & 2004 Computing Frontiers Conference & 2004 & ACM & N/A & N & CE1-05 & N/A & N/A \\ \hline

Fine-grained dynamic voltage and frequency scaling for precise energy and performance trade-off based on the ratio of off-chip access to on-chip computation times & Choi K., Soma R., Pedram M. & Proceedings - Design, Automation and Test in Europe Conference and Exhibition & 2004 & ACM & S & S & CI1-01 & N & CE2-02 \\ \hline

Power-aware scheduling for periodic real-time tasks & Aydin H., Melhem R., Mosse D., Mejia-Alvarez P. & IEEE Transactions on Computers & 2004 & IEEE & S & S & CI1-01 & N & CE2-02 \\ \hline

A dynamic voltage scaling algorithm for sporadic tasks & Qadi A., Goddard S., Farritor S. & Proceedings - Real-Time Systems Symposium & 2003 & ACM and IEEE & S & S & CI1-01 & N & CE2-02 \\ \hline

Collaborative operating system and compiler power management for real-time applications & Aboughazaleh N., Mosse D., Childers B., Melhem R., Craven M. & Real-Time TechNlogy and Applications - Proceedings & 2003 & IEEE & S & S & CI1-01 & S & CI2-01 \\ \hline

Dynamic voltage and frequency scaling for scientific applications & Hsu C.-H., Kremer U. & Lecture Ntes in Computer Science (including subseries Lecture Ntes in Artificial Intelligence and Lecture Ntes in Bioinformatics) & 2003 & Springer & N/A & N & CE1-05 & N/A & N/A \\ \hline

Exploring efficient operating points for voltage scaled embedded processor cores & Buss M., Givargis T., Dutt N. & Proceedings - Real-Time Systems Symposium & 2003 & ACM and IEEE & S & S & CI1-01 & S & CI2-01 \\ \hline

Multiversion scheduling in rechargeable energy-aware real-time systems & Rusu C., Melhem R., Mosse D. & Proceedings - Euromicro Conference on Real-Time Systems & 2003 & IEEE & N/A & N & CE1-06 & N/A & N/A \\ \hline

Pareto-Optimization-Based Run-Time Task Scheduling for Embedded Systems & Yang P., Catthoor F. & Hardware/Software Codesign - Proceedings of the International Workshop & 2003 & ACM & N/A & N & CE1-05 & N/A & N/A \\ \hline

Uncertainty-Based Scheduling: Energy-Efficient Ordering for Tasks with Variable Execution Time & Gruian F., Kuchcinski K. & Proceedings of the International Symposium on Low Power Electronics and Design & 2003 & IEEE & S & S & CI1-01 & N & CE2-02 \\ \hline

Energy management for real-time embedded applications with compiler support & AbouGhazaleh N., Childers B., Mosse D., Melhem R., Craven M. & Proceedings of the ACM SIGPLAN Conference on Languages, Compilers, and Tools for Embedded Systems (LCTES) & 2003 & ACM & S & S & CI1-01 & S & CI2-01 \\ \hline
\end{tabular}
}
\end{table}

\begin{table}[htp] 
\centering
\caption{Lista de publicações catalogadas após o 1\textordmasculine \, e 2\textordmasculine \, filtros - Parte 13.}
\resizebox{\columnwidth}{!}{%
\begin{tabular}{|>{\centering\arraybackslash}p{6cm}|>{\centering\arraybackslash}p{4cm}|>{\centering\arraybackslash}p{4cm}|c|>{\centering\arraybackslash}p{3cm}|c|c|>{\centering\arraybackslash}p{3cm}|c|>{\centering\arraybackslash}p{3cm}|}
\hline
\textbf{Título} & \textbf{Autores} & \textbf{Fonte da Publicação} & \textbf{Ano} & \textbf{Editora} & \textbf{Disp} & \textbf{1F} & \textbf{Critério} & \textbf{2F} & \textbf{Critério} \\ \hline

Application-Directed Voltage Scaling & Pouwelse J., Langendoen K., Sips H.J. & IEEE Transactions on Very Large Scale Integration (VLSI) Systems & 2003 & IEEE & N/A & N & CE1-05 & N/A & N/A \\ \hline

Maximizing the system value while satisfying time and energy constraints & Rusu C.A., Melhem R., Mosse D. & IBM Journal of Research and Development & 2003 & IEEE & S & S & CI1-01 & N & CE2-02 \\ \hline

Minimal energy fixed-priority scheduling for variable voltage processors & Quan G., Hu X.S. & IEEE Transactions on Computer-Aided Design of Integrated Circuits and Systems & 2003 & IEEE & S & S & CI1-01 & N & CE2-02 \\ \hline

Compile-time dynamic voltage scaling settings: Opportunities and limits & Xie F., MartoNsi M., Malik S. & Proceedings of the ACM SIGPLAN Conference on Programming Language Design and Implementation (PLDI) & 2003 & ACM & N/A & N & CE1-05 & N/A & N/A \\ \hline

The design, implementation, and evaluation of a compiler algorithm for CPU energy reduction & Hsu C.-H., Kremer U. & Proceedings of the ACM SIGPLAN Conference on Programming Language Design and Implementation (PLDI) & 2003 & ACM & N/A & N & CE1-05 & N/A & N/A \\ \hline

Energy management for real-time embedded applications with compiler support & AbouGhazaleh N., Childers B., Mosse D., Melhem R., Craven M. & ACM SIGPLAN Ntices & 2003 & ACM & S & N & CE1-07 & N & CE1-07 \\ \hline

Compile-time dynamic voltage scaling settings: Opportunities and limits & Xie F., MartoNsi M., Malik S. & ACM SIGPLAN Ntices & 2003 & ACM & N/A & N & CE1-05 & N/A & N/A \\ \hline

The design, implementation, and evaluation of a compiler algorithm for CPU energy reduction & Hsu C.-H., Kremer U. & ACM SIGPLAN Ntices & 2003 & ACM & N/A & N & CE1-05 & N/A & N/A \\ \hline

Variable voltage task scheduling algorithms for minimizing energy/power & Manzak A., Chakrabarti C. & IEEE Transactions on Very Large Scale Integration (VLSI) Systems & 2003 & IEEE & N/A & N & CE1-05 & N/A & N/A \\ \hline

An intra-task dynamic voltage scaling method for SoC design with hierarchical FSM and synchroNus dataflow model & Lee S., Yoo S., Choi K. & Proceedings of the International Symposium on Low Power Electronics and Design, Digest of Technical Papers & 2002 & ACM & S & S & CI1-01 & S & CI2-01 \\ \hline

Contents provider-assisted dynamic voltage scaling for low energy multimedia applications & Chung E.-Y., Benini L., De Micheli G. & Proceedings of the International Symposium on Low Power Electronics and Design, Digest of Technical Papers & 2002 & ACM and IEEE & N/A & N & CE1-05 & N/A & N/A \\ \hline

Energy-conserving feedback EDF scheduling for embedded systems with real-time constraints & Dudani A., Mueller F., Zhu Y. & Joint COnference on Languages, Compilers and Tools for Embedded Systems and Software and Compilers for Embedded Systems & 2002 & ACM & S & S & CI1-01 & N & CE2-02 \\ \hline

Energy-efficient duplex and TMR real-time systems & ElNzahy E.M., Melhem R., Mosse D. & Proceedings - Real-Time Systems Symposium & 2002 & IEEE & N/A & N & CE1-06 & N/A & N/A \\ \hline

Frame-based dynamic voltage and frequency scaling for a MPEG decoder & Choi K., Dantu K., Cheng W.-C., Pedram M. & IEEE/ACM International Conference on Computer-Aided Design, Digest of Technical Papers & 2002 & ACM & N/A & N & CE1-05 & N/A & N/A \\ \hline

Maximizing the system value while satisfying time and energy constraints & Rusu C., Melhem R., Mosse D. & Proceedings - Real-Time Systems Symposium & 2002 & IEEE & N/A & N & CE1-06 & N/A & N/A \\ \hline

Performance comparison of dynamic voltage scaling algorithms for hard real-time systems & Kim W., Shin D., Yun H.-S., Kim J., Min S.L. & Real-Time TechNlogy and Applications - Proceedings & 2002 & IEEE & N/A & N & CE1-07 & N/A & N/A \\ \hline

Low-energy intra-task voltage scheduling using static timing analysis & Shin D., Kim J., Lee S. & Proceedings - Design Automation Conference & 2001 & IEEE & S & S & CI1-01 & S & CI2-01 \\ \hline

Intra-task voltage scheduling for low-energy hard real-time applications & Shin D., Kim J., Lee S. & IEEE Design and Test of Computers & 2001 & IEEE & S & S & CI1-01 & S & CI2-01 \\ \hline

A profile-based energy-efficient intra-task voltage scheduling algorithm for hard real-time applications & Shin D., Kim J. & Proceedings of the International Symposium on Low Power Electronics and Design, Digest of Technical Papers & 2001 & ACM and IEEE & S & S & CI1-01 & S & CI2-01 \\ \hline

Compiler-directed dynamic voltage/frequency scheduling for energy reduction in microprocessors & Hsu C.-H., Kremer U., Hsiao M. & Proceedings of the International Symposium on Low Power Electronics and Design, Digest of Technical Papers & 2001 & ACM and IEEE & N/A & N & CE1-05 & N/A & N/A \\ \hline

\end{tabular}
}
\end{table}

\newpage
\begin{table}[htp] 
\centering
\caption{Lista de publicações catalogadas após o 1\textordmasculine \, e 2\textordmasculine \, filtros - Parte 14.}
\resizebox{\columnwidth}{!}{%
\begin{tabular}{|>{\centering\arraybackslash}p{6cm}|>{\centering\arraybackslash}p{4cm}|>{\centering\arraybackslash}p{4cm}|c|>{\centering\arraybackslash}p{3cm}|c|c|>{\centering\arraybackslash}p{3cm}|c|>{\centering\arraybackslash}p{3cm}|}
\hline
\textbf{Título} & \textbf{Autores} & \textbf{Fonte da Publicação} & \textbf{Ano} & \textbf{Editora} & \textbf{Disp} & \textbf{1F} & \textbf{Critério} & \textbf{2F} & \textbf{Critério} \\ \hline

Dynamic voltage scaling on a low-power microprocessor & Pouwelse J., Langendoen K., Sips H. & Proceedings of the Annual International Conference on Mobile Computing and Networking, MOBICOM & 2001 & ACM & N/A & N & CE1-05 & N/A & N/A \\ \hline

Dynamic voltage scheduling technique for low-power multimedia applications using buffers & Im C., Kim H., Ha S. & Proceedings of the International Symposium on Low Power Electronics and Design, Digest of Technical Papers & 2001 & ACM & N/A & N & CE1-05 & N/A & N/A \\ \hline

Hard real-time scheduling for low-energy using stochastic data and DVS processors & Gruian F. & Proceedings of the International Symposium on Low Power Electronics and Design, Digest of Technical Papers & 2001 & ACM & S & S & CI1-01 & N & CE2-02 \\ \hline

Low power system scheduling and synthesis & Jha N.K. & IEEE/ACM International Conference on Computer-Aided Design, Digest of Technical Papers & 2001 & IEEE & N/A & N & CE1-07 & N/A & N/A \\ \hline

Voltage scaling for energy minimization with QoS constraints & Manzak A., Chakrabarti C. & Proceedings - IEEE International Conference on Computer Design: VLSI in Computers and Processors & 2001 & IEEE & N/A & N & CE1-05 & N/A & N/A \\ \hline
\end{tabular}
}
\end{table}

\section{Base de Dados da Revisão Sistemática}
\label{section:Apendice_Base_Dados_Revisao_Sistematica}

Todas as informações coletadas e catalogadas após a execução do 2\textordmasculine \, filtro foram tabeladas em ordem cronológica e agrupada de acordo com os critérios de extração de dados definidos na Seção~\ref{subsection:Procedimentos_Extracao_Dados}. As tabelas a seguir mostram todos os dados extraídos de cada uma das 39 publicações selecionadas no 2\textordmasculine \, filtro.

\begin{table}[htp] 
\centering
\caption{Dados extraído da publicação de \cite{Shin_Kim_2001_a}.}
\resizebox{\columnwidth}{!}{%
\begin{tabular}{|p{3.8cm}|p{16cm}|}
\hline
\multicolumn{1}{|r|}{\textbf{ID}} & Shin\_Kim\_2001\_a \\ \hline
\multicolumn{1}{|r|}{\textbf{Título}} & A profile-based energy-efficient intra-task voltage scheduling algorithm for hard real-time applications \\ \hline
\multicolumn{1}{|r|}{\textbf{Autores}} & Shin D., Kim J. \\ \hline
\multicolumn{1}{|r|}{\textbf{Palavras-chave}} & Algorithms; Decoding; Delay circuits; Energy utilization; Probability; Voltage control; Dynamic voltage scaling; CMOS integrated circuits \\ \hline
\multicolumn{1}{|r|}{\textbf{Fonte da publicação}} & International Symposium on Low Electronics and Design (ISLPED'01) \\ \hline
\multicolumn{1}{|r|}{\textbf{Editora}} & ACM and IEEE \\ \hline
\multicolumn{1}{|r|}{\textbf{Ano da publicação}} & 2001 \\ \hline \hline

\multicolumn{2}{|c|}{\textbf{Resumo da publicação}} \\ \hline
\multicolumn{2}{|p{21cm}|}{Os autores foram os precursores no desenvolvimento de ferramentas para análise do WCET intra-tarefa em aplicações de tempo real. A principal finalidade dos algoritmos da ferramenta eram controlar a velocidade de execução da aplicação baseado nos caminhos de execução de caso médio (em inglês, \textit{Average-Case Execution Path - ACEP}), que são os caminhos mais frequentemente executados. Com essa abordagem os autores conseguiram provar que o algoritmo proposto é mais eficaz na redução do consumo de energia que o algoritmo original intraVS, chamado pelos autores de \textit{(RWEP)-based IntraVS} (em inglês, \textit{Remaining Worst-Case Execution Path-based Intra VS}), onde mesmo utilizando as ACEPs é possível satisfazer as restrições temporais da aplicação de tempo real. Esse método se baseia no perfil de comportamento da aplicação, através da análise dos caminhos de execução mais utilizados (ou \textit{hot paths}), chamado de \textit{(RAEP)-based IntraVS} (em inglês, \textit{Remaining Average-Case Execution Path-based IntraVS}). Sua principal contribuição está na exploração das probabilidades de cada caminho de execução da aplicação e garantir que as restrições temporais sejam respeitadas mesmo executando o pior caso. Os experimentos mostram que o \textit{RAEP-based} é 34\% mais eficiente energeticamente que o \textit{RWEP-based}.} \\ \hline \hline

\multicolumn{1}{|r|}{\textbf{Método(s) utilizados}} & DVFS Intra-Task; Control Flow Graph (CFG); \\ \hline
\multicolumn{1}{|r|}{\textbf{Ferramenta(s)}} & {Os autores utilizaram a ferramenta \textit{Automatic Voltage Scaler} (AVS) [Não disponível para Download]. \newline

Os autores utilizaram o seguinte \textit{benchmark}: \newline
- MPEG-4 \textit{decoder}} \\ \hline \hline

\multicolumn{2}{|c|}{\textbf{Impacto (positivo x negativo)}} \\ \hline
\multicolumn{2}{|p{21cm}|}{O artigo apresentou como ponto positivo uma nova metodologia que explora as probabilidades de cada caminho de execução da aplicação, tomando como base os \textit{hot paths} do caso médio (ACE), e provou que é possível reduzir o consumo de energia e respeitar as restrições temporais de todas as tarefas.. Como ponto negativo, a metodologia não dá suporte a preempções e não mostrou casos de teste para analisar o desempenho da aplicação diante da inserção de muitos pontos de escalonamento de tensão.} \\ \hline \hline

\multicolumn{2}{|c|}{\textbf{Validação do método}} \\ \hline
\multicolumn{2}{|p{21cm}|}{A validação da metodologia ocorreu através do uso de ferramentas e \textit{benchmarks} relatados na literatura. Além disso, compara os resultados experimentais com outros trabalhos de referência na área de baixo consumo de energia.} \\ \hline \hline

\multicolumn{2}{|c|}{\textbf{Limitações do método}} \\ \hline
\multicolumn{2}{|p{21cm}|}{A metodologia não dá suporte a preempções e não mostrou casos de teste para analisar o desempenho da aplicação diante da inserção de muitos pontos de escalonamento de tensão.} \\ \hline \hline

\multicolumn{2}{|c|}{\textbf{Integração de métodos}} \\ \hline
\multicolumn{2}{|p{21cm}|}{Não se aplica} \\ \hline \hline

\multicolumn{2}{|c|}{\textbf{Modo de aplicação do método}} \\ \hline
\multicolumn{2}{|p{21cm}|}{Offline} \\ \hline \hline

\multicolumn{2}{|c|}{\textbf{Perspectivas futuras}} \\ \hline
\multicolumn{2}{|p{21cm}|}{Não identificada} \\ \hline \hline

\multicolumn{2}{|c|}{\textbf{Comentários}} \\ \hline
\multicolumn{2}{|p{21cm}|}{A metodologia proposta não mostra evidências, no artigo, de que fornece suporte para recursos compartilhados e / ou preempções.} \\ \hline
\end{tabular}
}
\end{table}

\newpage
\begin{table}[htp] 
\centering
\caption{Dados extraído da publicação de \cite{Shin_Kim_Lee_2001_a}.}
\resizebox{\columnwidth}{!}{%
\begin{tabular}{|p{3.8cm}|p{16cm}|}
\hline
\multicolumn{1}{|r|}{\textbf{ID}} & Shin\_Kim\_Lee\_2001\_a \\ \hline
\multicolumn{1}{|r|}{\textbf{Título}} & Intra-task voltage scheduling for low-energy hard real-time applications \\ \hline
\multicolumn{1}{|r|}{\textbf{Autores}} & Shin D., Kim J., Lee S. \\ \hline
\multicolumn{1}{|r|}{\textbf{Palavras-chave}} & Dynamic power consumption; Dynamic voltage scaling technique; Intra-task voltage scheduling; Velocity saturation index; Algorithms; Capacitance; CMOS integrated circuits; Computer software; Performance; Portable equipment; Reliability; Threshold voltage; VLSI circuits; Electric power supplies to apparatus \\ \hline
\multicolumn{1}{|r|}{\textbf{Fonte da publicação}} & IEEE Design and Test of Computers \\ \hline
\multicolumn{1}{|r|}{\textbf{Editora}} & IEEE \\ \hline
\multicolumn{1}{|r|}{\textbf{Ano da publicação}} & 2001 \\ \hline \hline

\multicolumn{2}{|c|}{\textbf{Resumo da publicação}} \\ \hline
\multicolumn{2}{|p{21cm}|}{Os autores propõem um novo algoritmo de escalonamento de tensão intra-tarefa que controla a tensão de alimentação do processador durante a execução da tarefa, através da exploração dos tempos de folga. Esse método se baseia na análise do tempo de execução estático e na inserção de códigos, dentro do código fonte da aplicação, para a realização dos chaveamentos de tensões e frequências do processador, de forma que o consumo geral de energia seja reduzido. Esses códigos de chaveamento de tensão são definidos para cada um dos blocos de código selecionados a partir do grafo de fluxo de controle da aplicação (CFG). Dessa forma é possível definir as tensões e frequências para cada bloco de código, aproximando assim o tempo de execução ao deadline da tarefa, sempre respeitando as restrições temporais de todas as tarefas em execução. Neste trabalho os autores introduziram uma nova perspectiva para analisar as CFGs, que consiste em mapear os blocos de código por estruturas condicionais (chamado de \textit{B-types}) e por estruturas de repetição (chamado de \textit{L-types}), dessa forma é mais fácil analisar e predizer os cálculos do WCEC (em inglês, \textit{Worst Case Execution Cycle}) e RWCEC (em inglês, \textit{Remaining Worst Case Execution Cycle}). Todas essas analises foram introduzidas na ferramenta AVS (em inglês, \textit{Automatic Voltage Scaler}), desenvolvida pelos próprios autores. O único ponto negativo no estudo realizado é a falta de métricas para avaliar os reais impactos causados pela inserção de códigos adicionais dentro das aplicações.} \\ \hline \hline

\multicolumn{1}{|r|}{\textbf{Método(s) utilizados}} & DVFS Intra-Task; Control Flow Graph (CFG); \\ \hline
\multicolumn{1}{|r|}{\textbf{Ferramenta(s)}} & {Os autores utilizaram a ferramenta \textit{Automatic Voltage Scaler} (AVS) [Não disponível para Download]. \newline

Os autores utilizaram o seguinte \textit{benchmark}: \newline
- MPEG4 \textit{decoder}} \\ \hline \hline

\multicolumn{2}{|c|}{\textbf{Impacto (positivo x negativo)}} \\ \hline
\multicolumn{2}{|p{21cm}|}{O artigo apresentou como ponto positivo uma nova metodologia para a inserção automática de pontos de escalonamento de tensão, de forma que cada bloco de código extraído da aplicação possa ser executado com valores de tensão e frequência ideais. Como ponto negativo, a metodologia não dá suporte a preempções e não mostrou estudos de caso para avaliar o impacto da inserção de muitos pontos de chaveamento de tensão dentro do código fonte da aplicação.} \\ \hline \hline

\multicolumn{2}{|c|}{\textbf{Validação do método}} \\ \hline
\multicolumn{2}{|p{21cm}|}{A validação da metodologia ocorreu através do uso de ferramentas e \textit{benchmarks} relatados na literatura.} \\ \hline \hline

\multicolumn{2}{|c|}{\textbf{Limitações do método}} \\ \hline
\multicolumn{2}{|p{21cm}|}{A metodologia não dá suporte a preempções e não mostrou estudos de caso para avaliar o impacto da inserção de muitos pontos de chaveamento de tensão dentro do código fonte da aplicação.} \\ \hline \hline

\multicolumn{2}{|c|}{\textbf{Integração de métodos}} \\ \hline
\multicolumn{2}{|p{21cm}|}{Não se aplica} \\ \hline \hline

\multicolumn{2}{|c|}{\textbf{Modo de aplicação do método}} \\ \hline
\multicolumn{2}{|p{21cm}|}{Offline} \\ \hline \hline

\multicolumn{2}{|c|}{\textbf{Perspectivas futuras}} \\ \hline
\multicolumn{2}{|p{21cm}|}{Os autores pretendem realizar uma comparação quantitativa entre as abordagens intra-tarefa e inter-tarefa.} \\ \hline \hline

\multicolumn{2}{|c|}{\textbf{Comentários}} \\ \hline
\multicolumn{2}{|p{21cm}|}{A metodologia proposta não mostra evidências, no artigo, de que fornece suporte para recursos compartilhados e / ou preempções.} \\ \hline
\end{tabular}
}
\end{table}

\newpage
\begin{table}[htp] 
\centering
\caption{Dados extraído da publicação de \cite{Shin_Kim_Lee_2001_b}.}
\resizebox{\columnwidth}{!}{%
\begin{tabular}{|p{3.8cm}|p{16cm}|}
\hline
\multicolumn{1}{|r|}{\textbf{ID}} & Shin\_Kim\_Lee\_2001\_b \\ \hline
\multicolumn{1}{|r|}{\textbf{Título}} & Low-energy intra-task voltage scheduling using static timing analysis \\ \hline
\multicolumn{1}{|r|}{\textbf{Autores}} & Shin D., Kim J., Lee S. \\ \hline
\multicolumn{1}{|r|}{\textbf{Palavras-chave}} & Algorithms; Computer aided software engineering; Image coding; VLSI circuits; Voltage scheduling; Real time systems \\ \hline
\multicolumn{1}{|r|}{\textbf{Fonte da publicação}} & 38th Design Automation Conference; Las Vegas, NV; United States; \\ \hline
\multicolumn{1}{|r|}{\textbf{Editora}} & IEEE \\ \hline
\multicolumn{1}{|r|}{\textbf{Ano da publicação}} & 2001 \\ \hline \hline

\multicolumn{2}{|c|}{\textbf{Resumo da publicação}} \\ \hline
\multicolumn{2}{|p{21cm}|}{Os autores propõem uma nova metodologia para analise do WCET (em inglês, \textit{Worst Case Execution Time}) em aplicações de tempo real e tomaram com base o trabalho de \cite{Shin_Kim_Lee_2001_a}. Essa análise é feita em tempo de compilação, de modo \textit{offline}, utilizando o grafo de fluxo de controle da aplicação, onde o cálculo do WCET é feita para cada nó da CFG, enquanto que no trabalho anterior dos mesmos autores a estimativa do WCET era feita tendo como base o programa inteiro.} \\ \hline \hline

\multicolumn{1}{|r|}{\textbf{Método(s) utilizados}} & DVFS Intra-Task; Control Flow Graph (CFG); \\ \hline
\multicolumn{1}{|r|}{\textbf{Ferramenta(s)}} & {Os autores utilizaram a ferramenta \textit{Automatic Voltage Scaler} (AVS) [Ferramenta não localizada] \newline

Os autores utilizaram o seguinte \textit{benchmark}: \newline
- MPEG-4 \textit{encoder e decoder}} \\ \hline \hline

\multicolumn{2}{|c|}{\textbf{Impacto (positivo x negativo)}} \\ \hline
\multicolumn{2}{|p{21cm}|}{O artigo apresentou como ponto positivo uma nova metodologia para realizar análises estáticas dos tempos de execução da aplicação, visando mapear os tempos de folga e, principalmente, identificar as tensões e frequências ideais que devem ser aplicadas no processador para que a tarefa termine de executar o mais próximo possível do seu deadline, economizando energia. Como ponto negativo, a metodologia não dá suporte a preempções.} \\ \hline \hline

\multicolumn{2}{|c|}{\textbf{Validação do método}} \\ \hline
\multicolumn{2}{|p{21cm}|}{A validação da metodologia ocorreu através do uso de ferramentas e \textit{benchmarks} relatados na literatura.} \\ \hline \hline

\multicolumn{2}{|c|}{\textbf{Limitações do método}} \\ \hline
\multicolumn{2}{|p{21cm}|}{A metodologia não dá suporte a preempções.} \\ \hline \hline

\multicolumn{2}{|c|}{\textbf{Integração de métodos}} \\ \hline
\multicolumn{2}{|p{21cm}|}{Não se aplica} \\ \hline \hline

\multicolumn{2}{|c|}{\textbf{Modo de aplicação do método}} \\ \hline
\multicolumn{2}{|p{21cm}|}{Offline} \\ \hline \hline

\multicolumn{2}{|c|}{\textbf{Perspectivas futuras}} \\ \hline
\multicolumn{2}{|p{21cm}|}{O autor pretende incluir no método proposto outras informações, em tempo de execução, para melhor o algoritmo de escalonamento e a redução do consumo de energia do processador.} \\ \hline \hline

\multicolumn{2}{|c|}{\textbf{Comentários}} \\ \hline
\multicolumn{2}{|p{21cm}|}{A metodologia proposta não mostra evidências, no artigo, de que fornece suporte para recursos compartilhados e / ou preempções.} \\ \hline
\end{tabular}
}
\end{table}

\newpage
\begin{table}[htp] 
\centering
\caption{Dados extraído da publicação de \cite{Lee_Yoo_Choi_2002_a}.}
\resizebox{\columnwidth}{!}{%
\begin{tabular}{|p{3.8cm}|p{16cm}|}
\hline
\multicolumn{1}{|r|}{\textbf{ID}} & Lee\_Yoo\_Choi\_2002\_a \\ \hline
\multicolumn{1}{|r|}{\textbf{Título}} & An intra-task dynamic voltage scaling method for SoC design with hierarchical FSM and synchronous dataflow model \\ \hline
\multicolumn{1}{|r|}{\textbf{Autores}} & Lee S., Yoo S., Choi K. \\ \hline
\multicolumn{1}{|r|}{\textbf{Palavras-chave}} & Dynamic voltage scaling; Finite state machine; Formal model; Low power; Synchronous dataflow; Variable supply voltage \\ \hline
\multicolumn{1}{|r|}{\textbf{Fonte da publicação}} & Proceedings of the 2002 International Symposium on Low Power Electronics and Design; Monterey, CA;  \\ \hline
\multicolumn{1}{|r|}{\textbf{Editora}} & ACM \\ \hline
\multicolumn{1}{|r|}{\textbf{Ano da publicação}} & 2002 \\ \hline \hline

\multicolumn{2}{|c|}{\textbf{Resumo da publicação}} \\ \hline
\multicolumn{2}{|p{21cm}|}{Os autores propõem um método de escalonamento de tensão para o projeto de SoCs (em inglês, \textit{System on a Chips}) com hierarquia FSM (em inglês, \textit{Finite State Machine}) e modelo de dados síncrono. Essa técnica foi chamada de modelo HFSM-SDF (em inglês, \textit{Hierarchical FSM and Synchronous Dataflow Model}). Essa metodologia calcula o caminho de execução da aplicação em tempo de execução e utiliza muitos dos conceitos definidos por \cite{Shin_Kim_Lee_2001_a}, para calcular a carga de trabalho restante das tarefas de tempo real e assim aplicar sobre o processador as tensões e frequências ideais para que o consumo de energia seja o menor possível.} \\ \hline \hline

\multicolumn{1}{|r|}{\textbf{Método(s) utilizados}} & DVFS Intra-Task; Control Flow Graph (CFG); \\ \hline
\multicolumn{1}{|r|}{\textbf{Ferramenta(s)}} & {Os autores utilizaram a ferramenta ARMulator [Disponível em: \href{http://x-y-z.github.io/armulator/}{http://x-y-z.github.io/armulator/}]. \newline

Os autores utilizaram o seguinte \textit{benchmark}: \newline
- MPEG-4 \textit{natural video decoder} [Disponível em: \href{http://www.m4if.org/}{http://www.m4if.org/}].} \\ \hline \hline

\multicolumn{2}{|c|}{\textbf{Impacto (positivo x negativo)}} \\ \hline
\multicolumn{2}{|p{21cm}|}{O artigo apresentou como ponto positivo uma nova metodologia online para calcular a carga de trabalho restante da aplicação e as tensões e frequências ideais que devem ser aplicadas no processador. Como ponto negativo, a metodologia não leva em consideração múltiplas tarefas em execução.} \\ \hline \hline

\multicolumn{2}{|c|}{\textbf{Validação do método}} \\ \hline
\multicolumn{2}{|p{21cm}|}{A validação da metodologia ocorreu através do uso de ferramentas e \textit{benchmarks} relatados na literatura.} \\ \hline \hline

\multicolumn{2}{|c|}{\textbf{Limitações do método}} \\ \hline
\multicolumn{2}{|p{21cm}|}{A metodologia não leva em consideração um ambiente com múltiplas tarefas em execução.} \\ \hline \hline

\multicolumn{2}{|c|}{\textbf{Integração de métodos}} \\ \hline
\multicolumn{2}{|p{21cm}|}{Não se aplica} \\ \hline \hline

\multicolumn{2}{|c|}{\textbf{Modo de aplicação do método}} \\ \hline
\multicolumn{2}{|p{21cm}|}{Híbrida} \\ \hline \hline

\multicolumn{2}{|c|}{\textbf{Perspectivas futuras}} \\ \hline
\multicolumn{2}{|p{21cm}|}{Extender o método proposto para um ambiente com multi-tarefas e dar suporte a preempções.} \\ \hline \hline

\multicolumn{2}{|c|}{\textbf{Comentários}} \\ \hline
\multicolumn{2}{|p{21cm}|}{A metodologia proposta não mostra evidências, no artigo, de que fornece suporte para recursos compartilhados e / ou preempções.} \\ \hline
\end{tabular}
}
\end{table}

\newpage
\begin{table}[htp] 
\centering
\caption{Dados extraído da publicação de \cite{Aboughazaleh_2003_a}.}
\resizebox{\columnwidth}{!}{%
\begin{tabular}{|p{3.8cm}|p{16cm}|}
\hline
\multicolumn{1}{|r|}{\textbf{ID}} & Aboughazaleh\_2003\_a \\ \hline
\multicolumn{1}{|r|}{\textbf{Título}} & Energy management for real-time embedded applications with compiler support \\ \hline
\multicolumn{1}{|r|}{\textbf{Autores}} & AbouGhazaleh N., Childers B., Mosse D., Melhem R., Craven M. \\ \hline
\multicolumn{1}{|r|}{\textbf{Palavras-chave}} & Dynamic voltage scaling; Power management; Real-time \\ \hline
\multicolumn{1}{|r|}{\textbf{Fonte da publicação}} & Proceedings of the ACM SIGPLAN Conference on Languages, Compilers, and Tools for Embedded Systems (LCTES) \\ \hline
\multicolumn{1}{|r|}{\textbf{Editora}} & ACM \\ \hline
\multicolumn{1}{|r|}{\textbf{Ano da publicação}} & 2003 \\ \hline \hline

\multicolumn{2}{|c|}{\textbf{Resumo da publicação}} \\ \hline
\multicolumn{2}{|p{21cm}|}{Os autores propõem uma técnica que explora as variações dos tempos de execução em diferentes caminhos de execução da aplicação. Esta é uma abordagem híbrida que depende do compilador e do sistema operacional para melhor gerenciar o desempenho e a redução do consumo de energia do processador. O compilador então insere os chamados PMHs (em inglês, \textit{Power Management Hints}), que são trechos de código responsáveis por fornecer e coletar informações em tempo de execução da aplicação para o sistema operacional, além de estimar o desempenho da aplicação no pior caso. Dessa forma, o sistema operacional invoca os PMPs (em inglês, \textit{Power Management Points}) para realizar o chaveamento de tensão e frequência do processador com base nas informações passadas pelos PMHs.} \\ \hline \hline

\multicolumn{1}{|r|}{\textbf{Método(s) utilizados}} & DVFS Intra-Task; Control Flow Graph (CFG); \\ \hline
\multicolumn{1}{|r|}{\textbf{Ferramenta(s)}} & {Os autores utilizaram a ferramenta Simplescalar Micro-architecture Toolkit [Disponível em: \href{http://www.simplescalar.com/}{http://www.simplescalar.com/}]. \newline

Os autores utilizaram os seguintes \textit{benchmarks}: \newline
- MPEG-2 \textit{decoder} [Disponível em: \href{http://www.mpeg.org/}{http://www.mpeg.org/}]. \newline
- \textit{Automatic Target Recognition} (ATR) \textit{Application}.} \\ \hline \hline

\multicolumn{2}{|c|}{\textbf{Impacto (positivo x negativo)}} \\ \hline
\multicolumn{2}{|p{21cm}|}{O artigo apresentou como ponto positivo uma nova metodologia que visa integrar o compilador e o sistema operacional para trabalharem em conjunto na redução do consumo de energia do processador. Como ponto negativo, a metodologia foi modelada apenas para tarefas não preemptivas e durante a etapa de experimentações constatou-se que o desempenho da aplicação pode ser prejudicado dependendo da quantidade de instruções inseridas pelo compilador.} \\ \hline \hline

\multicolumn{2}{|c|}{\textbf{Validação do método}} \\ \hline
\multicolumn{2}{|p{21cm}|}{A validação da metodologia ocorreu através do uso de ferramentas e \textit{benchmarks} relatados na literatura. Além disso, compara os resultados experimentais com outros trabalhos de referência na área de baixo consumo de energia.} \\ \hline \hline

\multicolumn{2}{|c|}{\textbf{Limitações do método}} \\ \hline
\multicolumn{2}{|p{21cm}|}{A metodologia não dá suporte a preempções e pode prejudicar o desempenho das aplicações dependendo da quantidade de instruções inseridas pelo compilador.} \\ \hline \hline

\multicolumn{2}{|c|}{\textbf{Integração de métodos}} \\ \hline
\multicolumn{2}{|p{21cm}|}{Não se aplica} \\ \hline \hline

\multicolumn{2}{|c|}{\textbf{Modo de aplicação do método}} \\ \hline
\multicolumn{2}{|p{21cm}|}{Híbrida} \\ \hline \hline

\multicolumn{2}{|c|}{\textbf{Perspectivas futuras}} \\ \hline
\multicolumn{2}{|p{21cm}|}{O autor pretende incluir no seu método a extração de informações inter-tarefas.} \\ \hline \hline

\multicolumn{2}{|c|}{\textbf{Comentários}} \\ \hline
\multicolumn{2}{|p{21cm}|}{A metodologia proposta não mostra evidências, no artigo, de que fornece suporte para recursos compartilhados e / ou preempções.} \\ \hline
\end{tabular}
}
\end{table}

\begin{table}[htp] 
\centering
\caption{Dados extraído da publicação de \cite{Buss_Givargis_Dutt_2003_a}.}
\resizebox{\columnwidth}{!}{%
\begin{tabular}{|p{3.8cm}|p{16cm}|}
\hline
\multicolumn{1}{|r|}{\textbf{ID}} & Buss\_Givargis\_Dutt\_2003\_a \\ \hline
\multicolumn{1}{|r|}{\textbf{Título}} & Exploring efficient operating points for voltage scaled embedded processor cores \\ \hline
\multicolumn{1}{|r|}{\textbf{Autores}} & Buss M., Givargis T., Dutt N. \\ \hline
\multicolumn{1}{|r|}{\textbf{Palavras-chave}} & Algorithms; CMOS integrated circuits; Electric potential; Multimedia systems; Personal digital assistants; Reduced instruction set computing; Scheduling; Dynamic voltage scaled processors; Instruction-set tailoring; Register file sizing; Voltage scaled embedded processor cores; Embedded systems \\ \hline
\multicolumn{1}{|r|}{\textbf{Fonte da publicação}} & 24th IEEE International Real-Time Systems Symposium RTSS 2003 \\ \hline
\multicolumn{1}{|r|}{\textbf{Editora}} & ACM and IEEE \\ \hline
\multicolumn{1}{|r|}{\textbf{Ano da publicação}} & 2003 \\ \hline \hline

\multicolumn{2}{|c|}{\textbf{Resumo da publicação}} \\ \hline
\multicolumn{2}{|p{21cm}|}{Os autores propõem a exploração e seleção de potenciais pontos de escalonamento de tensão que possam atuar na diminuição eficiente do consumo de energia em aplicações de tempo real não críticos. A problemática desse método está em selecionar esses pontos de controle para atuar em conjunto com a técnica DVS intra-tarefa, proporcionando uma redução do consumo de energia do processador. Esse método se baseia basicamente em três passos, são eles: (1) fazer a análise estática da aplicação e atribuir um fator de desaceleração ideal para cada bloco; (2) computar as frequências de operação com base na análise da aplicação inteira; (3) reatribuir os fatores de aceleração para cada bloco, com base nas frequências de operação válidas e computadas no passo 2. Essa abordagem é muito semelhante a técnica dos coreanos \citep{Shin_Kim_Lee_2001_a}, onde a principal diferença esta na metodologia de definição dos fatores de desaceleração.} \\ \hline \hline

\multicolumn{1}{|r|}{\textbf{Método(s) utilizados}} & DVFS Intra-Task; Control Flow Graph (CFG); \\ \hline
\multicolumn{1}{|r|}{\textbf{Ferramenta(s)}} & {Os autores utilizaram a ferramenta WARTS / EEL (\textit{Executable Editing Library}) [Disponível em: \href{http://pages.cs.wisc.edu/~larus/eel.html}{http://pages.cs.wisc.edu/$\sim$larus/eel.html}]. \newline

Os autores utilizaram os seguintes \textit{benchmarks}: \newline
- MiBench [Disponível em: \href{http://www.eecs.umich.edu/mibench/}{http://www.eecs.umich.edu/mibench/}] \newline
- Powerstone \newline
- MediaBench [Disponível em: \href{http://cares.icsl.ucla.edu/MediaBench}{http://cares.icsl.ucla.edu/MediaBench}] \newline
- lame (MP3 \textit{encoding engine}) \newline
- jpeg (\textit{Image compression and decompression}) \newline
- compress (\textit{Data compression tool}) \newline
- engine (\textit{Engine controller}) \newline
- mpeg (\textit{Video encoder}) \newline
- madplay (MPEG \textit{audio decoder}) \newline
- tiff2ps (Tiff \textit{to postscript conversion}) \newline
- gsm (\textit{Speech transcoding}) \newline
- basicmath (\textit{Basic math operations}) \newline
- adpcm (\textit{Voice encoder}) \newline
- dijkstra (\textit{Shortest path algorithm})} \\ \hline \hline

\multicolumn{2}{|c|}{\textbf{Impacto (positivo x negativo)}} \\ \hline
\multicolumn{2}{|p{21cm}|}{O artigo apresentou como ponto positivo uma nova metodologia para realizar a analise e classificação de blocos de código quanto ao seu fator de desaceleração, que permite ter um controle mais refinado sobre as tensões e frequências ideais que devem ser aplicadas sobre o processador. Como ponto negativo, a metodologia não deixa claro se dar suporte há algumas estruturas de dados como \textit{loops} ocultos e recursões.} \\ \hline \hline

\multicolumn{2}{|c|}{\textbf{Validação do método}} \\ \hline
\multicolumn{2}{|p{21cm}|}{A validação da metodologia ocorreu através do uso de ferramentas e \textit{benchmarks} relatados na literatura.} \\ \hline \hline

\multicolumn{2}{|c|}{\textbf{Limitações do método}} \\ \hline
\multicolumn{2}{|p{21cm}|}{A metodologia não deixa claro se dar suporte há algumas estruturas de dados como \textit{loops} ocultos e recursões.} \\ \hline \hline

\multicolumn{2}{|c|}{\textbf{Integração de métodos}} \\ \hline
\multicolumn{2}{|p{21cm}|}{Não se aplica} \\ \hline \hline

\multicolumn{2}{|c|}{\textbf{Modo de aplicação do método}} \\ \hline
\multicolumn{2}{|p{21cm}|}{Offline} \\ \hline \hline

\multicolumn{2}{|c|}{\textbf{Perspectivas futuras}} \\ \hline
\multicolumn{2}{|p{21cm}|}{Os autores pretendem incluir os custos de tempo e energia para melhorar a eficiência do seu método, tornando assim a metodologia mais realista e preciso.} \\ \hline \hline

\multicolumn{2}{|c|}{\textbf{Comentários}} \\ \hline
\multicolumn{2}{|p{21cm}|}{A metodologia proposta não mostra evidências, no artigo, de que fornece suporte para recursos compartilhados e / ou preempções.} \\ \hline
\end{tabular}
}
\end{table}

\newpage
\begin{table}[htp] 
\centering
\caption{Dados extraído da publicação de \cite{Aboughazaleh_2003_b}.}
\resizebox{\columnwidth}{!}{%
\begin{tabular}{|p{3.8cm}|p{16cm}|}
\hline
\multicolumn{1}{|r|}{\textbf{ID}} & Aboughazaleh\_2003\_b \\ \hline
\multicolumn{1}{|r|}{\textbf{Título}} & Collaborative operating system and compiler power management for real-time applications \\ \hline
\multicolumn{1}{|r|}{\textbf{Autores}} & Aboughazaleh N., Mosse D., Childers B., Melhem R., Craven M. \\ \hline
\multicolumn{1}{|r|}{\textbf{Palavras-chave}} & Collaborative approach; Dynamic power consumption; Dynamic voltage scaling; Embedded application; Power management points; Real-time application; Reduce energy consumption; Temporal information \\ \hline
\multicolumn{1}{|r|}{\textbf{Fonte da publicação}} & 9th IEEE Real-Time and Embedded Technology and Applications Symposium, RTAS 2003 \\ \hline
\multicolumn{1}{|r|}{\textbf{Editora}} & IEEE \\ \hline
\multicolumn{1}{|r|}{\textbf{Ano da publicação}} & 2003 \\ \hline \hline

\multicolumn{2}{|c|}{\textbf{Resumo da publicação}} \\ \hline
\multicolumn{2}{|p{21cm}|}{Os autores tomaram como base o trabalho de \cite{Aboughazaleh_2003_a}, onde o foco principal da metodologia passou a ser a colaboração entre o compilador e o sistema operacional. O principal contribuição em relação ao trabalho anterior está no sistema operacional, que passa a monitorar periodicamente os chaveamentos de tensões e frequências do processador baseado nas informações providas pelos \textit{PMHs}.} \\ \hline \hline

\multicolumn{1}{|r|}{\textbf{Método(s) utilizados}} & DVFS Intra-Task; Control Flow Graph (CFG); \\ \hline
\multicolumn{1}{|r|}{\textbf{Ferramenta(s)}} & {Os autores utilizaram a ferramenta Sim-Outorder Simulator from SimpleScalar [Disponível em: \href{http://www.simplescalar.com/}{http://www.simplescalar.com/}]. \newline

Os autores utilizaram os seguintes \textit{benchmarks}: \newline
- MPEG2 \textit{decoder} [Disponível em: \href{http://www.mpeg.org/}{http://www.mpeg.org/}] \newline
- \textit{Automated Target Recognition} (ATR)} \\ \hline \hline

\multicolumn{2}{|c|}{\textbf{Impacto (positivo x negativo)}} \\ \hline
\multicolumn{2}{|p{21cm}|}{O artigo apresentou como ponto positivo uma metodologia cuja finalidade é criar um canal de comunicação entre o compilador e o sistema operacional, permitindo que ambos possam trabalhar em conjunto para a diminuição do consumo de energia do processador. Além disso, a metodologia utilizada dá suporte a preempções. Como ponto negativo, a metodologia insere uma quantidade excessiva de \textit{overheads} dentro da aplicação.} \\ \hline \hline

\multicolumn{2}{|c|}{\textbf{Validação do método}} \\ \hline
\multicolumn{2}{|p{21cm}|}{A validação da metodologia ocorreu através do uso de ferramentas e \textit{benchmarks} relatados na literatura. Além disso, compara os resultados experimentais com outros trabalhos de referência na área de baixo consumo de energia.} \\ \hline \hline

\multicolumn{2}{|c|}{\textbf{Limitações do método}} \\ \hline
\multicolumn{2}{|p{21cm}|}{A metodologia insere uma quantidade excessiva de \textit{overheads} dentro da aplicação.} \\ \hline \hline

\multicolumn{2}{|c|}{\textbf{Integração de métodos}} \\ \hline
\multicolumn{2}{|p{21cm}|}{Não se aplica} \\ \hline \hline

\multicolumn{2}{|c|}{\textbf{Modo de aplicação do método}} \\ \hline
\multicolumn{2}{|p{21cm}|}{Híbrida} \\ \hline \hline

\multicolumn{2}{|c|}{\textbf{Perspectivas futuras}} \\ \hline
\multicolumn{2}{|p{21cm}|}{Não identificada} \\ \hline \hline

\multicolumn{2}{|c|}{\textbf{Comentários}} \\ \hline
\multicolumn{2}{|p{21cm}|}{A metodologia dá suporte a preempções, através da chamada de interrupções do sistema operacional para adaptar as tensões e frequências do processador. Quanto aos recursos compartilhados, a metodologia proposta não dá suporte.} \\ \hline
\end{tabular}
}
\end{table}

\newpage
\begin{table}[htp] 
\centering
\caption{Dados extraído da publicação de \cite{Seo_Kim_Chung_2004_a}.}
\resizebox{\columnwidth}{!}{%
\begin{tabular}{|p{3.8cm}|p{16cm}|}
\hline
\multicolumn{1}{|r|}{\textbf{ID}} & Seo\_Kim\_Chung\_2004\_a \\ \hline
\multicolumn{1}{|r|}{\textbf{Título}} & Profile-based optimal intra-task voltage scheduling for hard real-time applications \\ \hline
\multicolumn{1}{|r|}{\textbf{Autores}} & Seo J., Kim T., Chung K.-S. \\ \hline
\multicolumn{1}{|r|}{\textbf{Palavras-chave}} & DVS; Intra-task voltage scheduling; Low energy design \\ \hline
\multicolumn{1}{|r|}{\textbf{Fonte da publicação}} & DAC '04 Proceedings of the 41st annual Design Automation Conference \\ \hline
\multicolumn{1}{|r|}{\textbf{Editora}} & ACM and IEEE \\ \hline
\multicolumn{1}{|r|}{\textbf{Ano da publicação}} & 2004 \\ \hline \hline

\multicolumn{2}{|c|}{\textbf{Resumo da publicação}} \\ \hline
\multicolumn{2}{|p{21cm}|}{Os autores propõem uma metodologia baseada no perfil de execução da tarefa, onde os níveis de tensão são definidos para cada bloco de código. Esse método tem como objetivo gerenciar melhor os \textit{overheads} de transição, que são totalmente ou parcialmente ignorados nos outros trabalhos presentes na literatura, e obter melhores níveis de redução do consumo de energia do processador. Essa técnica é chamada de ``ROEP-\textit{based technique}'' (ROEP - \textit{Remaining Optimal-Case Execution Path}), que é uma melhoria da metodologia \textit{RAEP-based} proposta por \citep{Shin_Kim_2001_a}, cujo principal foco está relacionado com desperdícios de energia, com as trocas excessivas de tensão e frequência do processador e com a diminuição dos \textit{overheads} inseridos dentro das aplicações. Seguindo a escala cronológica dos trabalhos catalogados nesta revisão sistemática, esta foi uma das primeiras abordagens a otimizar estes parâmetros.} \\ \hline \hline

\multicolumn{1}{|r|}{\textbf{Método(s) utilizados}} & DVFS Intra-Task; Control Flow Graph (CFG); \\ \hline
\multicolumn{1}{|r|}{\textbf{Ferramenta(s)}} & {Os autores não especificaram nenhuma ferramenta para a realização dos experimentos, mas relataram o uso dos seguintes \textit{benchmarks}: amotsa, dawson, gcf, gser, gsimp, hypser, igray, realft, rtnewt, trsec, sncndn e trapzd (OBS.: estes são \textit{benchmarks} retirados do livro "\textit{Numerical Recipes in C}").} \\ \hline \hline

\multicolumn{2}{|c|}{\textbf{Impacto (positivo x negativo)}} \\ \hline
\multicolumn{2}{|p{21cm}|}{O artigo apresentou como ponto positivo uma metodologia para gerenciar melhor os \textit{overheads} de transição, que são totalmente ou parcialmente ignorados nos trabalhos presentes na literatura. Como ponto negativo, a metodologia não deixa claro se dar suporte há alguns tipos de estruturas como \textit{loops} ocultos e recursões.} \\ \hline \hline

\multicolumn{2}{|c|}{\textbf{Validação do método}} \\ \hline
\multicolumn{2}{|p{21cm}|}{A validação da metodologia ocorreu através do uso de \textit{benchmarks} relatados na literatura. Além disso, compara os resultados experimentais com outros trabalhos de referência na área de baixo consumo de energia.} \\ \hline \hline

\multicolumn{2}{|c|}{\textbf{Limitações do método}} \\ \hline
\multicolumn{2}{|p{21cm}|}{A metodologia não deixa claro se dar suporte há alguns tipos de estruturas como \textit{loops} ocultos e recursões.} \\ \hline \hline

\multicolumn{2}{|c|}{\textbf{Integração de métodos}} \\ \hline
\multicolumn{2}{|p{21cm}|}{Não se aplica} \\ \hline \hline

\multicolumn{2}{|c|}{\textbf{Modo de aplicação do método}} \\ \hline
\multicolumn{2}{|p{21cm}|}{Offline} \\ \hline \hline

\multicolumn{2}{|c|}{\textbf{Perspectivas futuras}} \\ \hline
\multicolumn{2}{|p{21cm}|}{Os autores pretendem verificar os impactos da metodologia proposta quando aplicado em um contexto de escalonamento inter-tarefas.} \\ \hline \hline

\multicolumn{2}{|c|}{\textbf{Comentários}} \\ \hline
\multicolumn{2}{|p{21cm}|}{A metodologia proposta não mostra evidências, no artigo, de que fornece suporte para recursos compartilhados e / ou preempções.} \\ \hline
\end{tabular}
}
\end{table}

\newpage
\begin{table}[htp] 
\centering
\caption{Dados extraído da publicação de \cite{Shin_Kim_2005_a}.}
\resizebox{\columnwidth}{!}{%
\begin{tabular}{|p{3.8cm}|p{16cm}|}
\hline
\multicolumn{1}{|r|}{\textbf{ID}} & Shin\_Kim\_2005\_a \\ \hline
\multicolumn{1}{|r|}{\textbf{Título}} & Intra-task voltage scheduling on DVS-enabled hard real-time systems \\ \hline
\multicolumn{1}{|r|}{\textbf{Autores}} & Shin D., Kim J. \\ \hline
\multicolumn{1}{|r|}{\textbf{Palavras-chave}} & Dynamic voltage scaling; Low-power design; Power management; Real-time systems; Variable-voltage processor \\ \hline
\multicolumn{1}{|r|}{\textbf{Fonte da publicação}} & IEEE Transactions on Computer-Aided Design of Integrated Circuits and Systems \\ \hline
\multicolumn{1}{|r|}{\textbf{Editora}} & IEEE \\ \hline
\multicolumn{1}{|r|}{\textbf{Ano da publicação}} & 2005 \\ \hline \hline

\multicolumn{2}{|c|}{\textbf{Resumo da publicação}} \\ \hline
\multicolumn{2}{|p{21cm}|}{Os autores melhoraram a eficiência do método \textit{RAEP-based} proposto por eles mesmo em \cite{Shin_Kim_2001_a}. Nesta nova abordagem, a principal diferença está nas otimizações de \textit{overheads} para a realização das transições de tensão, que antes era feita de forma \textit{offline} e agora o método de atribuição de tensões passou a ser \textit{online} e mais eficiente. Os autores utilizaram os mesmos casos de teste para realização dos experimentos e fizeram alterações na ferramenta AVS para adaptá-la a nova abordagem. Um fato interessante a ser relatado é que os autores começaram a introduzir o conceito de ciclos salvos ou ciclos economizados (em inglês, \textit{Saved Cycles ou $C_{saved}$}), ou seja, são trechos de código que deixaram de ser executados dentro da aplicação. Esse conceito será melhor amadurecida em \cite{Shin_Kim_2005_b}.} \\ \hline \hline

\multicolumn{1}{|r|}{\textbf{Método(s) utilizados}} & DVFS Intra-Task; Control Flow Graph (CFG); \\ \hline
\multicolumn{1}{|r|}{\textbf{Ferramenta(s)}} & {Os autores utilizaram a ferramenta AVS (\textit{Automatic Voltage Scaler} - Não disponível para download) e o MPEG-4 \textit{enconder / decoder} como \textit{benchmarks}.} \\ \hline \hline

\multicolumn{2}{|c|}{\textbf{Impacto (positivo x negativo)}} \\ \hline
\multicolumn{2}{|p{21cm}|}{O artigo apresentou como ponto positivo uma metodologia que analisa e cria perfis de consumo de energia baseados nos caminhos de execução da aplicação, procurando minimizar os \textit{overheads} introduzidos no código fonte da aplicação. Como ponto negativo, a metodologia não deixa claro se dar suporte a recursões e \textit{loops} ocultos.} \\ \hline \hline

\multicolumn{2}{|c|}{\textbf{Validação do método}} \\ \hline
\multicolumn{2}{|p{21cm}|}{A validação da metodologia ocorreu através do uso de ferramenta e \textit{benchmarks} relatados na literatura. Além disso, compara os resultados experimentais com outros trabalhos de referência na área de baixo consumo de energia.} \\ \hline \hline

\multicolumn{2}{|c|}{\textbf{Limitações do método}} \\ \hline
\multicolumn{2}{|p{21cm}|}{A metodologia não deixa claro se dar suporte a alguns tipos de estruturas como: \textit{loops} ocultos e recursões.} \\ \hline \hline

\multicolumn{2}{|c|}{\textbf{Integração de métodos}} \\ \hline
\multicolumn{2}{|p{21cm}|}{Não se aplica} \\ \hline \hline

\multicolumn{2}{|c|}{\textbf{Modo de aplicação do método}} \\ \hline
\multicolumn{2}{|p{21cm}|}{Híbrida} \\ \hline \hline

\multicolumn{2}{|c|}{\textbf{Perspectivas futuras}} \\ \hline
\multicolumn{2}{|p{21cm}|}{O autor pretende incluir informações em tempo de execução, como por exemplo \textit{cache hit} e \textit{miss}, para melhorar a eficiência energética do método proposto.} \\ \hline \hline

\multicolumn{2}{|c|}{\textbf{Comentários}} \\ \hline
\multicolumn{2}{|p{21cm}|}{A metodologia proposta não mostra evidências, no artigo, de que fornece suporte para recursos compartilhados e / ou preempções.} \\ \hline
\end{tabular}
}
\end{table}

\newpage
\begin{table}[htp] 
\centering
\caption{Dados extraído da publicação de \cite{Yi_Yang_Chen_2005_a}.}
\resizebox{\columnwidth}{!}{%
\begin{tabular}{|p{3.8cm}|p{16cm}|}
\hline
\multicolumn{1}{|r|}{\textbf{ID}} & Yi\_Yang\_Chen\_2005\_a \\ \hline
\multicolumn{1}{|r|}{\textbf{Título}} & The optimal profile-guided greedy dynamic voltage scaling in real-time applications \\ \hline
\multicolumn{1}{|r|}{\textbf{Autores}} & Yi H., Yang X., Chen J. \\ \hline
\multicolumn{1}{|r|}{\textbf{Palavras-chave}} & Computer simulation; Energy utilization; Frequencies; Mathematical models; Program compilers; Real time systems; Scheduling; Dock frequency; Dynamic voltage scaling (DVS); Optimal real-time voltage scheduling; Voltage adjustment; Voltage measurement \\ \hline
\multicolumn{1}{|r|}{\textbf{Fonte da publicação}} & ICESS'05 Proceedings of the Second international conference on Embedded Software and Systems \\ \hline
\multicolumn{1}{|r|}{\textbf{Editora}} & Springer \\ \hline
\multicolumn{1}{|r|}{\textbf{Ano da publicação}} & 2005 \\ \hline \hline

\multicolumn{2}{|c|}{\textbf{Resumo da publicação}} \\ \hline
\multicolumn{2}{|p{21cm}|}{Os autores propõem um modelo analítico de escalonamento dinâmico de tensão "ganancioso", cujo o objetivo é encontrar as tensões ideais para as aplicações de tempo real, através da análise dos casos de execução mais frequentes, ou também chamados de \textit{Hot Path}, referenciados em \cite{Shin_Kim_2001_a}. Essas análises visam identificar os tempos de folga distribuídos pela aplicação e, em seguida, repassa os ganhos obtidos para o processador, minimizando o consumo do energia. Esse método foi chamado de OPTDVS (em inglês, \textit{Optimal Dinamic Voltage Scheduling}). Em outras palavras, esse método é um mecanismo de ajuste de tensão ganancioso guiado por perfil (ou \textit{profile-guided}) que se baseia nos \textit{hot paths} para definir o melhor perfil de consumo de energia para uma dada aplicação de tempo real.} \\ \hline \hline

\multicolumn{1}{|r|}{\textbf{Método(s) utilizados}} & DVFS Intra-Task; Control Flow Graph (CFG); \\ \hline
\multicolumn{1}{|r|}{\textbf{Ferramenta(s)}} & {Os autores não mencionam o uso de ferramentas e / ou \textit{benchmarks}.} \\ \hline \hline

\multicolumn{2}{|c|}{\textbf{Impacto (positivo x negativo)}} \\ \hline
\multicolumn{2}{|p{21cm}|}{O artigo apresentou como ponto positivo uma nova metodologia que realiza a analise dos \textit{hot paths} da aplicação e criam perfis de consumo de energia para cada caminho identificado, visando maximizar os ganhos obtidos com o uso da técnica DVFS intra-tarefa, através da identificação de tempos de folga na dentro da aplicação. Como ponto negativo, a metodologia insere muitos pontos de controle dentro da aplicação e não mostra um estudo sobre os impactos ocasionados pelos \textit{overheads} inseridos.} \\ \hline \hline

\multicolumn{2}{|c|}{\textbf{Validação do método}} \\ \hline
\multicolumn{2}{|p{21cm}|}{A validação da metodologia ocorreu através de provas matemáticas e comparações de desempenhos com outros trabalhos presentes na literatura.} \\ \hline \hline

\multicolumn{2}{|c|}{\textbf{Limitações do método}} \\ \hline
\multicolumn{2}{|p{21cm}|}{A metodologia não mostra estudos sobre os impactos dos \textit{overheads} inseridos estaticamente dentro da aplicação.} \\ \hline \hline

\multicolumn{2}{|c|}{\textbf{Integração de métodos}} \\ \hline
\multicolumn{2}{|p{21cm}|}{Não se aplica} \\ \hline \hline

\multicolumn{2}{|c|}{\textbf{Modo de aplicação do método}} \\ \hline
\multicolumn{2}{|p{21cm}|}{Offline} \\ \hline \hline

\multicolumn{2}{|c|}{\textbf{Perspectivas futuras}} \\ \hline
\multicolumn{2}{|p{21cm}|}{Não identificada} \\ \hline \hline

\multicolumn{2}{|c|}{\textbf{Comentários}} \\ \hline
\multicolumn{2}{|p{21cm}|}{A metodologia proposta não mostra evidências, no artigo, de que fornece suporte para recursos compartilhados e / ou preempções.} \\ \hline
\end{tabular}
}
\end{table}

\newpage
\begin{table}[htp] 
\centering
\caption{Dados extraído da publicação de \cite{Mohan_Mueller_Root_2005_a}.}
\resizebox{\columnwidth}{!}{%
\begin{tabular}{|p{3.8cm}|p{16cm}|}
\hline
\multicolumn{1}{|r|}{\textbf{ID}} & Mohan\_Mueller\_Root\_2005\_a \\ \hline
\multicolumn{1}{|r|}{\textbf{Título}} & ParaScale: Exploiting parametric timing analysis for real-time schedulers and dynamic voltage scaling \\ \hline
\multicolumn{1}{|r|}{\textbf{Autores}} & Mohan S., Mueller F., Hawkins W., Root M., Healy C., Whalley D. \\ \hline
\multicolumn{1}{|r|}{\textbf{Palavras-chave}} & Dynamic voltage scaling; Hard real-time systems; Lower-power consumption; Parametric -analysis; Parametric approach; Real-time schedulers; Static timing analysis; Worst-case execution time; Real time systems; Scheduling; Voltage stabilizing circuits \\ \hline
\multicolumn{1}{|r|}{\textbf{Fonte da publicação}} & Proceedings - Real-Time Systems Symposium \\ \hline
\multicolumn{1}{|r|}{\textbf{Editora}} & IEEE \\ \hline
\multicolumn{1}{|r|}{\textbf{Ano da publicação}} & 2005 \\ \hline \hline

\multicolumn{2}{|c|}{\textbf{Resumo da publicação}} \\ \hline
\multicolumn{2}{|p{21cm}|}{Os autores propõem uma nova técnica chamada \textit{ParaScale}, que permite fazer análises de tempo paramétrico em conjunto com o escalonamento. Essas análises permitem detectar dinamicamente os limites dos \textit{loops} e o limite inferior do WCET (em inglês, \textit{Wrost Case Execution Time}), em tempo real, durante o tempo de execução restante da tarefa. Portanto, o ganho desta metodologia está, principalmente, sobre os tempos de folga obtidos sobre as estruturas de repetição. Dentre os trabalhos catalogados nessa revisão sistemática este foi o primeiro a trabalhar com limites paramétricos de \textit{loops}, permitindo ter um melhor controle dos tempos de folga dentro de estruturas de repetição.} \\ \hline \hline

\multicolumn{1}{|r|}{\textbf{Método(s) utilizados}} & DVFS Intra-Task; \\ \hline
\multicolumn{1}{|r|}{\textbf{Ferramenta(s)}} & {Os autores utilizaram a ferramenta SimpleScalar [Disponível em: \href{http://www.simplescalar.com/}{http://www.simplescalar.com/}]. \newline

Os autores utilizaram os \textit{benchmarks}: \newline
- Adpcm (\textit{Adaptive Differential Pulse Code Modulation}) \newline
- Cnt (\textit{Sum and count of positive and negative numbers in an array}) \newline
- Lms (\textit{An LMS adaptive signal enhancement}) \newline
- Mm (\textit{Matrix Multiplication})} \\ \hline \hline

\multicolumn{2}{|c|}{\textbf{Impacto (positivo x negativo)}} \\ \hline
\multicolumn{2}{|p{21cm}|}{O artigo apresentou como ponto positivo uma nova metodologia para realizar a detecção dinâmica dos limites inferiores dos \textit{loops} e do WCET em tempo de execução. Como ponto negativo, a metodologia não mostra um estudo detalhado sobre os \textit{overheads} inseridos estaticamente nas tarefas.} \\ \hline \hline

\multicolumn{2}{|c|}{\textbf{Validação do método}} \\ \hline
\multicolumn{2}{|p{21cm}|}{A validação da metodologia ocorreu através do uso de ferramentas e \textit{benchmarks} relatados na literatura. Além disso, compara os resultados experimentais com outros trabalhos de referência na área de baixo consumo de energia.} \\ \hline \hline

\multicolumn{2}{|c|}{\textbf{Limitações do método}} \\ \hline
\multicolumn{2}{|p{21cm}|}{A metodologia proposta não mostra um estudo detalhado sobre os \textit{overheads} inseridos estaticamente nas tarefas.} \\ \hline \hline

\multicolumn{2}{|c|}{\textbf{Integração de métodos}} \\ \hline
\multicolumn{2}{|p{21cm}|}{Não se aplica} \\ \hline \hline

\multicolumn{2}{|c|}{\textbf{Modo de aplicação do método}} \\ \hline
\multicolumn{2}{|p{21cm}|}{Híbrida} \\ \hline \hline

\multicolumn{2}{|c|}{\textbf{Perspectivas futuras}} \\ \hline
\multicolumn{2}{|p{21cm}|}{Integrar ao método um mecanismo de detecção automática dos limites superiores dos loops, pois esses limites são definidos estaticamente no código fonte da aplicação.} \\ \hline \hline

\multicolumn{2}{|c|}{\textbf{Comentários}} \\ \hline
\multicolumn{2}{|p{21cm}|}{A metodologia proposta não mostra evidências, no artigo, de que fornece suporte para recursos compartilhados e / ou preempções.} \\ \hline
\end{tabular}
}
\end{table}

\newpage
\begin{table}[htp] 
\centering
\caption{Dados extraído da publicação de \cite{Yi_Yang_2005_a}.}
\resizebox{\columnwidth}{!}{%
\begin{tabular}{|p{3.8cm}|p{16cm}|}
\hline
\multicolumn{1}{|r|}{\textbf{ID}} & Yi\_Yang\_2005\_a \\ \hline
\multicolumn{1}{|r|}{\textbf{Título}} & Optimizing the configuration of dynamic voltage scaling points in real-time applications \\ \hline
\multicolumn{1}{|r|}{\textbf{Autores}} & Yi H., Yang X. \\ \hline
\multicolumn{1}{|r|}{\textbf{Palavras-chave}} & Computer simulation; Electric potential; Energy conservation; Energy efficiency; Energy utilization; Real time systems; Dynamic voltage scaling (DVS); Energy models; Real-time applications; Supply voltage; Program compilers \\ \hline
\multicolumn{1}{|r|}{\textbf{Fonte da publicação}} & 15th International Workshop, PATMOS 2005 \\ \hline
\multicolumn{1}{|r|}{\textbf{Editora}} & Springer \\ \hline
\multicolumn{1}{|r|}{\textbf{Ano da publicação}} & 2005 \\ \hline \hline

\multicolumn{2}{|c|}{\textbf{Resumo da publicação}} \\ \hline
\multicolumn{2}{|p{21cm}|}{Os autores propõem uma metodologia de configuração baseado em um método que constrói o padrão de execução de uma determinada aplicação, também chamado de \textit{Profile-Based Method} já relatado na publicação de \cite{Yi_Yang_Chen_2005_a}. O diferencial desta nova abordagem está em diminuir os \textit{overheads} inseridos pelo compilador no código fonte das aplicações. Esse processo é feito da seguinte forma: primeiro o compilador insere os pontos de escalonamento sem levar em consideração os \textit{overheads}; em seguida, todos os pontos de escalonamento são listados, já levando em consideração os \textit{overheads}; e por fim, os pontos que possuem maiores \textit{overheads} e / ou não trazem redução do consumo de energia para aplicação são excluídos.} \\ \hline \hline

\multicolumn{1}{|r|}{\textbf{Método(s) utilizados}} & DVFS Intra-Task; \\ \hline
\multicolumn{1}{|r|}{\textbf{Ferramenta(s)}} & {Os autores não especificaram o uso de ferramentas e / ou \textit{benchmarks}.} \\ \hline \hline

\multicolumn{2}{|c|}{\textbf{Impacto (positivo x negativo)}} \\ \hline
\multicolumn{2}{|p{21cm}|}{O artigo apresentou como ponto positivo uma nova metodologia para realizar a análise e aplicação de pontos de escalonamento de tensão (VSPs) em aplicações de tempo real. Como ponto negativo, a metodologia não deixa claro se dá suporte a alguns tipos de estruturas como \textit{loops} ocultos e recursões.} \\ \hline \hline

\multicolumn{2}{|c|}{\textbf{Validação do método}} \\ \hline
\multicolumn{2}{|p{21cm}|}{A validação da metodologia ocorreu através de casos de teste criados pelos próprios autores.} \\ \hline \hline

\multicolumn{2}{|c|}{\textbf{Limitações do método}} \\ \hline
\multicolumn{2}{|p{21cm}|}{A metodologia proposta não deixa claro quais são os tipos de estruturas suportadas.} \\ \hline \hline

\multicolumn{2}{|c|}{\textbf{Integração de métodos}} \\ \hline
\multicolumn{2}{|p{21cm}|}{Não se aplica} \\ \hline \hline

\multicolumn{2}{|c|}{\textbf{Modo de aplicação do método}} \\ \hline
\multicolumn{2}{|p{21cm}|}{Offline} \\ \hline \hline

\multicolumn{2}{|c|}{\textbf{Perspectivas futuras}} \\ \hline
\multicolumn{2}{|p{21cm}|}{Fornecer suporte a \textit{loops} ocultos e recursões.} \\ \hline \hline

\multicolumn{2}{|c|}{\textbf{Comentários}} \\ \hline
\multicolumn{2}{|p{21cm}|}{A metodologia proposta não mostra evidências, no artigo, de que fornece suporte para recursos compartilhados e / ou preempções.} \\ \hline
\end{tabular}
}
\end{table}

\newpage
\begin{table}[htp] 
\centering
\caption{Dados extraído da publicação de \cite{Shin_Kim_2005_b}.}
\resizebox{\columnwidth}{!}{%
\begin{tabular}{|p{3.8cm}|p{16cm}|}
\hline
\multicolumn{1}{|r|}{\textbf{ID}} & Shin\_Kim\_2005\_b \\ \hline
\multicolumn{1}{|r|}{\textbf{Título}} & Optimizing intra-task voltage scheduling using data flow analysis \\ \hline
\multicolumn{1}{|r|}{\textbf{Autores}} & Shin D., Kim J. \\ \hline
\multicolumn{1}{|r|}{\textbf{Palavras-chave}} & Control flows; Data flow; Data-flow information; Low power application; Optimization techniques; Slack time; Supply voltages; Voltage scheduling; Voltage-scaling; Algorithms; Computer aided design; Data flow analysis; Data transfer; Energy efficiency; Energy utilization; Motion Picture Experts Group standards; Scheduling; Optimization \\ \hline
\multicolumn{1}{|r|}{\textbf{Fonte da publicação}} & Proceedings of the Asia and South Pacific Design Automation Conference, ASP-DAC \\ \hline
\multicolumn{1}{|r|}{\textbf{Editora}} & IEEE \\ \hline
\multicolumn{1}{|r|}{\textbf{Ano da publicação}} & 2005 \\ \hline \hline

\multicolumn{2}{|c|}{\textbf{Resumo da publicação}} \\ \hline
\multicolumn{2}{|p{21cm}|}{Os autores propõem uma otimização na técnica intraDVS usando informações de fluxo de dados da aplicação de tempo real. A metodologia visa melhorar a eficiência energética antecipando os pontos de escalonamento de tensão (em inglês, \textit{Voltage Scaling Points} - VSPs), baseadas nos resultados de análises do fluxo de dados da aplicação. Essa técnica foi chamada de \textit{LaIntraDVS} (em inglês, Look Ahead IntraDVS). Em outras palavras, o método proposto antecipa os pontos de controle para maximizar os ganhos de energia da técnica intraDVS, como por exemplo: analisar uma estrutura de repetição e predizer quantas interações serão necessárias e aplicar as tensões e frequências ideais para essa bloco de código antes que ele seja realmente executado.} \\ \hline \hline

\multicolumn{1}{|r|}{\textbf{Método(s) utilizados}} & DVFS Intra-Task; \\ \hline
\multicolumn{1}{|r|}{\textbf{Ferramenta(s)}} & {Os autores não especificam a ferramenta utilizada nos experimentos. \newline Os autores utilizaram os seguintes \textit{benchmarks}: \newline - MPEG-4 \textit{video encoder} \newline - MPEG-4 \textit{video decoder}} \\ \hline \hline

\multicolumn{2}{|c|}{\textbf{Impacto (positivo x negativo)}} \\ \hline
\multicolumn{2}{|p{21cm}|}{O artigo apresentou como ponto positivo uma nova metodologia para antecipar (ou predizer) os pontos de escalonamento de tensão (VSPs) visando maximizar os ganhos da técnica intraDVS. Como ponto negativo, a metodologia fornece ganhos mais efetivos apenas quando as distâncias entre os pontos de escalonamento de tensão são grandes.} \\ \hline \hline

\multicolumn{2}{|c|}{\textbf{Validação do método}} \\ \hline
\multicolumn{2}{|p{21cm}|}{A validação da metodologia ocorreu através do uso de ferramentas e \textit{benchmarks} relatados na literatura. Além disso, compara os resultados experimentais com outros trabalhos de referência na área de baixo consumo de energia.} \\ \hline \hline

\multicolumn{2}{|c|}{\textbf{Limitações do método}} \\ \hline
\multicolumn{2}{|p{21cm}|}{A metodologia fornece ganhos mais efetivos apenas quando as distâncias entre os pontos de escalonamento de tensão são grandes.} \\ \hline \hline

\multicolumn{2}{|c|}{\textbf{Integração de métodos}} \\ \hline
\multicolumn{2}{|p{21cm}|}{Não se aplica} \\ \hline \hline

\multicolumn{2}{|c|}{\textbf{Modo de aplicação do método}} \\ \hline
\multicolumn{2}{|p{21cm}|}{Offline} \\ \hline \hline

\multicolumn{2}{|c|}{\textbf{Perspectivas futuras}} \\ \hline
\multicolumn{2}{|p{21cm}|}{Fornecer suporte a outros tipos de estruturas, como: \textit{loops} ocultos e recursões.} \\ \hline \hline

\multicolumn{2}{|c|}{\textbf{Comentários}} \\ \hline
\multicolumn{2}{|p{21cm}|}{- Este é um dos primeiros trabalhos a antecipar (ou predizer) pontos de escalonamento de tensão (VSPs). \newline - A metodologia proposta não mostra evidências, no artigo, de que fornece suporte para recursos compartilhados e / ou preempções.} \\ \hline
\end{tabular}
}
\end{table}

\newpage
\begin{table}[htp] 
\centering
\caption{Dados extraído da publicação de \cite{Seo_Kim_Dutt_2005_a}.}
\resizebox{\columnwidth}{!}{%
\begin{tabular}{|p{3.8cm}|p{16cm}|}
\hline
\multicolumn{1}{|r|}{\textbf{ID}} & Seo\_Kim\_Dutt\_2005\_a \\ \hline
\multicolumn{1}{|r|}{\textbf{Título}} & Optimal integration of inter-task and intra-task dynamic voltage scaling techniques for hard real-time applications \\ \hline
\multicolumn{1}{|r|}{\textbf{Autores}} & Seo J., Kim T., Dutt N.D. \\ \hline
\multicolumn{1}{|r|}{\textbf{Palavras-chave}} & Dynamic voltage scaling (DVS); Energy minimization; Polynomial time; Computation theory; Energy efficiency; Energy utilization; Optimal systems; Optimization; Polynomial approximation; Problem solving; Real time systems; Voltage control \\ \hline
\multicolumn{1}{|r|}{\textbf{Fonte da publicação}} & IEEE/ACM International Conference on Computer-Aided Design, Digest of Technical Papers, ICCAD \\ \hline
\multicolumn{1}{|r|}{\textbf{Editora}} & IEEE \\ \hline
\multicolumn{1}{|r|}{\textbf{Ano da publicação}} & 2005 \\ \hline \hline

\multicolumn{2}{|c|}{\textbf{Resumo da publicação}} \\ \hline
\multicolumn{2}{|p{21cm}|}{Os autores propõe uma nova técnica DVS que combinam as técnicas DVS intra-tarefa e inter-tarefa, chamada de \textit{DVS-intgr}. Essa metodologia examina os limites inferiores de consumo de energia baseado na técnica DVS intra-tarefa (parte dessa metodologia foi inspirada no trabalho de \cite{Shin_Kim_Lee_2001_a}) e com essas propriedades foram definidos os tempos de execução ideais de cada tarefa. Em seguida, as tarefas são divididas em vários grupos de trabalho de tal forma que cada tarefa possa ser executada dentro do limite preestabelecido para cada grupo, através da utilização da técnica DVS inter-tarefa melhorada para produzir o melhor escalonamento entre elas de forma que haja a redução no consumo de energia e garantindo que nenhum premissa temporal seja violada.} \\ \hline \hline

\multicolumn{1}{|r|}{\textbf{Método(s) utilizados}} & DVFS Intra-Task; DVFS Inter-Task; \\ \hline
\multicolumn{1}{|r|}{\textbf{Ferramenta(s)}} & {os autores utilizaram a ferramenta TGFF v3.0.} \\ \hline \hline

\multicolumn{2}{|c|}{\textbf{Impacto (positivo x negativo)}} \\ \hline
\multicolumn{2}{|p{21cm}|}{O artigo apresentou como ponto positivo uma nova metodologia para realizar a integração das técnicas DVS intra-tarefa com a inter-tarefa e introduziu o conceito de grupo de tarefas, com características semelhantes, a fim de facilitar o gerenciamento e ao mesmo tempo maximizando os resultados obtidos com o uso da técnica DVFS. Como ponto negativo, a metodologia não apresenta um estudo detalhado sobre os \textit{overheads} inseridos na aplicação.} \\ \hline \hline

\multicolumn{2}{|c|}{\textbf{Validação do método}} \\ \hline
\multicolumn{2}{|p{21cm}|}{A validação da metodologia ocorreu através do uso de ferramenta relatada na literatura. Além disso, compara os resultados experimentais com outros trabalhos de referência na área de baixo consumo de energia.} \\ \hline \hline

\multicolumn{2}{|c|}{\textbf{Limitações do método}} \\ \hline
\multicolumn{2}{|p{21cm}|}{A metodologia não mostra os impactos dos \textit{overheads} inseridos na aplicação (tanto intra-tarefa quanto inter-tarefa).} \\ \hline \hline

\multicolumn{2}{|c|}{\textbf{Integração de métodos}} \\ \hline
\multicolumn{2}{|p{21cm}|}{Integrou as técnica DVFS intra-tarefa com a técnica DVFS inter-tarefa} \\ \hline \hline

\multicolumn{2}{|c|}{\textbf{Modo de aplicação do método}} \\ \hline
\multicolumn{2}{|p{21cm}|}{Híbrida} \\ \hline \hline

\multicolumn{2}{|c|}{\textbf{Perspectivas futuras}} \\ \hline
\multicolumn{2}{|p{21cm}|}{Mostrar um estudo mais detalhado sobre os \textit{overheads} introduzidos pela metodologia.} \\ \hline \hline

\multicolumn{2}{|c|}{\textbf{Comentários}} \\ \hline
\multicolumn{2}{|p{21cm}|}{A metodologia proposta não mostra evidências, no artigo, de que fornece suporte para recursos compartilhados e / ou preempções.} \\ \hline
\end{tabular}
}
\end{table}

\newpage
\begin{table}[htp] 
\centering
\caption{Dados extraído da publicação de \cite{Cassandras_Zhuang_2005_a}.}
\resizebox{\columnwidth}{!}{%
\begin{tabular}{|p{3.8cm}|p{16cm}|}
\hline
\multicolumn{1}{|r|}{\textbf{ID}} & Cassandras\_Zhuang\_2005\_a \\ \hline
\multicolumn{1}{|r|}{\textbf{Título}} & Optimal dynamic voltage scaling for wireless sensor nodes with real-time constraints \\ \hline
\multicolumn{1}{|r|}{\textbf{Autores}} & Cassandras C.G., Zhuang S. \\ \hline
\multicolumn{1}{|r|}{\textbf{Palavras-chave}} & Product monitoring; Scalability; Voltage scaling; Computer simulation; Control equipment; Inventory control; Nonlinear programming; Real time systems; Sensors; Voltage control \\ \hline
\multicolumn{1}{|r|}{\textbf{Fonte da publicação}} & Proc. SPIE 5999, Intelligent Systems in Design and Manufacturing VI \\ \hline
\multicolumn{1}{|r|}{\textbf{Editora}} & SPIE - Digital Library \\ \hline
\multicolumn{1}{|r|}{\textbf{Ano da publicação}} & 2005 \\ \hline \hline

\multicolumn{2}{|c|}{\textbf{Resumo da publicação}} \\ \hline
\multicolumn{2}{|p{21cm}|}{Os autores propõe um controle intra-tarefa para minimizar o consumo de energia dentro do contexto de rede de sensores sem fio, processando tarefas de tempo real críticas. As variáveis de controle são basicamente os tempos de processamento das tarefas de tempo real, onde cada um desses tempos estão associados a diferentes níveis de tensão. O controle intra-tarefa é baseado na exploração das propriedades dos caminhos de execução ideais. Além disso os autores mostram em seus experimentos que soluções intra-tarefa minimizam mais energia que as soluções inter-tarefa.} \\ \hline \hline

\multicolumn{1}{|r|}{\textbf{Método(s) utilizados}} & DVFS Intra-Task; \\ \hline
\multicolumn{1}{|r|}{\textbf{Ferramenta(s)}} & {Os autores não mencionam o uso de nenhuma ferramenta e nenhum \textit{benchmark}.} \\ \hline \hline

\multicolumn{2}{|c|}{\textbf{Impacto (positivo x negativo)}} \\ \hline
\multicolumn{2}{|p{21cm}|}{O artigo apresentou como ponto positivo uma nova metodologia baseada na técnica DVFS intra-tarefa aplicada no contexto de redes de sensores sem fio, processando tarefas com restrições temporais. Como ponto negativo, a metodologia não mostra estudos sobre os impactos dos \textit{overheads} inseridos nas tarefas.} \\ \hline \hline

\multicolumn{2}{|c|}{\textbf{Validação do método}} \\ \hline
\multicolumn{2}{|p{21cm}|}{A validação da metodologia ocorreu através de casos de teste criados pelos próprios autores. Além disso, compara os resultados experimentais com outros trabalhos de referência na área de baixo consumo de energia.} \\ \hline \hline

\multicolumn{2}{|c|}{\textbf{Limitações do método}} \\ \hline
\multicolumn{2}{|p{21cm}|}{A metodologia não mostra um estudo sobre os \textit{overheads} inseridos estaticamente dentro das tarefas.} \\ \hline \hline

\multicolumn{2}{|c|}{\textbf{Integração de métodos}} \\ \hline
\multicolumn{2}{|p{21cm}|}{Não se aplica} \\ \hline \hline

\multicolumn{2}{|c|}{\textbf{Modo de aplicação do método}} \\ \hline
\multicolumn{2}{|p{21cm}|}{Híbrida} \\ \hline \hline

\multicolumn{2}{|c|}{\textbf{Perspectivas futuras}} \\ \hline
\multicolumn{2}{|p{21cm}|}{Mostrar um estudo mais detalhado sobre os overheads introduzidos pelo metodologia.} \\ \hline \hline

\multicolumn{2}{|c|}{\textbf{Comentários}} \\ \hline
\multicolumn{2}{|p{21cm}|}{A metodologia proposta não mostra evidências, no artigo, de que fornece suporte para recursos compartilhados e / ou preempções.} \\ \hline
\end{tabular}
}
\end{table}

\newpage
\begin{table}[htp] 
\centering
\caption{Dados extraído da publicação de \cite{Kumar_Manimaran_2005_a}.}
\resizebox{\columnwidth}{!}{%
\begin{tabular}{|p{3.8cm}|p{16cm}|}
\hline
\multicolumn{1}{|r|}{\textbf{ID}} & Kumar\_Manimaran\_2005\_a \\ \hline
\multicolumn{1}{|r|}{\textbf{Título}} & An intra-task DVS algorithm exploiting program path locality for real-time embedded systems \\ \hline
\multicolumn{1}{|r|}{\textbf{Autores}} & Kumar G.S.A., Manimaran G. \\ \hline
\multicolumn{1}{|r|}{\textbf{Palavras-chave}} & Computer simulation; Embedded systems; Energy utilization; Real time systems; Scheduling; Dynamic Voltage Scheduling (DVS); Frequently executed paths; Real-time embedded systems; Algorithms \\ \hline
\multicolumn{1}{|r|}{\textbf{Fonte da publicação}} & 12th International Conference on High Performance Computing, HiPC 2005; \\ \hline
\multicolumn{1}{|r|}{\textbf{Editora}} & Springer \\ \hline
\multicolumn{1}{|r|}{\textbf{Ano da publicação}} & 2005 \\ \hline \hline

\multicolumn{2}{|c|}{\textbf{Resumo da publicação}} \\ \hline
\multicolumn{2}{|p{21cm}|}{Os autores propõem um novo algoritmo DVS intra-tarefa de consumo de energia consciente cujo o objetivo central é explorar os caminhos mais comuns e frequentemente executados dentro de uma aplicação de tempo real. Esse algoritmo foi chamado de CHP (em inglês, \textit{Common Hot Path}). Essa metodologia considera todos os caminhos mais executados (ou \textit{hot-paths}), principio também utilizado nos trabalhos de \cite{Shin_Kim_2001_a}, \cite{Seo_Kim_Chung_2004_a}, \cite{Yi_Yang_Chen_2005_a} e \cite{Shin_Kim_2005_a}, e para cada um deles são atribuídas probabilidades que irão indicar os caminhos mais utilizados. Dessa forma, a metodologia consegue combinar todos os \textit{hot paths} em um único caminho base que é comum em comprimento com a maioria dos \textit{hot paths}, assim é possível descobrir qual o caminho que leva a melhores taxas de minimização do consumo de energia, pois nem sempre o caminho mais curto é o mais eficaz para minimização do consumo de energia.} \\ \hline \hline

\multicolumn{1}{|r|}{\textbf{Método(s) utilizados}} & DVFS Intra-Task; Control Flow Graph (CFG); \\ \hline
\multicolumn{1}{|r|}{\textbf{Ferramenta(s)}} & {Os autores não mencionam o uso de nenhuma ferramenta. \newline

Os autores utilizam o seguinte \textit{benchmark}: \newline
- MPEG-4 \textit{video decoder \& encoder}} \\ \hline \hline

\multicolumn{2}{|c|}{\textbf{Impacto (positivo x negativo)}} \\ \hline
\multicolumn{2}{|p{21cm}|}{O artigo apresentou como ponto positivo uma nova metodologia que obtém os caminhos de execução com menores consumo de energia, através do uso de probabilidade e caminhos mais executados (ou \textit{hot paths}). Como ponto negativo, a metodologia tem dificuldades para lhe dar com alguns tipos de estruturas, como \textit{loops} sem profundidade.} \\ \hline \hline

\multicolumn{2}{|c|}{\textbf{Validação do método}} \\ \hline
\multicolumn{2}{|p{21cm}|}{A validação da metodologia ocorreu através do uso de \textit{benchmarks} relatados na literatura. Além disso, compara os resultados experimentais com outros trabalhos de referência na área de baixo consumo de energia.} \\ \hline \hline

\multicolumn{2}{|c|}{\textbf{Limitações do método}} \\ \hline
\multicolumn{2}{|p{21cm}|}{A metodologia tem dificuldades para lhe dar com alguns tipos de estruturas, como por exemplo: \textit{loops} sem profundidade.} \\ \hline \hline

\multicolumn{2}{|c|}{\textbf{Integração de métodos}} \\ \hline
\multicolumn{2}{|p{21cm}|}{Não se aplica} \\ \hline \hline

\multicolumn{2}{|c|}{\textbf{Modo de aplicação do método}} \\ \hline
\multicolumn{2}{|p{21cm}|}{Offline} \\ \hline \hline

\multicolumn{2}{|c|}{\textbf{Perspectivas futuras}} \\ \hline
\multicolumn{2}{|p{21cm}|}{Os autores pretendem mostrar que utilizando informações \textit{offline}, como por exemplo a quantidade de interações de um \textit{loop}, é possível minimizar o consumo de energia.} \\ \hline \hline

\multicolumn{2}{|c|}{\textbf{Comentários}} \\ \hline
\multicolumn{2}{|p{21cm}|}{A metodologia proposta não mostra evidências, no artigo, de que fornece suporte para recursos compartilhados e / ou preempções.} \\ \hline
\end{tabular}
}
\end{table}

\newpage
\begin{table}[htp] 
\centering
\caption{Dados extraído da publicação de \cite{Gheorghita_2005_a}.}
\resizebox{\columnwidth}{!}{%
\begin{tabular}{|p{3.8cm}|p{16cm}|}
\hline
\multicolumn{1}{|r|}{\textbf{ID}} & Gheorghita\_2005\_a \\ \hline
\multicolumn{1}{|r|}{\textbf{Título}} & Intra-task scenario-aware voltage scheduling \\ \hline
\multicolumn{1}{|r|}{\textbf{Autores}} & Gheorghita S.V., Basten T., Corporaal H. \\ \hline
\multicolumn{1}{|r|}{\textbf{Palavras-chave}} & Dynamic Voltage Scheduling; Real-Time; Scenarios; WCET \\ \hline
\multicolumn{1}{|r|}{\textbf{Fonte da publicação}} & CASES 2005: International Conference on Compilers, Architecture, and Synthesis for Embedded Systems \\ \hline
\multicolumn{1}{|r|}{\textbf{Editora}} & ACM \\ \hline
\multicolumn{1}{|r|}{\textbf{Ano da publicação}} & 2005 \\ \hline \hline

\multicolumn{2}{|c|}{\textbf{Resumo da publicação}} \\ \hline
\multicolumn{2}{|p{21cm}|}{Os autores propõem uma abordagem proativa que visa melhorar a performance do algoritmo de escalonamento intra-tarefa, explorando os tempos de folga que aparecem em tempo de execução, em seguida repassa para o processador através trechos de código inseridos na aplicação original, chamados de pontos de escalonamento de tensão ou VSPs (em inglês, \textit{Voltage Scaling Points}). Essa abordagem consiste, basicamente, em quatro etapas: (1) identificar os parâmetros que poderiam ter um impacto sobre o tempo de execução da aplicação; (2) calcular o máximo de impacto destes parâmetros sobre o WCEC da aplicação; (3) particionar o aplicativo em possíveis cenários, considerando-se esses parâmetros, juntamente com o seu impacto, e selecionando apenas cenários que, isoladamente, reduzir o consumo de energia; por fim, (4) computar o escalonamento DVS para cada cenário selecionado no estágio 3 e combiná-los com o escalonamento global da aplicação de tempo real.} \\ \hline \hline

\multicolumn{1}{|r|}{\textbf{Método(s) utilizados}} & DVFS Intra-Task; Control Flow Graph (CFG); \\ \hline
\multicolumn{1}{|r|}{\textbf{Ferramenta(s)}} & {Os autores não mencionaram o uso de nenhuma ferramenta. \newline

Os autores utilizaram os seguintes \textit{benchmarks}: \newline
- MP3 \textit{decoder} \newline
- H.263 \textit{decoder}} \\ \hline \hline

\multicolumn{2}{|c|}{\textbf{Impacto (positivo x negativo)}} \\ \hline
\multicolumn{2}{|p{21cm}|}{O artigo apresentou como ponto positivo uma nova metodologia de análise de CFGs para explorar os tempos de folga que aparecem durante a execução da tarefa. Como ponto negativo, a metodologia tem dificuldades para lhe dar com alguns tipos de estruturas (exemplo: \textit{loops} sem profundidade, \textit{loops} ocultos e recursões).} \\ \hline \hline

\multicolumn{2}{|c|}{\textbf{Validação do método}} \\ \hline
\multicolumn{2}{|p{21cm}|}{A validação da metodologia ocorreu através do uso de \textit{benchmarks} relatados na literatura.} \\ \hline \hline

\multicolumn{2}{|c|}{\textbf{Limitações do método}} \\ \hline
\multicolumn{2}{|p{21cm}|}{A metodologia não prever alguns tipos de estruturas de dados, como por exemplo:  \textit{loops} sem profundidade, \textit{loops} ocultos e recursões.} \\ \hline \hline

\multicolumn{2}{|c|}{\textbf{Integração de métodos}} \\ \hline
\multicolumn{2}{|p{21cm}|}{Não se aplica} \\ \hline \hline

\multicolumn{2}{|c|}{\textbf{Modo de aplicação do método}} \\ \hline
\multicolumn{2}{|p{21cm}|}{Híbrido} \\ \hline \hline

\multicolumn{2}{|c|}{\textbf{Perspectivas futuras}} \\ \hline
\multicolumn{2}{|p{21cm}|}{Os autores pretendem introduzir uma abordagem probabilística para tentar prever quais blocos da aplicação serão executados e introduzir a técnica DVFS inter-tarefas para maximizar os ganhos da metodologia proposta neste trabalho.} \\ \hline \hline

\multicolumn{2}{|c|}{\textbf{Comentários}} \\ \hline
\multicolumn{2}{|p{21cm}|}{- Os autores inserem o conceito de abordagem proativa, ou seja, ela consegue prever para um determinado bloco de código quais serão as tensões e frequências ideais a serem utilizadas, por exemplo: se um \textit{loop} tem no máximo 50 interações e só executará 20, então será possível ajustar as tensões e frequências do processador antes de entrar no \textit{loop}, para atender apenas as interações que serão executadas. \newline - A metodologia proposta não mostra evidências, no artigo, de que fornece suporte para recursos compartilhados e / ou preempções.} \\ \hline
\end{tabular}
}
\end{table}

\newpage
\begin{table}[htp] 
\centering
\caption{Dados extraído da publicação de \cite{Yi_Yang_2006_a}.}
\resizebox{\columnwidth}{!}{%
\begin{tabular}{|p{3.8cm}|p{16cm}|}
\hline
\multicolumn{1}{|r|}{\textbf{ID}} & Yi\_Yang\_2006\_a \\ \hline
\multicolumn{1}{|r|}{\textbf{Título}} & Toward the optimal configuration of dynamic voltage scaling points in real-time applications \\ \hline
\multicolumn{1}{|r|}{\textbf{Autores}} & Yi H.-Z., Yang X.-J. \\ \hline
\multicolumn{1}{|r|}{\textbf{Palavras-chave}} & Energy-aware systems; Languages and compilers; Low-power design \\ \hline
\multicolumn{1}{|r|}{\textbf{Fonte da publicação}} & Journal of Computer Science and Technology \\ \hline
\multicolumn{1}{|r|}{\textbf{Editora}} & Springer \\ \hline
\multicolumn{1}{|r|}{\textbf{Ano da publicação}} & 2006 \\ \hline \hline

\multicolumn{2}{|c|}{\textbf{Resumo da publicação}} \\ \hline
\multicolumn{2}{|p{21cm}|}{Os autores apresentam uma metodologia de configuração ótima de pontos de escalonamento de tensão dinâmicos sem \textit{overheads} de escalonamento de tensão, onde tomaram como base os trabalhos de \cite{Aboughazaleh_2003_b} e \cite{Aboughazaleh_2003_a}. Com essa metodologia os autores conseguiram introduzir a menor quantidade necessária de pontos de escalonamento de tensão para melhor aproveitar os tempos de folga da aplicação e, além disso, provaram teoricamente todos os modelos e teoremas matemáticos definidos na metodologia, sempre visando a otimização ideal de energia.} \\ \hline \hline

\multicolumn{1}{|r|}{\textbf{Método(s) utilizados}} & DVFS Intra-Task; \\ \hline
\multicolumn{1}{|r|}{\textbf{Ferramenta(s)}} & {Os autores não especificaram a ferramenta utilizada e nem faz referência ao uso de algum \textit{benchmark}.} \\ \hline \hline

\multicolumn{2}{|c|}{\textbf{Impacto (positivo x negativo)}} \\ \hline
\multicolumn{2}{|p{21cm}|}{O artigo apresentou como ponto positivo uma nova heurística que visa inserir a menor quantidade de pontos de escalonamento de tensão (VSPs), garantindo que nenhum restrição de tempo real sejam violadas. Como ponto negativo, a metodologia não deixa claro os tipos de estruturas suportadas.} \\ \hline \hline

\multicolumn{2}{|c|}{\textbf{Validação do método}} \\ \hline
\multicolumn{2}{|p{21cm}|}{A validação da metodologia ocorreu através de provas matemáticas e pelo uso de casos de testes definidos pelos próprios autores.} \\ \hline \hline

\multicolumn{2}{|c|}{\textbf{Limitações do método}} \\ \hline
\multicolumn{2}{|p{21cm}|}{A metodologia não prever todos os tipos de estruturas de dados, como por exemplo: \textit{loops} ocultos e recursões.} \\ \hline \hline

\multicolumn{2}{|c|}{\textbf{Integração de métodos}} \\ \hline
\multicolumn{2}{|p{21cm}|}{Não se aplica} \\ \hline \hline

\multicolumn{2}{|c|}{\textbf{Modo de aplicação do método}} \\ \hline
\multicolumn{2}{|p{21cm}|}{Offline} \\ \hline \hline

\multicolumn{2}{|c|}{\textbf{Perspectivas futuras}} \\ \hline
\multicolumn{2}{|p{21cm}|}{Os autores propõem realizar estudos mais detalhados quando os \textit{overheads} de escalonamento de tensão forem muito grandes.} \\ \hline \hline

\multicolumn{2}{|c|}{\textbf{Comentários}} \\ \hline
\multicolumn{2}{|p{21cm}|}{A metodologia proposta não mostra evidências, no artigo, de que fornece suporte para recursos compartilhados e / ou preempções.} \\ \hline
\end{tabular}
}
\end{table}

\newpage
\begin{table}[htp] 
\centering
\caption{Dados extraído da publicação de \cite{Xian_Lu_2006_a}.}
\resizebox{\columnwidth}{!}{%
\begin{tabular}{|p{3.8cm}|p{16cm}|}
\hline
\multicolumn{1}{|r|}{\textbf{ID}} & Xian\_Lu\_2006\_a \\ \hline
\multicolumn{1}{|r|}{\textbf{Título}} & Dynamic voltage scaling for multitasking real-time systems with uncertain execution time \\ \hline
\multicolumn{1}{|r|}{\textbf{Autores}} & Xian C., Lu Y.-H. \\ \hline
\multicolumn{1}{|r|}{\textbf{Palavras-chave}} & Dynamic voltage scaling; Hard real-time; Low energy; Multitasking; Probability \\ \hline
\multicolumn{1}{|r|}{\textbf{Fonte da publicação}} & Proceedings of the ACM Great Lakes Symposium on VLSI, GLSVLSI \\ \hline
\multicolumn{1}{|r|}{\textbf{Editora}} & ACM \\ \hline
\multicolumn{1}{|r|}{\textbf{Ano da publicação}} & 2006 \\ \hline \hline

\multicolumn{2}{|c|}{\textbf{Resumo da publicação}} \\ \hline
\multicolumn{2}{|p{21cm}|}{Os autores propõem uma abordagem que visa integrar as técnicas de escalonamento de tensão intra-tarefa e inter-tarefa. O conceito principal do método proposto é que cada tarefa possa contribuir com informações individuais para que seja possível melhorar o escalonamento individual das demais tarefas em execução, sempre tomando como base as informações globais passadas pelas demais tarefas. Dessa forma, a abordagem é dividida, basicamente, em duas etapas: (1) É calculado estatisticamente o escalonamento de frequência ótimo para múltiplas tarefas periódicas utilizando o escalonamento EDF (em inglês, \textit{Earliest Deadline First}) para processadores que conseguem mudar suas frequências de forma contínua; e (2) para processadores que possuem uma faixa limitada de frequências discretas, é apresentado um algoritmo heurístico específico para construção do escalonamento de frequência baseado em informações de distribuição de probabilidade e restrições de escalonabilidade globais.} \\ \hline \hline

\multicolumn{1}{|r|}{\textbf{Método(s) utilizados}} & DVFS Intra-Task; DVFS Inter-Task; \\ \hline
\multicolumn{1}{|r|}{\textbf{Ferramenta(s)}} & {Os autores não utilizaram nenhuma ferramenta específica. \newline

Os autores utilizaram os \textit{benchmarks}: \newline
- MPEG \textit{video decoder} \newline
- MP3 \textit{audio decoder} \newline
- H263 \textit{video decoder} \newline
- GSM \textit{speech encoder}} \\ \hline \hline

\multicolumn{2}{|c|}{\textbf{Impacto (positivo x negativo)}} \\ \hline
\multicolumn{2}{|p{21cm}|}{O artigo apresentou como ponto positivo uma metodologia que integra as técnicas DVFS intra-tarefa com a técnica DVFS inter-tarefa, dando suporte a preempções. Como ponto negativo, a metodologia não prever os \textit{overheads} ocasioandos em tempo de execução.} \\ \hline \hline

\multicolumn{2}{|c|}{\textbf{Validação do método}} \\ \hline
\multicolumn{2}{|p{21cm}|}{A validação da metodologia ocorreu pelo uso de \textit{benchmarks} relatados na literatura. Além disso, compara os resultados experimentais com outros trabalhos de referência na área de baixo consumo de energia.} \\ \hline \hline

\multicolumn{2}{|c|}{\textbf{Limitações do método}} \\ \hline
\multicolumn{2}{|p{21cm}|}{A metodologia não prever os \textit{overheads} ocasionados em tempo de execução e nem possíveis interferências entre tarefas.} \\ \hline \hline

\multicolumn{2}{|c|}{\textbf{Integração de métodos}} \\ \hline
\multicolumn{2}{|p{21cm}|}{Integrou as técnicas DVFS intra-tarefa com a técnica DVFS inter-tarefa.} \\ \hline \hline

\multicolumn{2}{|c|}{\textbf{Modo de aplicação do método}} \\ \hline
\multicolumn{2}{|p{21cm}|}{Híbrida} \\ \hline \hline

\multicolumn{2}{|c|}{\textbf{Perspectivas futuras}} \\ \hline
\multicolumn{2}{|p{21cm}|}{Os autores propõem realizar análises de sensibilidade sobre os tamanhos das tarefas.} \\ \hline \hline

\multicolumn{2}{|c|}{\textbf{Comentários}} \\ \hline
\multicolumn{2}{|p{21cm}|}{A metodologia aborda modelo de tarefas preemptivos, mas não deixa claro como é feito esse suporte. Quanto aos recursos compartilhados, a metodologia proposta não dá suporte.} \\ \hline
\end{tabular}
}
\end{table}

\newpage
\begin{table}[htp] 
\centering
\caption{Dados extraído da publicação de \cite{Hong_Yoo_Choi_Kong_2006_a}.}
\resizebox{\columnwidth}{!}{%
\begin{tabular}{|p{3.8cm}|p{16cm}|}
\hline
\multicolumn{1}{|r|}{\textbf{ID}} & Hong\_Yoo\_Choi\_Kong\_2006\_a \\ \hline
\multicolumn{1}{|r|}{\textbf{Título}} & Runtime distribution-aware dynamic voltage scaling \\ \hline
\multicolumn{1}{|r|}{\textbf{Autores}} & Hong S., Yoo S., Jin H., Choi K.-M., Kong J.-T., Eo S.-K. \\ \hline
\multicolumn{1}{|r|}{\textbf{Palavras-chave}} & Basic blocks; Computer-aided design; Dynamic voltage scaling (DVS); Energy reductions; execution cycles; Experimental results; international conferences; Minimal energy; MPEG-4 decoder; Multimedia software; Performance level (PL); Run time; Software runtime; Worst case execution; Worst case execution cycles; BASIC (programming language); Decoding; Design; Energy policy; Multimedia systems; Voltage stabilizing circuits; Computer software \\ \hline
\multicolumn{1}{|r|}{\textbf{Fonte da publicação}} & IEEE/ACM International Conference on Computer-Aided Design, Digest of Technical Papers, ICCAD \\ \hline
\multicolumn{1}{|r|}{\textbf{Editora}} & IEEE \\ \hline
\multicolumn{1}{|r|}{\textbf{Ano da publicação}} & 2006 \\ \hline \hline

\multicolumn{2}{|c|}{\textbf{Resumo da publicação}} \\ \hline
\multicolumn{2}{|p{21cm}|}{Os autores propõem uma nova técnica de escalonamento de tensão (DVS) intra-tarefa que não visa apenas explorar as distribuições de tempo de execução da aplicação, mas também o fluxo de dados e a arquitetura. Em outras palavras, essa abordagem utiliza os dados da aplicação e da arquitetura para predizer o RWCEC e aplicar com antecedência as tensões e frequências ideais no processador. Portanto, com este trabalho os autores introduziram o conceito de perfil estatístico de ciclos de execução dentro da técnica DVS intra-tarefa ao invés de ciclos de execução no pior caso (WCEC).} \\ \hline \hline

\multicolumn{1}{|r|}{\textbf{Método(s) utilizados}} & DVFS Intra-Task; \\ \hline
\multicolumn{1}{|r|}{\textbf{Ferramenta(s)}} & {O autor utilizou a ferramenta MaxSimTechnology, ARM Co. Ltd [Não disponível para download (\href{http://www.arm.com/products/DevTools/}{http://www.arm.com/products/DevTools/})]. \newline

O autor utilizou os \textit{benchmarks}: \newline
- H.264 Decoder. \newline
- MPEG4 Decoder.} \\ \hline \hline

\multicolumn{2}{|c|}{\textbf{Impacto (positivo x negativo)}} \\ \hline
\multicolumn{2}{|p{21cm}|}{O artigo apresentou como ponto positivo uma metodologia que cria perfis estatísticos com base nos caminhos de execução da tarefa. Como ponto negativo, os autores não deixam claro quais são as estruturas de dados suportadas em sua metodologia.} \\ \hline \hline

\multicolumn{2}{|c|}{\textbf{Validação do método}} \\ \hline
\multicolumn{2}{|p{21cm}|}{A validação da metodologia ocorreu pelo uso de de ferramenta e \textit{benchmarks} descritos na literatura.} \\ \hline \hline

\multicolumn{2}{|c|}{\textbf{Limitações do método}} \\ \hline
\multicolumn{2}{|p{21cm}|}{A metodologia não deixa claro quais tipos de estruturas de dados suportadas, como por exemplo: recursões, \textit{loops} ocultos, invariantes de \textit{loops}, etc.} \\ \hline \hline

\multicolumn{2}{|c|}{\textbf{Integração de métodos}} \\ \hline
\multicolumn{2}{|p{21cm}|}{Não se aplica} \\ \hline \hline

\multicolumn{2}{|c|}{\textbf{Modo de aplicação do método}} \\ \hline
\multicolumn{2}{|p{21cm}|}{Offline} \\ \hline \hline

\multicolumn{2}{|c|}{\textbf{Perspectivas futuras}} \\ \hline
\multicolumn{2}{|p{21cm}|}{Os autores propõem melhorar a eficiência da metodologia proposta através da exploração das correlações entre os blocos de códigos arbitrários.} \\ \hline \hline

\multicolumn{2}{|c|}{\textbf{Comentários}} \\ \hline
\multicolumn{2}{|p{21cm}|}{A metodologia proposta não mostra evidências, no artigo, de que fornece suporte para recursos compartilhados e / ou preempções.} \\ \hline
\end{tabular}
}
\end{table}

\newpage
\begin{table}[htp] 
\centering
\caption{Dados extraído da publicação de \cite{Zhang_2006_a}.}
\resizebox{\columnwidth}{!}{%
\begin{tabular}{|p{3.8cm}|p{16cm}|}
\hline
\multicolumn{1}{|r|}{\textbf{ID}} & Zhang\_2006\_a \\ \hline
\multicolumn{1}{|r|}{\textbf{Título}} & Energy-efficient task scheduling algorithm for mobile terminal \\ \hline
\multicolumn{1}{|r|}{\textbf{Autores}} & Zhang L., Qi D. \\ \hline
\multicolumn{1}{|r|}{\textbf{Palavras-chave}} & Dynamic voltage scaling; Embedded system; Energy minimization; Hybird genetic algorithm; Task sheduling \\ \hline
\multicolumn{1}{|r|}{\textbf{Fonte da publicação}} & IET Conference Publications \\ \hline
\multicolumn{1}{|r|}{\textbf{Editora}} & IET Digital Library \\ \hline
\multicolumn{1}{|r|}{\textbf{Ano da publicação}} & 2006 \\ \hline \hline

\multicolumn{2}{|c|}{\textbf{Resumo da publicação}} \\ \hline
\multicolumn{2}{|p{21cm}|}{Os autores propõem um algoritmo de escalonamento de tarefas baseado em otimizações genéticas para diminuir o consumo de energia quando são especificados os deadlines e os ciclos de execução das tarefas. Esse algoritmo genético híbrido integra as técnicas inter e intra tarefas visando mensurar o pWCEC (em inglês, \textit{Probabilistic Worst-Case Execution Time}), a fim de encontrar o melhor coeficiente de escalonamento das tarefas de forma que todas as restrições temporais sejam obedecidas e ao mesmo tempo se obtenha uma minimização do consumo de energia do processador.} \\ \hline \hline

\multicolumn{1}{|r|}{\textbf{Método(s) utilizados}} & DVFS Intra-Task; DVFS Inter-Task; \\ \hline
\multicolumn{1}{|r|}{\textbf{Ferramenta(s)}} & {Os autores utilizaram a ferramenta: \newline
- Sim-Panalyzer \textit{performance / power simulator} [Ferramenta não disponível para download - Site fora do ar (\href{http://www.eecs.umich.edu/~panalyzer}{http://www.eecs.umich.edu/$\sim$panalyzer})] \newline

Os autores não utilizaram nenhum \textit{benchmark} específico. } \\ \hline \hline

\multicolumn{2}{|c|}{\textbf{Impacto (positivo x negativo)}} \\ \hline
\multicolumn{2}{|p{21cm}|}{O artigo apresentou como ponto positivo uma metodologia que aplica algoritmos genéticos para diminuição do consumo de energia em sistemas de tempo real. Como ponto negativo, a metodologia nem sempre consegue alcançar bons resultados de economia de energia, dependendo dos parâmetros iniciais das tarefas.} \\ \hline \hline

\multicolumn{2}{|c|}{\textbf{Validação do método}} \\ \hline
\multicolumn{2}{|p{21cm}|}{A validação da metodologia ocorreu pelo uso de de ferramentas e casos de teste definidos pelos próprios autores.} \\ \hline \hline

\multicolumn{2}{|c|}{\textbf{Limitações do método}} \\ \hline
\multicolumn{2}{|p{21cm}|}{O autor relata que em alguns casos é impossível alcançar frequências ideais no processador, pois o \textit{deadline} e outras restrições temporais podem ser violadas.} \\ \hline \hline

\multicolumn{2}{|c|}{\textbf{Integração de métodos}} \\ \hline
\multicolumn{2}{|p{21cm}|}{Integrou as técnica DVFS intra-task com a técnica DVFS inter-task.} \\ \hline \hline

\multicolumn{2}{|c|}{\textbf{Modo de aplicação do método}} \\ \hline
\multicolumn{2}{|p{21cm}|}{Híbrida} \\ \hline \hline

\multicolumn{2}{|c|}{\textbf{Perspectivas futuras}} \\ \hline
\multicolumn{2}{|p{21cm}|}{Os autores propõem que os trabalhos futuros devem introduzir estimadores dinâmicos para a quantidade de ciclos de execução.} \\ \hline \hline

\multicolumn{2}{|c|}{\textbf{Comentários}} \\ \hline
\multicolumn{2}{|p{21cm}|}{A metodologia proposta não mostra evidências, no artigo, de que fornece suporte para recursos compartilhados e / ou preempções.} \\ \hline
\end{tabular}
}
\end{table}

\newpage
\begin{table}[htp] 
\centering
\caption{Dados extraído da publicação de \cite{Yi_Chen_Yang_2006_a}.}
\resizebox{\columnwidth}{!}{%
\begin{tabular}{|p{3.8cm}|p{16cm}|}
\hline
\multicolumn{1}{|r|}{\textbf{ID}} & Yi\_Chen\_Yang\_2006\_a \\ \hline
\multicolumn{1}{|r|}{\textbf{Título}} & Static WCET analysis based compiler-directed DVS energy optimization in real-time applications \\ \hline
\multicolumn{1}{|r|}{\textbf{Autores}} & Yi H., Chen J., Yang X. \\ \hline
\multicolumn{1}{|r|}{\textbf{Palavras-chave}} & Compiler; Low-power; Real-time; WCET \\ \hline
\multicolumn{1}{|r|}{\textbf{Fonte da publicação}} & 11th Asia-Pacific Conference on Advances in Computer Systems Architecture, ACSAC 2006; \\ \hline
\multicolumn{1}{|r|}{\textbf{Editora}} & Springer \\ \hline
\multicolumn{1}{|r|}{\textbf{Ano da publicação}} & 2006 \\ \hline \hline

\multicolumn{2}{|c|}{\textbf{Resumo da publicação}} \\ \hline
\multicolumn{2}{|p{21cm}|}{Os autores propõem uma ferramenta chamada HEPTANE, cuja função é realizar a análise estática do WCET (em inglês, \textit{Worst Case Execution Time}), inserir os códigos da técnica DVFS e definir o perfil de consumo de energia da aplicação. Essa ferramenta trabalha em conjunto com o simulador de energia e performance chamado \textit{Sim-Panalyzer}, que roda em um ambiente \textit{RTLPower} (em inglês, \textit{Real-Time Low Power}), cuja função é simular o ambiente de experimentação para rodar os casos de teste criados pelos autores. Analisando de forma mais incisiva o trabalho, não ficou claro como a ferramenta HEPTANE trata as invariante de \textit{loops}, na definição do perfil de consumo de energia da aplicação.} \\ \hline \hline

\multicolumn{1}{|r|}{\textbf{Método(s) utilizados}} & DVFS Intra-Task; Control flow graph (CFG); \\ \hline
\multicolumn{1}{|r|}{\textbf{Ferramenta(s)}} & {O autor utilizou um ambiente nomeado de RTLPower (\textit{Real-Time Low Power}), que utiliza as seguintes ferramentas: \newline
- HEPTANE WCET \textit{analysis tool} [Enviar e-mail para \href{puaut@irisa.fr}{puaut@irisa.fr} solicitando a ferramenta]. \newline
- Sim-Panalyzer \textit{performance / power simulator} [Ferramenta não disponível para download - Site fora do ar (\href{http://www.eecs.umich.edu/~panalyzer}{http://www.eecs.umich.edu/$\sim$panalyzer})] \newline

Os \textit{benchmarks} utilizados foram: \newline
- SNU-RT \textit{benchmark} [Disponível em: \href{http://www.cprover.org/goto-cc/examples/snu.html}{http://www.cprover.org/goto-cc/examples/snu.html}] \newline
- \textit{Fast Fourier Transform} (fft1k) \newline
- \textit{Matrix Multiplication} (matmul)} \\ \hline \hline

\multicolumn{2}{|c|}{\textbf{Impacto (positivo x negativo)}} \\ \hline
\multicolumn{2}{|p{21cm}|}{O artigo apresentou como ponto positivo uma metodologia que realiza a análise estática do WCET (\textit{Worst Case Execution Time}) em aplicações de tempo real, uma vez que a maioria das soluções presentes na literatura leva em consideração o WCEC (\textit{Worst Case Execution Cycle}). Como ponto negativo, a metodologia não deixa claro como são feitas as análises de estruturas como \textit{loops} e \textit{loops} ocultos.} \\ \hline \hline

\multicolumn{2}{|c|}{\textbf{Validação do método}} \\ \hline
\multicolumn{2}{|p{21cm}|}{A validação da metodologia ocorreu pelo uso de de ferramentas e \textit{benchmarks} relatados na literatura. Além disso, compara os resultados experimentais com outros trabalhos de referência na área de baixo consumo de energia.} \\ \hline \hline

\multicolumn{2}{|c|}{\textbf{Limitações do método}} \\ \hline
\multicolumn{2}{|p{21cm}|}{A metodologia não prever a análise estática de recursões.} \\ \hline \hline

\multicolumn{2}{|c|}{\textbf{Integração de métodos}} \\ \hline
\multicolumn{2}{|p{21cm}|}{Não se aplica} \\ \hline \hline

\multicolumn{2}{|c|}{\textbf{Modo de aplicação do método}} \\ \hline
\multicolumn{2}{|p{21cm}|}{Offline} \\ \hline \hline

\multicolumn{2}{|c|}{\textbf{Perspectivas futuras}} \\ \hline
\multicolumn{2}{|p{21cm}|}{Não identificada} \\ \hline \hline

\multicolumn{2}{|c|}{\textbf{Comentários}} \\ \hline
\multicolumn{2}{|p{21cm}|}{A metodologia proposta não mostra evidências, no artigo, de que fornece suporte para recursos compartilhados e / ou preempções.} \\ \hline
\end{tabular}
}
\end{table}

\newpage
\begin{table}[htp] 
\centering
\caption{Dados extraído da publicação de \cite{Shin_Kim_2007_a}.}
\resizebox{\columnwidth}{!}{%
\begin{tabular}{|p{3.8cm}|p{16cm}|}
\hline
\multicolumn{1}{|r|}{\textbf{ID}} & Shin\_Kim\_2007\_a \\ \hline
\multicolumn{1}{|r|}{\textbf{Título}} & Optimizing intratask voltage scheduling using profile and data-flow information \\ \hline
\multicolumn{1}{|r|}{\textbf{Autores}} & Shin D., Kim J. \\ \hline
\multicolumn{1}{|r|}{\textbf{Palavras-chave}} & Dynamic-voltage scaling; Low-power design; Power management; Real-time systems; Variable-voltage processor \\ \hline
\multicolumn{1}{|r|}{\textbf{Fonte da publicação}} & IEEE Transactions on Computer-Aided Design of Integrated Circuits and Systems \\ \hline
\multicolumn{1}{|r|}{\textbf{Editora}} & IEEE \\ \hline
\multicolumn{1}{|r|}{\textbf{Ano da publicação}} & 2007 \\ \hline \hline

\multicolumn{2}{|c|}{\textbf{Resumo da publicação}} \\ \hline
\multicolumn{2}{|p{21cm}|}{Os autores propõem duas melhorias sobre a técnica IntraDVS. A primeira delas é uma melhoria da técnica chamada RAEP-IntraDVS (em inglês, \textit{Remaining Average-case Execution Path}), que visa otimizar o escalonamento de tensão através de análises das informações da aplicação, levando em consideração o caminho de execução de caso médio remanescente. A outra melhoria é sobre a técnica LaIntraDVS, citada no trabalho \cite{Shin_Kim_2005_b}, que leva em consideração as informações do fluxo de dados para gerar otimizações sobre os pontos de chaveamento de tensão (em inglês, \textit{Voltage-Scaling Points} - VSPs), principalmente através da predição das VSPs antes de estruturas condicionais e \textit{loops}.} \\ \hline \hline

\multicolumn{1}{|r|}{\textbf{Método(s) utilizados}} & DVFS Intra-Task; Control flow graph (CFG); \\ \hline
\multicolumn{1}{|r|}{\textbf{Ferramenta(s)}} & {Os autores não especificaram a ferramenta utilizada para a realização dos experimentos, mas deixaram claro que utilizaram o MPEG-4 \textit{encoder} como \textit{benchmark}.} \\ \hline \hline

\multicolumn{2}{|c|}{\textbf{Impacto (positivo x negativo)}} \\ \hline
\multicolumn{2}{|p{21cm}|}{O artigo apresentou como ponto positivo uma metodologia que reduz os \textit{overheads} sobre a técnica DVS intra-tarefas através do melhor gerenciamento dos pontos de escalonamento de tensão. Como ponto negativo, a metodologia não claro como ela faz para definir VSPs diante de invariantes de \textit{loops}.} \\ \hline \hline

\multicolumn{2}{|c|}{\textbf{Validação do método}} \\ \hline
\multicolumn{2}{|p{21cm}|}{A validação da metodologia ocorreu pelo uso de \textit{benchmark} presente na literatura. Além disso, compara os resultados experimentais com outros trabalhos de referência na área de baixo consumo de energia.} \\ \hline \hline

\multicolumn{2}{|c|}{\textbf{Limitações do método}} \\ \hline
\multicolumn{2}{|p{21cm}|}{A metodologia não deixou claro no artigo como faz para lhe dar com invariantes de \textit{loops} ao definir as VSPs.} \\ \hline \hline

\multicolumn{2}{|c|}{\textbf{Integração de métodos}} \\ \hline
\multicolumn{2}{|p{21cm}|}{Não se aplica} \\ \hline \hline

\multicolumn{2}{|c|}{\textbf{Modo de aplicação do método}} \\ \hline
\multicolumn{2}{|p{21cm}|}{Offline} \\ \hline \hline

\multicolumn{2}{|c|}{\textbf{Perspectivas futuras}} \\ \hline
\multicolumn{2}{|p{21cm}|}{Os autores visam melhorar sua metodologia adicionando características para lhe dar com energia de fulga (\textit{leakage power}) e energia dinâmica (\textit{dynamic power}) do processador.} \\ \hline \hline

\multicolumn{2}{|c|}{\textbf{Comentários}} \\ \hline
\multicolumn{2}{|p{21cm}|}{A metodologia proposta não mostra evidências, no artigo, de que fornece suporte para recursos compartilhados e / ou preempções.} \\ \hline
\end{tabular}
}
\end{table}

\newpage
\begin{table}[htp] 
\centering
\caption{Dados extraído da publicação de \cite{Neishaburi_2007_a}.}
\resizebox{\columnwidth}{!}{%
\begin{tabular}{|p{3.8cm}|p{16cm}|}
\hline
\multicolumn{1}{|r|}{\textbf{ID}} & Neishaburi\_2007\_a \\ \hline
\multicolumn{1}{|r|}{\textbf{Título}} & System level voltage scheduling technique using UML-RT model \\ \hline
\multicolumn{1}{|r|}{\textbf{Autores}} & Neishaburi M.H., Daneshtalab M., Nabi M., Mohammadi S. \\ \hline
\multicolumn{1}{|r|}{\textbf{Palavras-chave}} & Intra-task voltage scheduling; JPEG encoder system; System level voltage scheduling; UML-RT model; Data flow analysis; Data transfer; Encoding (symbols); Energy utilization; Voltage control; Scheduling algorithms \\ \hline
\multicolumn{1}{|r|}{\textbf{Fonte da publicação}} & 2007 IEEE/ACS International Conference on Computer Systems and Applications, AICCSA 2007 \\ \hline
\multicolumn{1}{|r|}{\textbf{Editora}} & IEEE \\ \hline
\multicolumn{1}{|r|}{\textbf{Ano da publicação}} & 2007 \\ \hline \hline

\multicolumn{2}{|c|}{\textbf{Resumo da publicação}} \\ \hline
\multicolumn{2}{|p{21cm}|}{Os autores apresentam uma otimização sobre o escalonamento de tensões intra-tarefa, através da análise do fluxo de dados e do fluxo de controle da aplicação. A partir dessa análise, a metodologia é capaz de antecipar os pontos de escalonamento de tensão (em inglês, \textit{Voltage Scaling Points} - VSP), enquanto que a técnica DVFS intra-tarefa tradicional apenas localiza os pontos de controle. Essa metodologia permite adicionar menos \textit{overheads} no código fonte da aplicação.} \\ \hline \hline

\multicolumn{1}{|r|}{\textbf{Método(s) utilizados}} & DVFS Intra-Task; Control flow graph (CFG); \\ \hline
\multicolumn{1}{|r|}{\textbf{Ferramenta(s)}} & {Os autores utilizaram a ferramenta UML-RT e o JPEG \textit{encoder} como \textit{benchmark}.} \\ \hline \hline

\multicolumn{2}{|c|}{\textbf{Impacto (positivo x negativo)}} \\ \hline
\multicolumn{2}{|p{21cm}|}{O artigo apresentou como ponto positivo uma metodologia para analisar os grafos de fluxo de controle da aplicação e diminuir os \textit{overheads} inseridos dentro das tarefas. Como ponto negativo, a metodologia tem dificuldade para lhe dar alguns tipos de estruturas, como por exemplo: \textit{loops} e recursões.} \\ \hline \hline

\multicolumn{2}{|c|}{\textbf{Validação do método}} \\ \hline
\multicolumn{2}{|p{21cm}|}{A validação da metodologia ocorreu pelo uso de ferramenta e \textit{benchmark} presentes na literatura.} \\ \hline \hline

\multicolumn{2}{|c|}{\textbf{Limitações do método}} \\ \hline
\multicolumn{2}{|p{21cm}|}{A metodologia possui dificuldade para lhe dar com funções recursivas (profundidade desconhecida), volta de saltos incondicionais (\textit{loops} ocultos) e chamadas de funções dinâmicas.} \\ \hline \hline

\multicolumn{2}{|c|}{\textbf{Integração de métodos}} \\ \hline
\multicolumn{2}{|p{21cm}|}{Não se aplica} \\ \hline \hline

\multicolumn{2}{|c|}{\textbf{Modo de aplicação do método}} \\ \hline
\multicolumn{2}{|p{21cm}|}{Offline} \\ \hline \hline

\multicolumn{2}{|c|}{\textbf{Perspectivas futuras}} \\ \hline
\multicolumn{2}{|p{21cm}|}{Os autores pretendem melhorar a metodologia para lhe dar melhor com \textit{loops} e chamadas recursivas.} \\ \hline \hline

\multicolumn{2}{|c|}{\textbf{Comentários}} \\ \hline
\multicolumn{2}{|p{21cm}|}{A metodologia proposta não mostra evidências, no artigo, de que fornece suporte para recursos compartilhados e/ou preempções.} \\ \hline
\end{tabular}
}
\end{table}

\newpage
\begin{table}[htp] 
\centering
\caption{Dados extraído da publicação de \cite{Chen_Hsieh_Lai_2008_b}.}
\resizebox{\columnwidth}{!}{%
\begin{tabular}{|p{3.8cm}|p{16cm}|}
\hline
\multicolumn{1}{|r|}{\textbf{ID}} & Chen\_Hsieh\_Lai\_2008\_b \\ \hline
\multicolumn{1}{|r|}{\textbf{Título}} & Efficient algorithms for periodic real-time tasks to optimal discrete voltage schedules \\ \hline
\multicolumn{1}{|r|}{\textbf{Autores}} & Chen D.-R., Hsieh S.-M., Lai M.-F. \\ \hline
\multicolumn{1}{|r|}{\textbf{Palavras-chave}} & Dynamic voltage scaling; Hard real-time systems; Jitter control; Power-aware scheduling \\ \hline
\multicolumn{1}{|r|}{\textbf{Fonte da publicação}} & IPDPS Miami 2008 - Proceedings of the 22nd IEEE International Parallel and Distributed Processing Symposium, Program and CD-ROM \\ \hline
\multicolumn{1}{|r|}{\textbf{Editora}} & IEEE \\ \hline
\multicolumn{1}{|r|}{\textbf{Ano da publicação}} & 2008 \\ \hline \hline

\multicolumn{2}{|c|}{\textbf{Resumo da publicação}} \\ \hline
\multicolumn{2}{|p{21cm}|}{Os autores propõem uma metodologia que visa minimizar o consumo de energia através da análise do fluxo de dados da aplicação, tanto do ponto de vista inter quanto intra tarefa. Essa abordagem consiste basicamente de três fases: (1) primeiramente é feita a transformação harmônica dos períodos de todas as tarefas, em seguida é feita a validação e compartilhamento dos tempo de folga entre as demais tarefas, utilizando um escalonamento definidos pelos autores de \textit{Jitterless Schedule}, que são interferências causadas pela chegada de sucessivas instâncias de uma mesma tarefa.; (2) o próximo passo é calcular a utilização total dado os novos parâmetros das tarefas definidos no passo 1; por último (3) é feita a computação das características de cada tarefa, tais como o início e fim relativos, com o objetivo de ajustar as tensões e frequência, evitando que restrições temporais venham ser violadas. } \\ \hline \hline

\multicolumn{1}{|r|}{\textbf{Método(s) utilizados}} & DVFS Intra-Task; DVFS Inter-Task; \\ \hline
\multicolumn{1}{|r|}{\textbf{Ferramenta(s)}} & {Os autores não especificaram a ferramenta utilizada para a realização dos experimentos e não utilizaram nenhum \textit{benchmark}.} \\ \hline \hline

\multicolumn{2}{|c|}{\textbf{Impacto (positivo x negativo)}} \\ \hline
\multicolumn{2}{|p{21cm}|}{O artigo apresentou como ponto positivo uma metodologia que propõe um escalonamento de tarefas com menor \textit{jitter}. Como ponto negativo, os autores não aplicam nenhum \textit{benchmark} para avaliação dos reais ganhos da metodologia proposta.} \\ \hline \hline

\multicolumn{2}{|c|}{\textbf{Validação do método}} \\ \hline
\multicolumn{2}{|p{21cm}|}{A validação da metodologia ocorreu pelo uso de provas formais e comparações de desempenho de complexidade computacional com outros trabalhos presentes na literatura.} \\ \hline \hline

\multicolumn{2}{|c|}{\textbf{Limitações do método}} \\ \hline
\multicolumn{2}{|p{21cm}|}{Os autores não deixam claro sobre os \textit{overheads} ocasionados pelo uso de sua metodologia.} \\ \hline \hline

\multicolumn{2}{|c|}{\textbf{Integração de métodos}} \\ \hline
\multicolumn{2}{|p{21cm}|}{Integrou as técnicas DVFS intra-tarefa com a técnica DVFS inter-tarefa} \\ \hline \hline

\multicolumn{2}{|c|}{\textbf{Modo de aplicação do método}} \\ \hline
\multicolumn{2}{|p{21cm}|}{Offline} \\ \hline \hline

\multicolumn{2}{|c|}{\textbf{Perspectivas futuras}} \\ \hline
\multicolumn{2}{|p{21cm}|}{Não identificada} \\ \hline \hline

\multicolumn{2}{|c|}{\textbf{Comentários}} \\ \hline
\multicolumn{2}{|p{21cm}|}{A metodologia proposta não mostra evidências, no artigo, de que fornece suporte para recursos compartilhados e / ou preempções.} \\ \hline
\end{tabular}
}
\end{table}

\newpage
\begin{table}[htp] 
\centering
\caption{Dados extraído da publicação de \cite{Oh_Kim_Kim_Kyung_2008_a}.}
\resizebox{\columnwidth}{!}{%
\begin{tabular}{|p{3.8cm}|p{16cm}|}
\hline
\multicolumn{1}{|r|}{\textbf{ID}} & Oh\_Kim\_Kim\_Kyung\_2008\_a \\ \hline
\multicolumn{1}{|r|}{\textbf{Título}} & Task partitioning algorithm for intra-task dynamic voltage scaling \\ \hline
\multicolumn{1}{|r|}{\textbf{Autores}} & Oh S., Kim J., Kim S., Kyung C.-M. \\ \hline
\multicolumn{1}{|r|}{\textbf{Palavras-chave}} & Dynamic voltage scaling; International symposium; Voltage switching; Axial flow; Boolean functions; Electric load forecasting; Technical presentations; Electric network analysis \\ \hline
\multicolumn{1}{|r|}{\textbf{Fonte da publicação}} & Proceedings - IEEE International Symposium on Circuits and Systems \\ \hline
\multicolumn{1}{|r|}{\textbf{Editora}} & IEEE \\ \hline
\multicolumn{1}{|r|}{\textbf{Ano da publicação}} & 2008 \\ \hline \hline

\multicolumn{2}{|c|}{\textbf{Resumo da publicação}} \\ \hline
\multicolumn{2}{|p{21cm}|}{Os autores propõem um novo algoritmo de particionamento de tarefas baseado na técnica DVS intra-tarefa, onde o seu principal objetivo é dividir de maneira mais eficiente os blocos de código da aplicação de forma que seja possível diminuir o número de chaveamento de tensões e frequências do processador. Essa abordagem, primeiramente, divide o código fonte da aplicação em um número máximo de seções de código. Em seguida, são calculados os ciclos de execução de cada nó, por meio de simulações estáticas, e as penalidades das predições que falharam. Essas penalidades são utilizadas como uma medida para decidir se determinados nós deverão ser agrupados ou não. Com essa metodologia os autores conseguiram reduzir o número de chaveamentos de tensão e, consequentemente, minimizar o consumo de energia do processador.} \\ \hline \hline

\multicolumn{1}{|r|}{\textbf{Método(s) utilizados}} & DVFS Intra-Task; \\ \hline
\multicolumn{1}{|r|}{\textbf{Ferramenta(s)}} & {Os autores não especificaram a ferramenta utilizada para a realização dos experimentos, somente o \textit{hardware} utilizado. \newline

Os autores utilizaram o \textit{benchmark}: \newline
- H.264 \textit{Decoder} } \\ \hline \hline

\multicolumn{2}{|c|}{\textbf{Impacto (positivo x negativo)}} \\ \hline
\multicolumn{2}{|p{21cm}|}{O artigo apresentou como ponto positivo uma metodologia para analisar a divisão eficiente dos blocos de código de uma aplicação. Como ponto negativo, os autores não levam em consideração as preempções durante a realização dos experimentos. } \\ \hline \hline

\multicolumn{2}{|c|}{\textbf{Validação do método}} \\ \hline
\multicolumn{2}{|p{21cm}|}{A validação da metodologia ocorreu pelo uso de \textit{benchmark} presente na literatura. Além disso, compara os resultados experimentais com outros trabalhos de referência na área de baixo consumo de energia.} \\ \hline \hline

\multicolumn{2}{|c|}{\textbf{Limitações do método}} \\ \hline
\multicolumn{2}{|p{21cm}|}{O metodologia apresentada não leva em consideração preempções.} \\ \hline \hline

\multicolumn{2}{|c|}{\textbf{Integração de métodos}} \\ \hline
\multicolumn{2}{|p{21cm}|}{Não se aplica.} \\ \hline \hline

\multicolumn{2}{|c|}{\textbf{Modo de aplicação do método}} \\ \hline
\multicolumn{2}{|p{21cm}|}{Offline} \\ \hline \hline

\multicolumn{2}{|c|}{\textbf{Perspectivas futuras}} \\ \hline
\multicolumn{2}{|p{21cm}|}{Não identificada} \\ \hline \hline

\multicolumn{2}{|c|}{\textbf{Comentários}} \\ \hline
\multicolumn{2}{|p{21cm}|}{A metodologia proposta não mostra evidências, no artigo, de que fornece suporte para recursos compartilhados e / ou preempções.} \\ \hline
\end{tabular}
}
\end{table}

\newpage
\begin{table}[htp] 
\centering
\caption{Dados extraído da publicação de \cite{Zitterell_2008_a}.}
\resizebox{\columnwidth}{!}{%
\begin{tabular}{|p{3.8cm}|p{16cm}|}
\hline
\multicolumn{1}{|r|}{\textbf{ID}} & Zitterell\_2008\_a \\ \hline
\multicolumn{1}{|r|}{\textbf{Título}} & Improving energy-efficient real-time scheduling by exploiting code instrumentation \\ \hline
\multicolumn{1}{|r|}{\textbf{Autores}} & Zitterell T., Scholl C. \\ \hline
\multicolumn{1}{|r|}{\textbf{Palavras-chave}} & Code instrumentation; Discrete frequencies; Dynamic frequency; Energy efficient; Frequency-scaling; Idle time; Library functions; Light weight; Low frequency; Real time scheduling; Save energy; Simulation framework; Slack time; Task information; Task-scheduling; Voltage-scaling; Computer science; Energy efficiency; Information technology; Program compilers; Response time (computer systems); Scheduling; Real time systems \\ \hline
\multicolumn{1}{|r|}{\textbf{Fonte da publicação}} & Proceedings of the International Multiconference on Computer Science and Information Technology, IMCSIT 2008 \\ \hline
\multicolumn{1}{|r|}{\textbf{Editora}} & IEEE \\ \hline
\multicolumn{1}{|r|}{\textbf{Ano da publicação}} & 2008 \\ \hline \hline

\multicolumn{2}{|c|}{\textbf{Resumo da publicação}} \\ \hline
\multicolumn{2}{|p{21cm}|}{Os autores propõem um escalonamento mais eficiente de energia para processadores com frequências discretas, chamado de ItcaEDF (em inglês, \textit{Intra-Task Characteristics Aware EDF}). Ela se baseia na integração das técnicas inter e intra tarefas para diminuição dos tempos ociosos do processador e dos tempos de folgas das tarefas. No algoritmo intra-tarefa, os autores focam na quantidade de ciclos economizados e em um contador de ciclos, que possibilita contabilizar os diferentes caminhos dentro de um \textit{loop}, permitindo diminuir os níveis de frequência do processador de acordo com as invariantes do \textit{loop}. Quanto ao algoritmo inter-tarefa, os autores implementam um conjunto de bibliotecas que permitem as tarefas compartilhar informações umas com as outras, contribuindo para um melhor escalonamento global e diminuição do tempo ocioso do processador.} \\ \hline \hline

\multicolumn{1}{|r|}{\textbf{Método(s) utilizados}} & DVFS Intra-Task; DVFS Inter-Task; Control Flow Graph (CFG); \\ \hline
\multicolumn{1}{|r|}{\textbf{Ferramenta(s)}} & {Os autores não especificaram a ferramenta utilizada para a realização dos experimentos.} \\ \hline \hline

\multicolumn{2}{|c|}{\textbf{Impacto (positivo x negativo)}} \\ \hline
\multicolumn{2}{|p{21cm}|}{O artigo apresentou como ponto positivo uma metodologia que explora invariantes de \textit{loops} para realizar o melhor controle da quantidade de ciclos a serem executados. Como ponto negativo, os autores não deixam claro os reais impactos gerados pelo uso da sua metodologia, principalmente do ponto de visto dos \textit{overheads}, uma vez que a solução proposta é uma abordagem \textit{online}.} \\ \hline \hline

\multicolumn{2}{|c|}{\textbf{Validação do método}} \\ \hline
\multicolumn{2}{|p{21cm}|}{A validação da metodologia ocorreu pelo uso de casos de teste gerados pelos próprios autores. Além disso, comparam os resultados experimentais com outros trabalhos de referência na área de baixo consumo de energia.} \\ \hline \hline

\multicolumn{2}{|c|}{\textbf{Limitações do método}} \\ \hline
\multicolumn{2}{|p{21cm}|}{O método não deixa claro sobre os \textit{overheads} gerados pelas inclusões das bibliotecas de compartilhamento de dados entre tarefas em execução.} \\ \hline \hline

\multicolumn{2}{|c|}{\textbf{Integração de métodos}} \\ \hline
\multicolumn{2}{|p{21cm}|}{Integrou a técnica DVFS intra-tarefas com a técnica DVFS inter-tarefa} \\ \hline \hline

\multicolumn{2}{|c|}{\textbf{Modo de aplicação do método}} \\ \hline
\multicolumn{2}{|p{21cm}|}{Online} \\ \hline \hline

\multicolumn{2}{|c|}{\textbf{Perspectivas futuras}} \\ \hline
\multicolumn{2}{|p{21cm}|}{Não identificado} \\ \hline \hline

\multicolumn{2}{|c|}{\textbf{Comentários}} \\ \hline
\multicolumn{2}{|p{21cm}|}{A metodologia proposta não mostra evidências, no artigo, de que fornece suporte para recursos compartilhados e / ou preempções.} \\ \hline
\end{tabular}
}
\end{table}

\newpage
\begin{table}[htp] 
\centering
\caption{Dados extraído da publicação de \cite{Chen_2008_a}.}
\resizebox{\columnwidth}{!}{%
\begin{tabular}{|p{3.8cm}|p{16cm}|}
\hline
\multicolumn{1}{|r|}{\textbf{ID}} & Chen\_2008\_a \\ \hline
\multicolumn{1}{|r|}{\textbf{Título}} & Expected energy consumption minimization in DVS systems with discrete frequencies \\ \hline
\multicolumn{1}{|r|}{\textbf{Autores}} & Chen J.-J. \\ \hline
\multicolumn{1}{|r|}{\textbf{Palavras-chave}} & Dynamic voltage scaling; Energy-efficient scheduling; Expected energy consumption minimization; Probability \\ \hline
\multicolumn{1}{|r|}{\textbf{Fonte da publicação}} & Proceedings of the ACM Symposium on Applied Computing \\ \hline
\multicolumn{1}{|r|}{\textbf{Editora}} & ACM \\ \hline
\multicolumn{1}{|r|}{\textbf{Ano da publicação}} & 2008 \\ \hline \hline

\multicolumn{2}{|c|}{\textbf{Resumo da publicação}} \\ \hline
\multicolumn{2}{|p{21cm}|}{O autor apresenta uma nova abordagem para minimizar o consumo de energia utilizando funções de densidade de probabilidade com base nas cargas de trabalho das tarefas de tempo real. Para o escalonamento intra-tarefa foi feito um algoritmo eficiente para obter a frequência ideal para uma única tarefa, de modo que o consumo de energia seja minimizado. Enquanto que o algoritmo de escalonamento inter-tarefa, chamado \textit{M-Greedy}, foi desenvolvido com base em uma abordagem de programação linear cuja finalidade é obter as melhores soluções para as tarefas de tempo real baseada em quadros, visando diminuir os tempos de folga.} \\ \hline \hline

\multicolumn{1}{|r|}{\textbf{Método(s) utilizados}} & DVFS Intra-Task; DVFS Inter-Task; \\ \hline
\multicolumn{1}{|r|}{\textbf{Ferramenta(s)}} & {O autor não utilizou nenhuma ferramenta específica. No entanto, utilizou os \textit{benchmarks}: \newline
- mpegplay (MPEG \textit{video decoder}) \newline
- madplay (MP3 \textit{audio decoder}) \newline
- tmndec (H.263 \textit{video decoder}) \newline
- toast (GSM \textit{speech encoder})} \\ \hline \hline

\multicolumn{2}{|c|}{\textbf{Impacto (positivo x negativo)}} \\ \hline
\multicolumn{2}{|p{21cm}|}{O artigo apresentou como ponto positivo uma metodologia que visa minimizar o consumo de energia através da aplicação de funções de densidade de probabilidade com base nas cargas de trabalho das tarefas de tempo real. Como ponto negativo, o autor leva em consideração uma única tarefa na aplicação da sua metodologia offline.} \\ \hline \hline

\multicolumn{2}{|c|}{\textbf{Validação do método}} \\ \hline
\multicolumn{2}{|p{21cm}|}{A validação da metodologia ocorreu através do uso de \textit{benchmarks} relatados na literatura. Além disso, compara os resultados experimentais com outros trabalhos de referência na área de baixo consumo de energia.} \\ \hline \hline

\multicolumn{2}{|c|}{\textbf{Limitações do método}} \\ \hline
\multicolumn{2}{|p{21cm}|}{A metodologia proposta leva em consideração apenas uma única tarefa.} \\ \hline \hline

\multicolumn{2}{|c|}{\textbf{Integração de métodos}} \\ \hline
\multicolumn{2}{|p{21cm}|}{Integrou a técnica DVFS intra-tarefa com a técnica DVFS inter-tarefa} \\ \hline \hline

\multicolumn{2}{|c|}{\textbf{Modo de aplicação do método}} \\ \hline
\multicolumn{2}{|p{21cm}|}{Híbrida} \\ \hline \hline

\multicolumn{2}{|c|}{\textbf{Perspectivas futuras}} \\ \hline
\multicolumn{2}{|p{21cm}|}{O autor pretende expandir a metodologia proposta para outros dispositivos periféricos.} \\ \hline \hline

\multicolumn{2}{|c|}{\textbf{Comentários}} \\ \hline
\multicolumn{2}{|p{21cm}|}{A artigo comenta que as tarefas podem ser preemptadas por outras de maior prioridade, mas não deixa claro como é feito esse tratamento. Quanto aos recursos compartilhados, a metodologia não dá suporte.} \\ \hline
\end{tabular}
}
\end{table}

\newpage
\begin{table}[htp] 
\centering
\caption{Dados extraído da publicação de \cite{Chen_Hsieh_Lai_2008_a}.}
\resizebox{\columnwidth}{!}{%
\begin{tabular}{|p{3.8cm}|p{16cm}|}
\hline
\multicolumn{1}{|r|}{\textbf{ID}} & Chen\_Hsieh\_Lai\_2008\_a \\ \hline
\multicolumn{1}{|r|}{\textbf{Título}} & Efficient algorithms for jitterless real-time tasks to DVS schedules \\ \hline
\multicolumn{1}{|r|}{\textbf{Autores}} & Chen D.-R., Hsieh S.-M., Lai M.-F. \\ \hline
\multicolumn{1}{|r|}{\textbf{Palavras-chave}} & Canonical schedules; Dvs algorithms; DVS schedules; Efficient algorithms; Finish times; Minimum energies; Periodic tasks; Real-time tasks; Variable voltages; Voltage stabilizing circuits \\ \hline
\multicolumn{1}{|r|}{\textbf{Fonte da publicação}} & Parallel and Distributed Computing, Applications and Technologies, PDCAT Proceedings \\ \hline
\multicolumn{1}{|r|}{\textbf{Editora}} & IEEE \\ \hline
\multicolumn{1}{|r|}{\textbf{Ano da publicação}} & 2008 \\ \hline \hline

\multicolumn{2}{|c|}{\textbf{Resumo da publicação}} \\ \hline
\multicolumn{2}{|p{21cm}|}{Os autores propõem um algoritmo intra-tarefa e um inter-tarefa para diminuir o consumo de energia durante o escalonamento das tarefas. Essa metodologia tem como finalidade diminuir os \textit{overheads} e os tempos de folga entre as tarefas, dando mais previsibilidade e otimizações \textit{offline} para o escalonamento. Para facilitar a geração do escalonamento, as tarefas com períodos arbitrários são transformados em períodos harmônicos para que os tempos de preempção, início e término de cada tarefa possam ser facilmente derivados, principalmente para tratar o que o autor chama de \textit{Jitterless Schedule}. Essa abordagem foi desenvolvida a partir do trabalho de \cite{Chen_Hsieh_Lai_2008_b}.} \\ \hline \hline

\multicolumn{1}{|r|}{\textbf{Método(s) utilizados}} & DVFS Intra-Task; DVFS Inter-Task; \\ \hline
\multicolumn{1}{|r|}{\textbf{Ferramenta(s)}} & {Os autores não utilizaram nenhuma ferramenta ou benchmark.} \\ \hline \hline

\multicolumn{2}{|c|}{\textbf{Impacto (positivo x negativo)}} \\ \hline
\multicolumn{2}{|p{21cm}|}{O artigo apresentou como ponto positivo uma metodologia para lidar com preempções e interferências entre tarefas. Como ponto negativo, os autores não utilizaram nenhum \textit{benchmark} para a realização dos seus experimentos. } \\ \hline \hline

\multicolumn{2}{|c|}{\textbf{Validação do método}} \\ \hline
\multicolumn{2}{|p{21cm}|}{A validação da metodologia ocorreu pelo uso de casos de testes gerados pelos próprios autores. Além disso, comparam os resultados experimentais com outros trabalhos de referência na área de baixo consumo de energia.} \\ \hline \hline

\multicolumn{2}{|c|}{\textbf{Limitações do método}} \\ \hline
\multicolumn{2}{|p{21cm}|}{Os autores apresentaram apenas comparações de complexidade em relação aos trabalhos presentes na literatura, faltou mostrar o real ganho da metodologia proposta em níveis de consumo de energia.} \\ \hline \hline

\multicolumn{2}{|c|}{\textbf{Integração de métodos}} \\ \hline
\multicolumn{2}{|p{21cm}|}{Integrou a técnica DVFS intra-tarefa com a técnica DVFS inter-tarefa} \\ \hline \hline

\multicolumn{2}{|c|}{\textbf{Modo de aplicação do método}} \\ \hline
\multicolumn{2}{|p{21cm}|}{Offline} \\ \hline \hline

\multicolumn{2}{|c|}{\textbf{Perspectivas futuras}} \\ \hline
\multicolumn{2}{|p{21cm}|}{Não identificada} \\ \hline \hline

\multicolumn{2}{|c|}{\textbf{Comentários}} \\ \hline
\multicolumn{2}{|p{21cm}|}{A metodologia proposta não mostra evidências, no artigo, de que fornece suporte para recursos compartilhados e / ou preempções.} \\ \hline
\end{tabular}
}
\end{table}

\newpage
\begin{table}[htp] 
\centering
\caption{Dados extraído da publicação de \cite{He_2008_a}.}
\resizebox{\columnwidth}{!}{%
\begin{tabular}{|p{3.8cm}|p{16cm}|}
\hline
\multicolumn{1}{|r|}{\textbf{ID}} & He\_2008\_a \\ \hline
\multicolumn{1}{|r|}{\textbf{Título}} & Stochastic voltage scheduling of fixed-priority tasks with preemption thresholds \\ \hline
\multicolumn{1}{|r|}{\textbf{Autores}} & He X., Jia Y., Wa H. \\ \hline
\multicolumn{1}{|r|}{\textbf{Palavras-chave}} & Embedded systems; Energy management; Energy policy; Integrated circuits; Mobile computing; Scheduling; Voltage stabilizing circuits; Wireless networks; Wireless telecommunication systems; Clock speeds; Dynamic voltage scaling; Energy consumptions; Energy gains; Power managements; Preemption thresholds; Priority tasks; Responsiveness.; Scarce resources; Scheduling policies; Stochastic datums; Supply voltages; Variable execution times; Voltage scheduling; Real time systems \\ \hline
\multicolumn{1}{|r|}{\textbf{Fonte da publicação}} & 2008 International Conference on Wireless Communications, Networking and Mobile Computing, WiCOM 2008 \\ \hline
\multicolumn{1}{|r|}{\textbf{Editora}} & IEEE \\ \hline
\multicolumn{1}{|r|}{\textbf{Ano da publicação}} & 2008 \\ \hline \hline

\multicolumn{2}{|c|}{\textbf{Resumo da publicação}} \\ \hline
\multicolumn{2}{|p{21cm}|}{Os autores exploram os tempos de execução variáveis de tarefas, dentro da política de escalonamento FPPT (em inglês, \textit{Fixed-Priority scheduling with Preemption Threshold} - Escalonamento com Prioridade Fixa com Limite de Preempção). Essa política de escalonamento, executando em um processador com tensão variável, permite reduzir os custos com preempções desnecessárias das tarefas. Então, os autores desenvolveram um algoritmo para analisar todas as possibilidades de carga de trabalho para cada tarefa. Em seguida, utilizou esses dados estocásticos para definir as tensões e frequências do processador de acordo com o tamanho da tarefa e sua distribuição de probabilidade, com o intuito de minimizar o consumo de energia no caso médio.} \\ \hline \hline

\multicolumn{1}{|r|}{\textbf{Método(s) utilizados}} & DVFS Intra-Task; FPPT scheduling policy; \\ \hline
\multicolumn{1}{|r|}{\textbf{Ferramenta(s)}} & {Os autores utilizaram a ferramenta SPAK-0.3 (Disponível em: \href{http://www.cs.utah.edu/~regehr/spak/}{http://www.cs.utah.edu/$\sim$regehr/spak/}).} \\ \hline \hline

\multicolumn{2}{|c|}{\textbf{Impacto (positivo x negativo)}} \\ \hline
\multicolumn{2}{|p{21cm}|}{O artigo apresentou como ponto positivo uma metodologia que visa otimizar custos com preempções. Como ponto negativo, os autores não deixam muito claro no artigo como são feitos os tratamentos de preempção entre as tarefas em execução.} \\ \hline \hline

\multicolumn{2}{|c|}{\textbf{Validação do método}} \\ \hline
\multicolumn{2}{|p{21cm}|}{A validação da metodologia ocorreu pelo uso de ferramentas de simulação e de casos de testes gerados pelos próprios autores.} \\ \hline \hline

\multicolumn{2}{|c|}{\textbf{Limitações do método}} \\ \hline
\multicolumn{2}{|p{21cm}|}{A metodologia proposta não deixa muito claro como são feitos os tratamentos de preempção entre as tarefas em execução.} \\ \hline \hline

\multicolumn{2}{|c|}{\textbf{Integração de métodos}} \\ \hline
\multicolumn{2}{|p{21cm}|}{Integrou a política de escalonamento FPPT com a técnica DVFS intra-tarefa;} \\ \hline \hline

\multicolumn{2}{|c|}{\textbf{Modo de aplicação do método}} \\ \hline
\multicolumn{2}{|p{21cm}|}{Offline} \\ \hline \hline

\multicolumn{2}{|c|}{\textbf{Perspectivas futuras}} \\ \hline
\multicolumn{2}{|p{21cm}|}{Os autores pretendem aplicar essa metodologia em algum RTOS (em inglês, \textit{Real Time Operating System}).} \\ \hline \hline

\multicolumn{2}{|c|}{\textbf{Comentários}} \\ \hline
\multicolumn{2}{|p{21cm}|}{A metodologia proposta mostra indícios de que dá suporte a preempções, mas não ficou muito claro no artigo como é feito. Quanto aos recursos compartilhados, a metodologia não dá suporte.} \\ \hline
\end{tabular}
}
\end{table}

\newpage
\begin{table}[htp] 
\centering
\caption{Dados extraído da publicação de \cite{Yang_2009_a}.}
\resizebox{\columnwidth}{!}{%
\begin{tabular}{|p{3.8cm}|p{16cm}|}
\hline
\multicolumn{1}{|r|}{\textbf{ID}} & Yang\_2009\_a \\ \hline
\multicolumn{1}{|r|}{\textbf{Título}} & Energy efficient intra-task dynamic voltage scaling for realistic CPUs of mobile devices \\ \hline
\multicolumn{1}{|r|}{\textbf{Autores}} & Yang C.-C., Wang K., Lin M.-H., Lin P. \\ \hline
\multicolumn{1}{|r|}{\textbf{Palavras-chave}} & CPU; Dynamic voltage scaling; Energy efficient; Intra-task; Mobile device; Real time \\ \hline
\multicolumn{1}{|r|}{\textbf{Fonte da publicação}} & Journal of Information Science and Engineering \\ \hline
\multicolumn{1}{|r|}{\textbf{Editora}} & Scopus (Elsevier) \\ \hline
\multicolumn{1}{|r|}{\textbf{Ano da publicação}} & 2009 \\ \hline \hline

\multicolumn{2}{|c|}{\textbf{Resumo da publicação}} \\ \hline
\multicolumn{2}{|p{21cm}|}{Os autores propõem dois algoritmos de escalonamento dinâmico de tensão (DVS) intra-tarefa para CPU's. O algoritmo OSRC (em inglês, \textit{Optimal Schedule for Realistic CPUs}) tem por objetivo aplicar conceitos de programação dinâmica dentro da CFG da aplicação para identificar os caminhos ideais com menor consumo de energia, tendo como referência a especificação de uma \textit{CPU} realista, que são processadores que possuem um conjunto limitado de níveis de tensão e frequência \citep{Yang_2009_a}. O segundo algoritmo, chamado LO-OSRC (em inglês, \textit{Low Overhead Optimal Schedule for Realistic CPUs}), já leva em consideração o potencial de computação das tarefas e os \textit{overheads} de transição, permitindo apenas uma mudanção de tensão de frequência da CPU por tarefa. Dessa forma, os autores mostram em seus experimentos que seus algoritmos conseguem obter ganhos acima de 10\% em comparação com outros trabalhos presentes na literatura.} \\ \hline \hline

\multicolumn{1}{|r|}{\textbf{Método(s) utilizados}} & DVFS Intra-Task; Control Flow Graph (CFG); Stochastic Method; \\ \hline
\multicolumn{1}{|r|}{\textbf{Ferramenta(s)}} & {Os autores específicam os \textit{hardwares} utilizados, mas não deixaram claro as ferramentas ou softwares utilizados no experimentos e nem fazem referências quanto ao uso de algum \textit{benchmark}; } \\ \hline \hline

\multicolumn{2}{|c|}{\textbf{Impacto (positivo x negativo)}} \\ \hline
\multicolumn{2}{|p{21cm}|}{O artigo apresentou como ponto positivo a integração de métodos estocásticos a técnica DVFS intra-tarefa. Como ponto negativo, os autores utilizaram somente dois modelos de tarefas para validar o seu método, onde o correto deveria ser utilizar várias modelos de tarefas diferentes visto que sua metodologia se basea em processos não determinísticos (métodos estocásticos).} \\ \hline \hline

\multicolumn{2}{|c|}{\textbf{Validação do método}} \\ \hline
\multicolumn{2}{|p{21cm}|}{A validação da metodologia ocorreu pelo uso de casos de testes gerados pelos próprios autores e, além disso, comparam os resultados experimentais obtidos com outras abordagens presentes na literatura.} \\ \hline \hline

\multicolumn{2}{|c|}{\textbf{Limitações do método}} \\ \hline
\multicolumn{2}{|p{21cm}|}{Os autores levaram em consideração apenas dois modelos de tarefas, criados por eles mesmo, para validar uma abordagem baseada em modelos estocásticos.} \\ \hline \hline

\multicolumn{2}{|c|}{\textbf{Integração de métodos}} \\ \hline
\multicolumn{2}{|p{21cm}|}{Integrou métodos estocásticos com a técnica DVFS intra-tarefa;} \\ \hline \hline

\multicolumn{2}{|c|}{\textbf{Modo de aplicação do método}} \\ \hline
\multicolumn{2}{|p{21cm}|}{Offline} \\ \hline \hline

\multicolumn{2}{|c|}{\textbf{Perspectivas futuras}} \\ \hline
\multicolumn{2}{|p{21cm}|}{Não identificada} \\ \hline \hline

\multicolumn{2}{|c|}{\textbf{Comentários}} \\ \hline
\multicolumn{2}{|p{21cm}|}{A metodologia proposta não mostra evidências, no artigo, de que fornece suporte para recursos compartilhados e / ou preempções.} \\ \hline
\end{tabular}
}
\end{table}

\newpage
\begin{table}[htp] 
\centering
\caption{Dados extraído da publicação de \cite{Ishihara_2009_a}.}
\resizebox{\columnwidth}{!}{%
\begin{tabular}{|p{3.8cm}|p{16cm}|}
\hline
\multicolumn{1}{|r|}{\textbf{ID}} & Ishihara\_2009\_a \\ \hline
\multicolumn{1}{|r|}{\textbf{Título}} & Real-time power management for a multi-performance processor \\ \hline
\multicolumn{1}{|r|}{\textbf{Autores}} & Ishihara T. \\ \hline
\multicolumn{1}{|r|}{\textbf{Palavras-chave}} & Embedded systems; Low-power design; Microprocessor; Real-time systems \\ \hline
\multicolumn{1}{|r|}{\textbf{Fonte da publicação}} & 2009 International SoC Design Conference, ISOCC 2009 \\ \hline
\multicolumn{1}{|r|}{\textbf{Editora}} & IEEE \\ \hline
\multicolumn{1}{|r|}{\textbf{Ano da publicação}} & 2009 \\ \hline \hline

\multicolumn{2}{|c|}{\textbf{Resumo da publicação}} \\ \hline
\multicolumn{2}{|p{21cm}|}{O autor propõe uma metodologia baseada em uma nova arquitetura contendo vários núcleos de processamento, chamada \textit{Architecture of Multi-Performance Processor}, onde cada núcleo trabalha em uma frequência e tensão específica. Dessa forma, o processador não perde tempo chaveando tensões e frequência, que são em média na casa das centenas de microsegundos. Essa arquitetura quando integrada a técnica DVFS intra-tarefa permite maximizar os ganhos de economiza de energia, através da diminuição dos  \textit{overheads} da técnica DVFS, permitindo fazer chaveamentos de tensão e frequência na casa dos 1.5 microsegundos e dissipando apenas 10 nano-joules. Essa metodologia, reduziu cerca de 25\% de energia em comparação com a técnica DVS convencional do processador.} \\ \hline \hline

\multicolumn{1}{|r|}{\textbf{Método(s) utilizados}} & DVFS Intra-Task; \\ \hline
\multicolumn{1}{|r|}{\textbf{Ferramenta(s)}} & {O autor utilizou a ferramenta RTL Simulation. \newline

O autor utilizou os Benchmarks: \newline
- Adpcm decoder (\textit{Adaptive Differential Pulse Code Modulation}) \newline
- JPEG encoder \newline
- MPEG2 encoder } \\ \hline \hline

\multicolumn{2}{|c|}{\textbf{Impacto (positivo x negativo)}} \\ \hline
\multicolumn{2}{|p{21cm}|}{O artigo apresentou como ponto positivo uma nova proposta de arquitetura contendo vários processadores trabalhando em tensões e frequências diferentes entre si, com o intuito de reduzir os \textit{overheads} da técnica DVFS. Como pontos negativos, temos nesta proposta uma arquitetura mais cara, pois necessita de mais componentes como memórias secundárias, caches para cada núcleo de processamento e de um software mais específico e robusto para realizar todo esse gerenciamento.} \\ \hline \hline

\multicolumn{2}{|c|}{\textbf{Validação do método}} \\ \hline
\multicolumn{2}{|p{21cm}|}{A validação da metodologia ocorreu pelo uso de ferramentas de simulação e \textit{benchmarks} presentes na literatura.} \\ \hline \hline

\multicolumn{2}{|c|}{\textbf{Limitações do método}} \\ \hline
\multicolumn{2}{|p{21cm}|}{Não mostrou nenhum resultado envolvendo várias tarefas e nem levou em consideração preempções.} \\ \hline \hline

\multicolumn{2}{|c|}{\textbf{Integração de métodos}} \\ \hline
\multicolumn{2}{|p{21cm}|}{Não se aplica} \\ \hline \hline

\multicolumn{2}{|c|}{\textbf{Modo de aplicação do método}} \\ \hline
\multicolumn{2}{|p{21cm}|}{Offline} \\ \hline \hline

\multicolumn{2}{|c|}{\textbf{Perspectivas futuras}} \\ \hline
\multicolumn{2}{|p{21cm}|}{Os autores visam melhorar o sistema de cache da arquitetura proposta, pois hoje não está prevendo cache de dados, somente cache de instruções.} \\ \hline \hline

\multicolumn{2}{|c|}{\textbf{Comentários}} \\ \hline
\multicolumn{2}{|p{21cm}|}{A metodologia proposta não mostra evidências, no artigo, de que fornece suporte para recursos compartilhados e / ou preempções.} \\ \hline
\end{tabular}
}
\end{table}

\newpage
\begin{table}[htp] 
\centering
\caption{Dados extraído da publicação de \cite{Mohan_2010_a}.}
\resizebox{\columnwidth}{!}{%
\begin{tabular}{|p{3.8cm}|p{16cm}|}
\hline
\multicolumn{1}{|r|}{\textbf{ID}} & Mohan\_2010\_a \\ \hline
\multicolumn{1}{|r|}{\textbf{Título}} & Parametric timing analysis and its application to dynamic voltage scaling \\ \hline
\multicolumn{1}{|r|}{\textbf{Autores}} & Mohan S., Mueller F., Root M., Hawkins W., Healy C., Whalley D., Vivancos E. \\ \hline
\multicolumn{1}{|r|}{\textbf{Palavras-chave}} & Dynamic voltage scaling; Real-time systems; Timing analysis; Worst-case execution time \\ \hline
\multicolumn{1}{|r|}{\textbf{Fonte da publicação}} & Transactions on Embedded Computing Systems \\ \hline
\multicolumn{1}{|r|}{\textbf{Editora}} & ACM \\ \hline
\multicolumn{1}{|r|}{\textbf{Ano da publicação}} & 2010 \\ \hline \hline

\multicolumn{2}{|c|}{\textbf{Resumo da publicação}} \\ \hline
\multicolumn{2}{|p{21cm}|}{Os autores propõem um metodologia que remove as restrições sobre as invariantes de \textit{loops} através de analises paramétricas, com o objetivo de maximizar a identificação dos tempos de folga das tarefas e minimizar o consumo de energia do processador. Dessa forma, os ganhos dessa abordagem está diretamente relacionado com a redução do número de interações dos \textit{loops} mapeados dentro das aplicações.} \\ \hline \hline

\multicolumn{1}{|r|}{\textbf{Método(s) utilizados}} & DVFS Intra-Task; DVFS Inter-Task; Control Flow Graph (CFG); Loop-Invariant; \\ \hline
\multicolumn{1}{|r|}{\textbf{Ferramenta(s)}} & {Os autores utilizaram as ferramentas: \newline
- SimpleScalar processor simulator \newline
- Preprocesses gcc-generated PISA assembly \newline
- FAST framework (Apenas citada pelos autores) \newline
- VISA framework (Apenas citada pelos autores) \newline

Os autores utilizaram os \textit{benchmarks} (todos eles pertencem ao C Benchmark): \newline
- adpcm (\textit{Adaptive Differential Pulse Code Modulation}) \newline
- cnt (\textit{Sum and count of positive and negative numbers in an array}) \newline
- lms (\textit{An LMS adaptive signal enhancement}) \newline
- mm (\textit{Matrix Multiplication}) } \\ \hline \hline

\multicolumn{2}{|c|}{\textbf{Impacto (positivo x negativo)}} \\ \hline
\multicolumn{2}{|p{21cm}|}{O artigo apresentou como ponto positivo uma metodologia que mostra como aplicar técnicas de invariante de \textit{loop} no contexto de sistemas de tempo real para diminuir o consumo de energia. Como ponto negativo, os autores relatam que precisa melhorar sua metodologia para lidar com casos especiais que tenha um número menor de iterações dentro dos \textit{loops}.} \\ \hline \hline

\multicolumn{2}{|c|}{\textbf{Validação do método}} \\ \hline
\multicolumn{2}{|p{21cm}|}{A validação da metodologia ocorreu pelo uso de ferramentas de simulação e \textit{benchmarks} presentes na literatura. Além disso, compara os resultados experimentais com outros trabalhos de referência na área de baixo consumo de energia.} \\ \hline \hline

\multicolumn{2}{|c|}{\textbf{Limitações do método}} \\ \hline
\multicolumn{2}{|p{21cm}|}{A metodologia proposta tem dificuldades para lhe dar, em alguns casos, com um número menor de iterações dentro dos \textit{loops}.} \\ \hline \hline

\multicolumn{2}{|c|}{\textbf{Integração de métodos}} \\ \hline
\multicolumn{2}{|p{21cm}|}{Foram integradas as técnicas: \textit{inter-task}, \textit{intra-task} e \textit{loop-invariant}} \\ \hline \hline

\multicolumn{2}{|c|}{\textbf{Modo de aplicação do método}} \\ \hline
\multicolumn{2}{|p{21cm}|}{Híbrida} \\ \hline \hline

\multicolumn{2}{|c|}{\textbf{Perspectivas futuras}} \\ \hline
\multicolumn{2}{|p{21cm}|}{Os autores visam melhorar seu método para lidar com um número menor de iterações dentro dos \textit{loops}.} \\ \hline \hline

\multicolumn{2}{|c|}{\textbf{Comentários}} \\ \hline
\multicolumn{2}{|p{21cm}|}{A metodologia mostra indícios de que dá suporte a preempções entre tarefas, mas não deixa claro como é feito. Quanto aos recursos compartilhados, a metodologia não dá suporte.} \\ \hline
\end{tabular}
}
\end{table}

\newpage
\begin{table}[htp] 
\centering
\caption{Dados extraído da publicação de \cite{Tatematsu_2011_a}.}
\resizebox{\columnwidth}{!}{%
\begin{tabular}{|p{3.8cm}|p{16cm}|}
\hline
\multicolumn{1}{|r|}{\textbf{ID}} & Tatematsu\_2011\_a \\ \hline
\multicolumn{1}{|r|}{\textbf{Título}} & Checkpoint extraction using execution traces for intra-task DVFS in embedded systems \\ \hline
\multicolumn{1}{|r|}{\textbf{Autores}} & Tatematsu T., Takase H., Zeng G., Tomiyama H., Takada H. \\ \hline
\multicolumn{1}{|r|}{\textbf{Palavras-chave}} & DVFS; Embedded system; Execution trace; Low energy \\ \hline
\multicolumn{1}{|r|}{\textbf{Fonte da publicação}} & Proceedings - 2011 6th IEEE International Symposium on Electronic Design, Test and Application, DELTA 2011 \\ \hline
\multicolumn{1}{|r|}{\textbf{Editora}} & IEEE \\ \hline
\multicolumn{1}{|r|}{\textbf{Ano da publicação}} & 2011 \\ \hline \hline

\multicolumn{2}{|c|}{\textbf{Resumo da publicação}} \\ \hline
\multicolumn{2}{|p{21cm}|}{Os autores propõem uma metodologia que analisa o código fonte da aplicação e lista todos os possíveis locais para a inserção de pontos de controle (também chamado pelos autores de \textit{checkpoints}). Em seguida, todos esses pontos são analisado e os que não trazem ganhos energéticos são removidos. Por fim, a metodologia compara essa listagem de pontos de controle com uma tabela RWCEC (em inglês, \textit{Remaining Worst Case Execution Cycles}), também extraída da aplicação, para então calcular as tensões e frequências que deverão ser utilizadas no processador.} \\ \hline \hline

\multicolumn{1}{|r|}{\textbf{Método(s) utilizados}} & DVFS Intra-Task; Control Flow Graph (CFG); \\ \hline
\multicolumn{1}{|r|}{\textbf{Ferramenta(s)}} & {Os autores utilizaram a ferramenta SimpleScalarARM (Disponível em: \href{http://www.simplescalar.com/}{http://www.simplescalar.com/}) e o benchmark MediaBench (Disponível em: \href{http://dl.acm.org/citation.cfm?id=266800.266832}{http://dl.acm.org/citation.cfm?id=266800.266832}).} \\ \hline \hline

\multicolumn{2}{|c|}{\textbf{Impacto (positivo x negativo)}} \\ \hline
\multicolumn{2}{|p{21cm}|}{O artigo apresentou como ponto positivo uma nova metologia que visa maximizar os ganhos energéticos da inserção de \textit{checkpoints}, através da análise e remoção dos pontos que não trazem ganhos energéticos para a aplicação. Como ponto negativo, a metodologia mostrou, durante a fase de experimentação, que a inserção de alguns \textit{checkpoints} na aplicação podem ocasionar a perda de \textit{deadlines}.} \\ \hline \hline

\multicolumn{2}{|c|}{\textbf{Validação do método}} \\ \hline
\multicolumn{2}{|p{21cm}|}{A validação da metodologia ocorreu pelo uso de \textit{benchmarks} presentes na literatura.} \\ \hline \hline

\multicolumn{2}{|c|}{\textbf{Limitações do método}} \\ \hline
\multicolumn{2}{|p{21cm}|}{Os resultados experimentais mostram que a metodologia proposta pode ocasionar a violação das restrições temporais da tarefas.} \\ \hline \hline

\multicolumn{2}{|c|}{\textbf{Integração de métodos}} \\ \hline
\multicolumn{2}{|p{21cm}|}{Não se aplica} \\ \hline \hline

\multicolumn{2}{|c|}{\textbf{Modo de aplicação do método}} \\ \hline
\multicolumn{2}{|p{21cm}|}{\textit{Offline}} \\ \hline \hline

\multicolumn{2}{|c|}{\textbf{Perspectivas futuras}} \\ \hline
\multicolumn{2}{|p{21cm}|}{Os autores visam integrar métodos para lidar com as perdas de deadlines e melhorar a eficiência do algoritmo de extração de \textit{checkpoints}.} \\ \hline \hline

\multicolumn{2}{|c|}{\textbf{Comentários}} \\ \hline
\multicolumn{2}{|p{21cm}|}{A metodologia proposta não mostra evidências, no artigo, de que fornece suporte para recursos compartilhados e / ou preempções.} \\ \hline
\end{tabular}
}
\end{table}

\newpage
\begin{table}[htp] 
\centering
\caption{Dados extraído da publicação de \cite{Takase_2011_a}.}
\resizebox{\columnwidth}{!}{%
\begin{tabular}{|p{3.8cm}|p{16cm}|}
\hline
\multicolumn{1}{|r|}{\textbf{ID}} & Takase\_2011\_a \\ \hline
\multicolumn{1}{|r|}{\textbf{Título}} & An integrated optimization framework for reducing the energy consumption of embedded real-time applications \\ \hline
\multicolumn{1}{|r|}{\textbf{Autores}} & Takase H., Zeng G., Gauthier L., Kawashima H., Atsumi N., Tatematsu T., Kobayashi Y., Kohara S., Koshiro T., Ishihara T., Tomiyama H., Takada H. \\ \hline
\multicolumn{1}{|r|}{\textbf{Palavras-chave}} & development environment; embedded application; energy optimization \\ \hline
\multicolumn{1}{|r|}{\textbf{Fonte da publicação}} & Proceedings of the International Symposium on Low Power Electronics and Design					 \\ \hline
\multicolumn{1}{|r|}{\textbf{Editora}} & IEEE \\ \hline
\multicolumn{1}{|r|}{\textbf{Ano da publicação}} & 2011 \\ \hline \hline

\multicolumn{2}{|c|}{\textbf{Resumo da publicação}} \\ \hline
\multicolumn{2}{|p{21cm}|}{Os autores desenvolvem um \textit{framework} com o objetivo de melhor realizar o chaveamento entre performance e consumo de energia do processador. As configurações ótimas do processador são definidas de acordo com cada etapa de execução da tarefa. Além disso, esse \textit{framework} aplica técnicas de otimização sobre a alocação de memória da aplicação, visando diminuir \textit{overheads} de IO (em inglês, \textit{Input and Output}). Dessa forma, todas as características e comportamento da aplicação são analisados tando do ponto de vista inter quanto intra tarefa. Os resultados dessa análise, resultam na otimização de energia em tempo de execução de acordo com o comportamento da aplicação. Os resultados experimentais utilizando um sistema de vídeo conferencia conseguiram reduzir em média o consumo de energia em 44.9\% em comparação com outros estudos de caso criados pelo próprio autor.} \\ \hline \hline

\multicolumn{1}{|r|}{\textbf{Método(s) utilizados}} & DVFS Intra-Task; DVFS Inter-Task; Control Flow Graph (CFG); \\ \hline
\multicolumn{1}{|r|}{\textbf{Ferramenta(s)}} & {Os autores implementaram um \textit{framework}, mas não deixa claro o nome da ferramenta criada. \newline


Lista dos \textit{benchmarks} utilizados: \newline
- Xvid MPEG-4 video codec \newline
- FFmpeg library } \\ \hline \hline

\multicolumn{2}{|c|}{\textbf{Impacto (positivo x negativo)}} \\ \hline
\multicolumn{2}{|p{21cm}|}{O artigo apresentou como ponto positivo uma metodologia que visa realizar o melhor gerenciamento entre performance e consumo de energia do processador, além disso aplica técnicas de otimização para melhorar o processo de alocação de memória da aplicação. Como pontos negativos, os autores não mostraram nenhum estudo sobre os \textit{overheads} inseridos pelo \textit{framework}.} \\ \hline \hline

\multicolumn{2}{|c|}{\textbf{Validação do método}} \\ \hline
\multicolumn{2}{|p{21cm}|}{A validação da abordagem ocorreu por meio do uso de casos de testes criados pelos próprios autores.} \\ \hline \hline

\multicolumn{2}{|c|}{\textbf{Limitações do método}} \\ \hline
\multicolumn{2}{|p{21cm}|}{A metodologia não mostra estudos detalhados sobre os \textit{overheads} gerados dentro das tarefas.} \\ \hline \hline

\multicolumn{2}{|c|}{\textbf{Integração de métodos}} \\ \hline
\multicolumn{2}{|p{21cm}|}{Não se aplica} \\ \hline \hline

\multicolumn{2}{|c|}{\textbf{Modo de aplicação do método}} \\ \hline
\multicolumn{2}{|p{21cm}|}{Híbrida} \\ \hline \hline

\multicolumn{2}{|c|}{\textbf{Perspectivas futuras}} \\ \hline
\multicolumn{2}{|p{21cm}|}{Os autores pretendem estender o \textit{framework} para ambientes com mais de um processador.} \\ \hline \hline

\multicolumn{2}{|c|}{\textbf{Comentários}} \\ \hline
\multicolumn{2}{|p{21cm}|}{O artigo mostra indícios de que dá suporte a preempções, mas não deixa claro no artigo como é feito, e não dá suporte a recursos compartilhados.} \\ \hline
\end{tabular}
}
\end{table}

\newpage
\begin{table}[htp] 
\centering
\caption{Dados extraído da publicação de \cite{Yuan_2011_a}.}
\resizebox{\columnwidth}{!}{%
\begin{tabular}{|p{3.8cm}|p{16cm}|}
\hline
\multicolumn{1}{|r|}{\textbf{ID}} & Yuan\_2011\_a \\ \hline
\multicolumn{1}{|r|}{\textbf{Título}} & TALk: A temperature-aware leakage minimization technique for real-time systems \\ \hline
\multicolumn{1}{|r|}{\textbf{Autores}} & Yuan L., Leventhal S.R., Gu J., Qu G. \\ \hline
\multicolumn{1}{|r|}{\textbf{Palavras-chave}} & Dynamic voltage scaling; low power; real-time systems; temperature-aware design \\ \hline
\multicolumn{1}{|r|}{\textbf{Fonte da publicação}} & IEEE Transactions on Computer-Aided Design of Integrated Circuits and Systems \\ \hline
\multicolumn{1}{|r|}{\textbf{Editora}} & IEEE \\ \hline
\multicolumn{1}{|r|}{\textbf{Ano da publicação}} & 2011 \\ \hline \hline

\multicolumn{2}{|c|}{\textbf{Resumo da publicação}} \\ \hline
\multicolumn{2}{|p{21cm}|}{Os autores propõem um algoritmo de escalonamento intra-tarefa que visa diminuir a temperatura do processador e minimizar o consumo de energia em sistema de tempo real. Essa técnica foi chamada de \textit{TALk} (em inglês, \textit{Temperature- Aware Leakage}). A ideia básica do algoritmo é aumentar a frequência quando a temperatura do chip estiver baixa ou quando a carga de trabalho for alta e colocar o processador em baixo consumo de energia quando a temperatura do chip estiver alta ou com carga de trabalho leve. Para fazer isso, o algoritmo \textit{TALk} foi dividida em duas partes: (1) A \textit{Offline}, que usa métodos de programação dinâmica para alcançar os melhores níveis de economia de energia e de temperatura; (2) O \textit{Online} tem como objetivo determinar o modo de operação do processador com base na sua temperatura corrente e na quantidade de ciclos remanescentes das tarefas em execução. Com esse algoritmo os autores conseguiram melhorar a economia de energia em cerca de 18\% em comparação com a técnica DVS tradicional.} \\ \hline \hline

\multicolumn{1}{|r|}{\textbf{Método(s) utilizados}} & DVFS Intra-Task; \\ \hline
\multicolumn{1}{|r|}{\textbf{Ferramenta(s)}} & {Os autores não mencionaram nenhuma ferramenta específica, mas utilizaram 4 \textit{benchmarks} para geração dos 11 casos de teste utilizados para realização dos experimentos. \newline

Lista dos \textit{benchmarks} utilizados pelos autores:\newline
- MPEG4 \textit{media encoding} (Disponível em: \href{http://ieeexplore.ieee.org/xpl/articleDetails.jsp? arnumber=914596}{http://ieeexplore.ieee.org/xpl/articleDetails.jsp? arnumber=914596})\newline
- \textit{Hartstone suite} (Disponível em: \href{http://ieeexplore.ieee.org/xpl/articleDetails.jsp? arnumber=766516}{http://ieeexplore.ieee.org/xpl/articleDetails.jsp? arnumber=766516})\newline
- \textit{ADSL standard's initialization sequences} (Disponível em: \href{http://ieeexplore.ieee.org/xpl/ articleDetails.jsp?arnumber=953271}{http://ieeexplore.ieee.org/xpl/ articleDetails.jsp?arnumber=953271})\newline
- Os outros foram gerados artificialmente pelos autores.} \\ \hline \hline

\multicolumn{2}{|c|}{\textbf{Impacto (positivo x negativo)}} \\ \hline
\multicolumn{2}{|p{21cm}|}{O artigo apresentou como ponto positivo uma metodologia que visa diminuir o consumo de energia do processador e ao mesmo tempo controlar a sua temperatura. Como ponto negativo, os autores falam que para alguns \textit{benchmarks} o algoritmo \textit{offline} pode violar as restrições temporais de algumas tarefas.} \\ \hline \hline

\multicolumn{2}{|c|}{\textbf{Validação do método}} \\ \hline
\multicolumn{2}{|p{21cm}|}{A validação da abordagem ocorreu por meio do uso de \textit{benchmarks} presentes na literatura. Além disso, compara os resultados experimentais com outros trabalhos de referência na área de baixo consumo de energia.} \\ \hline \hline

\multicolumn{2}{|c|}{\textbf{Limitações do método}} \\ \hline
\multicolumn{2}{|p{21cm}|}{Os autores mencionam que o algoritmo \textit{offline} pode violar as restrições temporais de algumas tarefas.} \\ \hline \hline

\multicolumn{2}{|c|}{\textbf{Integração de métodos}} \\ \hline
\multicolumn{2}{|p{21cm}|}{Não se aplica} \\ \hline \hline

\multicolumn{2}{|c|}{\textbf{Modo de aplicação do método}} \\ \hline
\multicolumn{2}{|p{21cm}|}{Híbrida} \\ \hline \hline

\multicolumn{2}{|c|}{\textbf{Perspectivas futuras}} \\ \hline
\multicolumn{2}{|p{21cm}|}{Os autores pretendem melhorar o algoritmo \textit{offline} para resolver o problema de violação de restrições temporais de algumas tarefas.} \\ \hline \hline

\multicolumn{2}{|c|}{\textbf{Comentários}} \\ \hline
\multicolumn{2}{|p{21cm}|}{A metodologia proposta não mostra evidências, no artigo, de que fornece suporte para recursos compartilhados e / ou preempções.} \\ \hline
\end{tabular}
}
\end{table}

\newpage
\begin{table}[htp] 
\centering
\caption{Dados extraído da publicação de \cite{Cohen_2012_a}.}
\resizebox{\columnwidth}{!}{%
\begin{tabular}{|p{3.8cm}|p{16cm}|}
\hline
\multicolumn{1}{|r|}{\textbf{ID}} & Cohen\_2012\_a \\ \hline
\multicolumn{1}{|r|}{\textbf{Título}} & A Car Racing Based Strategy for the Dynamic Voltage and Frequency Scaling Technique \\ \hline
\multicolumn{1}{|r|}{\textbf{Autores}} & David Cohen, Eduardo Valentin, Raimundo Barreto, Horácio Oliveira, and Lucas Cordeiro \\ \hline
\multicolumn{1}{|r|}{\textbf{Palavras-chave}} & Dynamic voltage and frequency scaling; Energy consumption optimization; Low energy consumption; Real-time application; Real-time tasks; Worst-case execution; Buffer storage; Energy utilization; Industrial electronics \\ \hline
\multicolumn{1}{|r|}{\textbf{Fonte da publicação}} & IEEE International Symposium on Industrial Electronics \\ \hline
\multicolumn{1}{|r|}{\textbf{Editora}} & IEEE \\ \hline
\multicolumn{1}{|r|}{\textbf{Ano da publicação}} & 2012 \\ \hline \hline

\multicolumn{2}{|c|}{\textbf{Resumo da publicação}} \\ \hline
\multicolumn{2}{|p{21cm}|}{Os autores apresentam uma nova política de escalonamento de tarefas de tempo real, que leva em consideração preempções. Esse nova metodologia consegue economizar energia mesmo executando as tarefas no seu pior caso. Em resumo, o autor faz uma analogia entre escalonador e suas tarefas, com uma corrida de carros, onde o objetivo da corrida é que todos os carros (uma analogia as tarefas) cheguem juntos no final, utilizando as menores velocidades (uma analogia a tensões e frequências do processador) sem que nenhuma premissa temporal seja violada. Essa nova política de escalonamento foi experimentada apenas em ambiente simulado e utilizando casos de teste gerados pelos próprios autores.} \\ \hline \hline

\multicolumn{1}{|r|}{\textbf{Método(s) utilizados}} & DVFS Intra-Task; Control flow graph (CFG); \\ \hline
\multicolumn{1}{|r|}{\textbf{Ferramenta(s)}} & {O autor não utilizou nenhuma ferramenta, mas nas referências é citada a ferramenta COPPER Framework.} \\ \hline \hline

\multicolumn{2}{|c|}{\textbf{Impacto (positivo x negativo)}} \\ \hline
\multicolumn{2}{|p{21cm}|}{O artigo apresentou como ponto positivo uma nova política de escalonamento para aplicações de tempo real, dando suporte a preempções. Como impacto negativo, a metodologia foi valida apenas com simulações, não foi utilizado nenhum \textit{benchmark} ou casos de testes reais para validação do método proposto.} \\ \hline \hline

\multicolumn{2}{|c|}{\textbf{Validação do método}} \\ \hline
\multicolumn{2}{|p{21cm}|}{A validação da abordagem ocorreu por meio de exemplos propostos pelos próprios autores, onde estes exemplos exploraram várias tarefas em execução e preempções.} \\ \hline \hline

\multicolumn{2}{|c|}{\textbf{Limitações do método}} \\ \hline
\multicolumn{2}{|p{21cm}|}{O método proposto foi aplicado apenas em ambiente simulado. Talvez seja necessário fazer um estudo mais detalhado para saber o real impacto dos \textit{overheads} gerados pelo uso da metodologia.} \\ \hline \hline

\multicolumn{2}{|c|}{\textbf{Integração de métodos}} \\ \hline
\multicolumn{2}{|p{21cm}|}{Não se aplica} \\ \hline \hline

\multicolumn{2}{|c|}{\textbf{Modo de aplicação do método}} \\ \hline
\multicolumn{2}{|p{21cm}|}{Online} \\ \hline \hline

\multicolumn{2}{|c|}{\textbf{Perspectivas futuras}} \\ \hline
\multicolumn{2}{|p{21cm}|}{Os autores pretendem realizar experimentos com estudos de casos reais da industria e implementar o método proposto em uma plataforma real, através da inclusão de uma nova camada entre a aplicação e o sistema operacional.} \\ \hline \hline

\multicolumn{2}{|c|}{\textbf{Comentários}} \\ \hline
\multicolumn{2}{|p{21cm}|}{A metodologia deixa clara que dá suporte a preempções. Quanto aos recursos compartilhados, eles são previstos apenas no teste de escalonabilidade. No entanto, não mostram nos resultados experimentos modelos de tarefas que compartilhem recursos entre si.} \\ \hline
\end{tabular}
}
\end{table}

\newpage
\begin{table}[htp] 
\centering
\caption{Dados extraído da publicação de \cite{Seo_Seo_Kim_2012_a}.}
\resizebox{\columnwidth}{!}{%
\begin{tabular}{|p{3.8cm}|p{16cm}|}
\hline
\multicolumn{1}{|r|}{\textbf{ID}} & Seo\_Seo\_Kim\_2012\_a \\ \hline
\multicolumn{1}{|r|}{\textbf{Título}} & Algorithms for combined inter- and intra-task dynamic voltage scaling \\ \hline
\multicolumn{1}{|r|}{\textbf{Autores}} & Seo H., Seo J., Kim T. \\ \hline
\multicolumn{1}{|r|}{\textbf{Palavras-chave}} & dynamic voltage scaling (DVS); embedded systems; power saving \\ \hline
\multicolumn{1}{|r|}{\textbf{Fonte da publicação}} & Computer Journal \\ \hline
\multicolumn{1}{|r|}{\textbf{Editora}} & Oxford University Press on behalf of The British Computer Society \\ \hline
\multicolumn{1}{|r|}{\textbf{Ano da publicação}} & 2012 \\ \hline \hline

\multicolumn{2}{|c|}{\textbf{Resumo da publicação}} \\ \hline
\multicolumn{2}{|p{21cm}|}{Os autores apresentam uma técnica de baixo consumo de energia que se baseia na combinação simultânea entre inter-tarefa e intra-tarefa, também chamada DVS combinado (em inglês, \textit{Combined} DVS - CDVS). Essa nova abordagem leva em consideração o estado do sistema dormindo (em inglês, \textit{Sleep State} - CDVS-S) e não dormindo (em inglês, \textit{No Sleep State} - CDVS-NS). Ela consiste basicamente de 4 etapas: (1) Aplicar a técnica CDVS-NS para determinar os intervalos de execução das tarefas de modo que o consumo total de energia seja minimizado, sem estados de dormindo (\textit{sleep state}); (2) Realizar a análise estática dos blocos de código da tarefa, a fim de identificar os tempos ociosos e as tensões e frequências que deverão ser utilizadas; (3) Combinar os intervalos de tempo salvos na segunda etapa com o maior tempo possível no qual o sistema possa estar no estado ocioso (\textit{idle state}) de forma eficiente; e por último (4) Monitorar dinamicamente todas as instâncias das tarefas em execução que concluíram sua execução antes do prazo final e, em seguida, irá decidir se coloca a tarefa em estado ociosa (\textit{idle state}) ou em estado de dormindo (\textit{sleep state}), dependendo do que for mais econômico energeticamente. Em geral, essa metodologia conseguiu reduzir o consumo de energia em média de 7\% com a técnica CDVS-S e de 12\% com a técnica CDVS-NS em comparação com outros trabalhos presentes na literatura.} \\ \hline \hline

\multicolumn{1}{|r|}{\textbf{Método(s) utilizados}} & DVFS \textit{Intra-Task}; DVFS \textit{Inter-Task}; \\ \hline
\multicolumn{1}{|r|}{\textbf{Ferramenta(s)}} & {Os autores utilizaram as ferramentas TGFF v 3.0 (usando C++ e suas bibliotecas padrões) e MiBench \textit{benchmarks}.} \\ \hline \hline

\multicolumn{2}{|c|}{\textbf{Impacto (positivo x negativo)}} \\ \hline
\multicolumn{2}{|p{21cm}|}{O artigo apresentou como ponto positivo uma nova metodologia que combina as técnicas inter e intra tarefas para melhor gerenciar os estados de dormindo (\textit{Sleep State}) e não dormindo (\textit{No Sleep State}) do sistema. Como ponto negativo, os autores não mostraram estudos para avaliar os \textit{overheads} ocasionados para realizar a troca de contexto entre o modo dormindo e não dormindo.} \\ \hline \hline

\multicolumn{2}{|c|}{\textbf{Validação do método}} \\ \hline
\multicolumn{2}{|p{21cm}|}{A validação da metodologia ocorreu por meio de \textit{benchmarks}, provas matemáticas e comparações dos resultados com outros trabalhos presentes na literatura.} \\ \hline \hline

\multicolumn{2}{|c|}{\textbf{Limitações do método}} \\ \hline
\multicolumn{2}{|p{21cm}|}{Os autores não mostraram um estudo para avaliar os \textit{overheads} ocasionados para realizar a troca de contexto entre o sistema no modo dormindo e não dormindo.} \\ \hline \hline

\multicolumn{2}{|c|}{\textbf{Integração de métodos}} \\ \hline
\multicolumn{2}{|p{21cm}|}{Foi realizado a integração das técnicas DVFS inter e intra tarefas.} \\ \hline \hline

\multicolumn{2}{|c|}{\textbf{Modo de aplicação do método}} \\ \hline
\multicolumn{2}{|p{21cm}|}{Híbrido.} \\ \hline \hline

\multicolumn{2}{|c|}{\textbf{Perspectivas futuras}} \\ \hline
\multicolumn{2}{|p{21cm}|}{Os autores pretendem criar um modelo de economia de energia que leve em consideração outros componentes do sistema com as memórias, interfaces e comportamento das tarefas.} \\ \hline \hline

\multicolumn{2}{|c|}{\textbf{Comentários}} \\ \hline
\multicolumn{2}{|p{21cm}|}{A metodologia proposta não mostra evidências, no artigo, de que fornece suporte para recursos compartilhados e / ou preempções.} \\ \hline
\end{tabular}
}
\end{table}

\newpage
\begin{table}[htp] 
\centering
\caption{Dados extraído da publicação de \cite{Awan_Petters_2012_a}.}
\resizebox{\columnwidth}{!}{%
\begin{tabular}{|p{3.8cm}|p{16cm}|}
\hline
\multicolumn{1}{|r|}{\textbf{ID}} & Awan\_Petters\_2012\_a \\ \hline
\multicolumn{1}{|r|}{\textbf{Título}} & Online Intra-Task Device Scheduling for Hard Real-Time Systems \\ \hline
\multicolumn{1}{|r|}{\textbf{Autores}} & Muhammad Ali Awan, Stefan M. Petters \\ \hline
\multicolumn{1}{|r|}{\textbf{Palavras-chave}} & Battery life; Device scheduling; Energy gain; Hard real-time systems; I/O Device; Large parts; Power saving mechanism; Schedulability; Technology enhancements; Embedded systems; Energy resources; Energy utilization; Interactive computer systems; Scheduling algorithms; Real time systems \\ \hline
\multicolumn{1}{|r|}{\textbf{Fonte da publicação}} & 7th IEEE International Symposium on Industrial Embedded Systems, SIES 2012 - Conference Proceedings \\ \hline
\multicolumn{1}{|r|}{\textbf{Editora}} & IEEE \\ \hline
\multicolumn{1}{|r|}{\textbf{Ano da publicação}} & 2012 \\ \hline \hline

\multicolumn{2}{|c|}{\textbf{Resumo da publicação}} \\ \hline
\multicolumn{2}{|p{21cm}|}{Os autores propõem um algoritmo (online) de escalonamento intra-tarefa, cuja principal funcionalidade é ligar e desligar dispositivos do hardware, permitindo que eles sejam utilizados somente quando necessário. Essa metodologia se aplica a sistemas de tempo real crítico e funciona basicamente explorando os tempos de folga entre as execuções das tarefas, a fim de realizar o melhor gerenciamento dos acionamentos e desligamento dos dispositivos,  melhorando assim a performance de economia de energia do sistema. Os experimentos mostram um ganho de economia de energia acima dos 90\% em comparação com outras técnicas presentes na literatura.} \\ \hline \hline

\multicolumn{1}{|r|}{\textbf{Método(s) utilizados}} & DVFS \textit{Intra-Task}; \\ \hline
\multicolumn{1}{|r|}{\textbf{Ferramenta(s)}} & {Os autores utilizaram o simulador SPARTS V 2.0 (\textit{Simulator for Power Aware and Real-Time System}).} \\ \hline \hline

\multicolumn{2}{|c|}{\textbf{Impacto (positivo x negativo)}} \\ \hline
\multicolumn{2}{|p{21cm}|}{O artigo apresentou como impacto positivo uma metodologia para realizar o melhor gerenciamento dos dispositivos do sistema sobre demanda, permitindo assim acioná-los somente quando necessário. No entanto, o impacto negativo dessa abordagem está nos \textit{overheads} adicionados nas premissas temporais de cada tarefa.} \\ \hline \hline

\multicolumn{2}{|c|}{\textbf{Validação do método}} \\ \hline
\multicolumn{2}{|p{21cm}|}{A validação da abordagem ocorreu por meio de exemplos propostos pelos próprios autores.} \\ \hline \hline

\multicolumn{2}{|c|}{\textbf{Limitações do método}} \\ \hline
\multicolumn{2}{|p{21cm}|}{O método proposto foi aplicado apenas em ambiente simulado, prevê o uso de um único dispositivo e não leva em consideração o compartilhamento de dispositivos entre tarefas.} \\ \hline \hline

\multicolumn{2}{|c|}{\textbf{Integração de métodos}} \\ \hline
\multicolumn{2}{|p{21cm}|}{Não se aplica} \\ \hline \hline

\multicolumn{2}{|c|}{\textbf{Modo de aplicação do método}} \\ \hline
\multicolumn{2}{|p{21cm}|}{Online} \\ \hline \hline

\multicolumn{2}{|c|}{\textbf{Perspectivas futuras}} \\ \hline
\multicolumn{2}{|p{21cm}|}{Os autores pretendem melhorar a aplicabilidade deste método em sistemas mais versáteis, permitindo que seja possível compartilhar dispositivos entre tarefas.} \\ \hline \hline

\multicolumn{2}{|c|}{\textbf{Comentários}} \\ \hline
\multicolumn{2}{|p{21cm}|}{A metodologia proposta dá suporte a recursos compartilhados e em relação ao suporte a preempções, o artigo mostra algumas deduções formais, mas não deixa claro se dá suporte.} \\ \hline
\end{tabular}
}
\end{table}

\end{document}